\documentclass[,man,floatsintext]{apa6}
\usepackage{lmodern}
\usepackage{amssymb,amsmath}
\usepackage{ifxetex,ifluatex}
\usepackage{fixltx2e} 
\ifnum 0\ifxetex 1\fi\ifluatex 1\fi=0 
  \usepackage[T1]{fontenc}
  \usepackage[utf8]{inputenc}
\else 
  \ifxetex
    \usepackage{mathspec}
  \else
    \usepackage{fontspec}
  \fi
  \defaultfontfeatures{Ligatures=TeX,Scale=MatchLowercase}
\fi
\IfFileExists{upquote.sty}{\usepackage{upquote}}{}
\IfFileExists{microtype.sty}{%
\usepackage{microtype}
\UseMicrotypeSet[protrusion]{basicmath} 
}{}
\usepackage{hyperref}
\PassOptionsToPackage{usenames,dvipsnames}{color} 
\hypersetup{unicode=true,
            pdftitle={How to capitalize on a priori contrasts in linear (mixed) models: A tutorial},
            pdfauthor={Daniel J. Schad, Shravan Vasishth, Sven Hohenstein, \& Reinhold Kliegl},
            pdfkeywords={Contrasts, null hypothesis significance testing, linear models, a priori hypotheses},
            colorlinks=true,
            linkcolor=Maroon,
            citecolor=Blue,
            urlcolor=blue,
            breaklinks=true}
\usepackage{color}
\usepackage{fancyvrb}

\DefineVerbatimEnvironment{Highlighting}{Verbatim}{commandchars=\\\{\}}
\usepackage{framed}
\definecolor{shadecolor}{RGB}{248,248,248}
\newenvironment{Shaded}{\begin{snugshade}}{\end{snugshade}}

\newcommand{\CommentTok}[1]{\textcolor[rgb]{0.56,0.35,0.01}{\textit{#1}}}

\newcommand{\ControlFlowTok}[1]{\textcolor[rgb]{0.13,0.29,0.53}{\textbf{#1}}}
\newcommand{\DataTypeTok}[1]{\textcolor[rgb]{0.13,0.29,0.53}{#1}}
\newcommand{\DecValTok}[1]{\textcolor[rgb]{0.00,0.00,0.81}{#1}}

\newcommand{\FloatTok}[1]{\textcolor[rgb]{0.00,0.00,0.81}{#1}}

\newcommand{\KeywordTok}[1]{\textcolor[rgb]{0.13,0.29,0.53}{\textbf{#1}}}
\newcommand{\NormalTok}[1]{#1}
\newcommand{\OperatorTok}[1]{\textcolor[rgb]{0.81,0.36,0.00}{\textbf{#1}}}
\newcommand{\OtherTok}[1]{\textcolor[rgb]{0.56,0.35,0.01}{#1}}

\newcommand{\StringTok}[1]{\textcolor[rgb]{0.31,0.60,0.02}{#1}}

\usepackage{graphicx,grffile}
\makeatletter
\def\maxwidth{\ifdim\Gin@nat@width>\linewidth\linewidth\else\Gin@nat@width\fi}
\def\maxheight{\ifdim\Gin@nat@height>\textheight\textheight\else\Gin@nat@height\fi}
\makeatother
\setkeys{Gin}{width=\maxwidth,height=\maxheight,keepaspectratio}
\IfFileExists{parskip.sty}{%
\usepackage{parskip}
}{
\setlength{\parindent}{0pt}
\setlength{\parskip}{6pt plus 2pt minus 1pt}
}
\providecommand{\tightlist}{%
  \setlength{\itemsep}{0pt}\setlength{\parskip}{0pt}}
\ifx\paragraph\undefined\else
\let\oldparagraph\paragraph
\renewcommand{\paragraph}[1]{\oldparagraph{#1}\mbox{}}
\fi
\ifx\subparagraph\undefined\else
\let\oldsubparagraph\subparagraph
\renewcommand{\subparagraph}[1]{\oldsubparagraph{#1}\mbox{}}
\fi

\let\rmarkdownfootnote\footnote%
\def\footnote{\protect\rmarkdownfootnote}

  \title{How to capitalize on \emph{a priori} contrasts in linear (mixed) models: A tutorial}
    \author{Daniel J. Schad\textsuperscript{1}, Shravan Vasishth\textsuperscript{1}, Sven Hohenstein\textsuperscript{1}, \& Reinhold Kliegl\textsuperscript{1}}
    \date{}
  
\shorttitle{A tutorial on contrast coding}
\affiliation{
\vspace{0.5cm}
\textsuperscript{1} University of Potsdam, Germany}
\usepackage{csquotes}
\usepackage{upgreek}
\usepackage{longtable}
\usepackage{lscape}
\usepackage{multirow}
\usepackage{tabularx}
\usepackage[flushleft]{threeparttable}
\usepackage{threeparttablex}

\makeatletter
\newcommand\LastLTentrywidth{1em}
\newlength\longtablewidth
\newcommand{\getlongtablewidth}{\begingroup \ifcsname LT@\roman{LT@tables}\endcsname \global\longtablewidth=0pt \renewcommand{\LT@entry}[2]{\global\advance\longtablewidth by ##2\relax\gdef\LastLTentrywidth{##2}}\@nameuse{LT@\roman{LT@tables}} \fi \endgroup}
\usepackage{amsmath}
\usepackage[american]{babel}
\usepackage[utf8]{inputenc}
\usepackage[T1]{fontenc}
\usepackage[backend=biber,useprefix=true,style=apa,url=false,doi=false,sorting=nyt,eprint=false]{biblatex}
\DeclareLanguageMapping{american}{american-apa}
\usepackage{fancyvrb}
\usepackage{newfloat}
\usepackage{graphicx}
\usepackage{caption}
\usepackage{listings}
\usepackage{glossaries}
\makeglossaries
\usepackage[svgnames]{xcolor}
\DeclareFloatingEnvironment[fileext=los, listname=List of Schemes, name=Listing, placement=!htbp, within=section]{listing}
\usepackage{placeins}

\authornote{

Correspondence concerning this article should be addressed to Daniel J. Schad, University of Potsdam, Germany. Phone: +49-331-977-2309. Fax: +49-331-977-2087. E-mail: \href{mailto:danieljschad@gmail.com}{\nolinkurl{danieljschad@gmail.com}}}

\abstract{
Factorial experiments in research on memory, language, and in other areas are often analyzed using analysis of variance (ANOVA). However, for effects with more than one numerator degrees of freedom, e.g., for experimental factors with more than two levels, the ANOVA omnibus F-test is not informative about the source of a main effect or interaction. Because researchers typically have specific hypotheses about which condition means differ from each other, a priori contrasts (i.e., comparisons planned before the sample means are known) between specific conditions or combinations of conditions are the appropriate way to represent such hypotheses in the statistical model. Many researchers have pointed out that contrasts should be ``tested instead of, rather than as a supplement to, the ordinary `omnibus' F test'' (Hays, 1973, p.~601). In this tutorial, we explain the mathematics underlying different kinds of contrasts (i.e., treatment, sum, repeated, polynomial, custom, nested, interaction contrasts), discuss their properties, and demonstrate how they are applied in the R System for Statistical Computing (R Core Team, 2018). In this context, we explain the generalized inverse which is needed to compute the coefficients for contrasts that test hypotheses that are not covered by the default set of contrasts. A detailed understanding of contrast coding is crucial for successful and correct specification in linear models (including linear mixed models). Contrasts defined a priori yield far more useful confirmatory tests of experimental hypotheses than standard omnibus F-test. Reproducible code is available from \url{https://osf.io/7ukf6/}.

}

\begin{document}
\maketitle

\hypertarget{introduction}{%
\section{Introduction}\label{introduction}}

Whenever an experimental factor comprises more than two levels, analysis of variance (ANOVA) F-statistics provide very little information about the source of an effect or interaction involving this factor. \textcolor{black}{For example, let's assume an experiment with three groups of subjects. Let's also assume that an ANOVA shows that the main effect of group is significant. This of course leaves unclear which groups differ from each other and how they differ. However, scientists typically} have \emph{a priori} expectations about the pattern of means. That is, we usually have specific expectations about which groups differ from each other.
One potential strategy is to follow up on these results using t-tests. However, this approach does not consider all the data in each test and therefore loses statistical power, does not generalize well to more complex models (e.g., linear mixed-effects models), and is subject to problems of multiple comparisons. In this paper, we will show how to test specific hypotheses directly in a regression model, which gives much more control over the analysis.
Specifically, we show how planned comparisons between specific conditions \textcolor{black}{(groups)} or clusters of conditions, \textcolor{black}{are} implemented as contrasts\textcolor{black}{. This is} a very effective way to align expectations with the statistical model. Indeed, if \textcolor{black}{planned comparisons, implemented in contrasts, are} defined \emph{a priori}, and are not defined after the results are known, planned comparisons should be \enquote{tested \emph{instead of}, rather than as a supplement to, the ordinary \enquote{omnibus} F test.} (Hays, 1973, p. 601).

Every contrast consumes exactly one degree of freedom. Every degree of freedom in the ANOVA source-of-variance table can be spent to test a specific hypothesis about a difference between means or a difference between clusters of means.\\
\textcolor{black}{Linear mixed-effects models} (LMMs) (Baayen, Davidson, \& Bates, 2008; Bates, Kliegl, Vasishth, \& Baayen, 2015; Bates, Maechler, Bolker, \& Walker, 2014; Kliegl, Masson, \& Richter, 2010; Pinheiro \& Bates, 2000) \textcolor{black}{are a great tool and represent an important development in statistical practice in psychology and linguistics. LMMs are often taken to replace more traditional ANOVA analyses. However, LMMs also present some challenges. One key challenge is about how to incorporate categorical effects from factors with discrete levels into LMMs. One approach to analyzing factors is to do model comparison; this is akin to the ANOVA omnibus test, and again leaves it unclear which groups differ from which others.
\textcolor{black}{An alternative approach is to base analyses on contrasts, which allow us to code factors as independent variables in linear regression models.}
The present paper explains how to understand and use contrast coding to test particular comparisons between conditions of your experiment (for a Glossary of key terms see Appendix}~\ref{app:Glossary}).
\textcolor{black}{Such knowledge about contrast specification is required if analysis of factors is to be based on LMMs} instead of ANOVAs. Arguably, in the R System for Statistical Computing (R Core Team, 2018), an understanding of contrast specification is \textcolor{black}{therefore} a necessary pre-condition for the proper use of LMMs.

To model differences between categories/groups/cells/conditions, regression models (such as multiple regression, logistic regression and linear mixed models) specify a set of contrasts (i.e., which groups are compared to which baselines or groups). There are several ways to specify such contrasts mathematically, and as discussed below, which of these is more useful depends on the hypotheses about the expected pattern of means. If the analyst does not provide the specification explicitly, R will pick a default specification on its own, which may not align very well with the hypotheses that the analyst intends to test. Therefore, LMMs effectively demand that the user implement planned comparisons---perhaps almost as intended by Hays (1973). Obtaining a statistic roughly equivalent to an ordinary \enquote{omnibus} F test requires the extra effort of model comparison. This tutorial provides a practical introduction to contrast coding for factorial experimental designs that will allow scientists to express their hypotheses within the statistical model.

\hypertarget{prerequisites-for-this-tutorial}{%
\subsection{Prerequisites for this tutorial}\label{prerequisites-for-this-tutorial}}

The reader is assumed to be familiar with R, and with the foundational ideas behind frequentist statistics, specifically null hypothesis significance testing.
Some notational conventions:
True parameters are referred to with Greek letters (e.g., \(\mu\), \(\beta\)), and estimates of these parameters have a hat (\(\hat\cdot\)) on the parameter (e.g., \(\hat\mu\), \(\hat\beta\)). The mean of several parameters is written as \(\mu\) or as \(\bar{\mu}\).

Example analyses of simulated data-sets are presented using linear models (LMs) in R. We use simulated data rather than real data-sets because this allows full control over the dependent variable.

To guide the reader, we provide a preview of the structure of the paper:

\begin{itemize}
\item
First, basic concepts of different contrasts are explained, \textcolor{black}{using a factor with two levels to explain the concepts}. After a demonstration of the default  contrast setting in R, \textsc{treatment contrasts}, a commonly used contrast for factorial designs, and \textsc{sum contrasts}, are illustrated. 
\item 
We then demonstrate how the generalized inverse (see Fieller, 2016, sec. 8.3) is used to convert what we call the  hypothesis matrix to create a contrast matrix. We demonstrate this workflow of going from a hypothesis matrix to a contrast matrix for \textsc{sum contrasts} \textcolor{black}{using an example data-set involving one factor with three levels. We also demonstrate the workflow for} \textsc{repeated contrasts} \textcolor{black}{using an example data-set involving one factor with four levels}.
\item 
After showing how contrasts implement hypotheses as predictors in a multiple regression model, \textcolor{black}{we introduce \textsc{polynomial contrasts} and \textsc{custom contrasts}}. 
\item 
\textcolor{black}{Then we discuss what makes a good contrast. Here, we introduce two important concepts: centering and orthogonality of contrasts, and their implications for contrast coding.}
\item 
We provide additional information about the generalized inverse, and how it is used to switch between the hypothesis matrix and the contrast matrix. \textcolor{black}{The section on the matrix notation (Hypothesis matrix and contrast matrix in matrix form) can optionally be skipped on first reading.}
\item 
\textcolor{black}{Then we discuss an effect size measure for contrasts.}
\item
The next section \textcolor{black}{discusses designs with two factors. We} compare regression models with analysis of variance (ANOVA) in simple $2 \times 2$ designs, and look at contrast centering, nested effects, and at a priori interaction contrasts. 
\end{itemize}

Throughout the tutorial, we show how contrasts are implemented and applied in R and how they relate to hypothesis testing via the generalized inverse.

\begin{figure}

{\centering \includegraphics[width=44.3in]{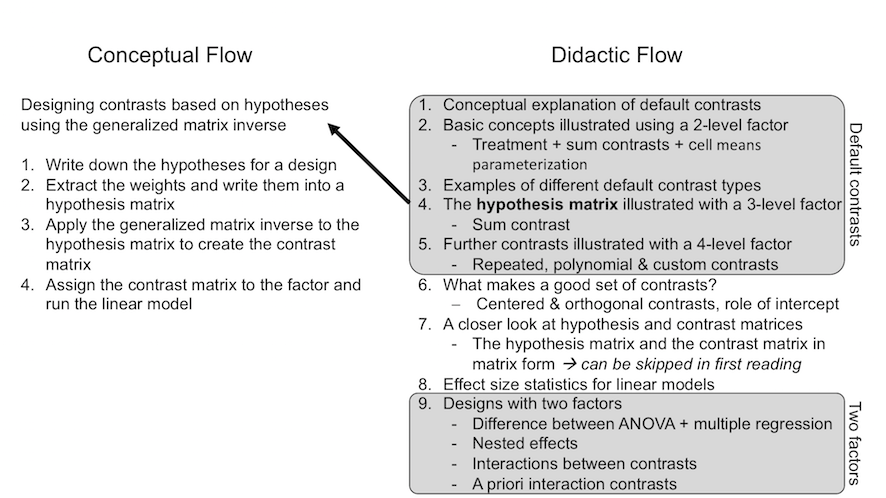} 

}

\caption{Conceptual and didactic flow. The left panel, conceptual flow, shows how to use the hypothesis matrix and the generalized inverse to construct custom contrast matrices for a given set of hypotheses. The right panel, didactic flow, shows an overview of the sections of this paper. The first grey box indicates the sections that introduce a set of default contrasts (treatment, sum, repeated, polynomial, and custom contrasts); the second grey box indicates sections that introduce contrasts for designs with two factors.}\label{fig:Flow}
\end{figure}

\textcolor{black}{An overview of the conceptual and didactic flow is provided in Figure} \ref{fig:Flow}.

\FloatBarrier

\hypertarget{conceptual-explanation-of-default-contrasts}{%
\section{Conceptual explanation of default contrasts}\label{conceptual-explanation-of-default-contrasts}}

What are examples for different contrast specifications? One contrast in widespread use is the \textsc{treatment contrast}.
As suggested by its name, the \textcolor{black}{\textsc{treatment contrast}} is often used in intervention studies, where one or several intervention groups receive some treatment, which are compared to a control group. For example, two treatment groups may obtain (a) psychotherapy and (b) pharmacotherapy, and they may be compared to \textcolor{black}{(c)} a control group of patients waiting for treatment. This implies one factor with three levels. In this setting, a \textsc{treatment contrast} for this factor makes two comparisons: it tests \textcolor{black}{(1)} whether the psychotherapy group is better than the control group, and \textcolor{black}{(2)} whether the pharmacotherapy group is better than the control group. That is, each treatment condition is compared to the same control group or baseline condition. An example in research on memory and language may be a priming study, where two different kinds of priming conditions (e.g., phonological versus orthographic priming) are each compared to a control condition without priming.

A second contrast of widespread use is the \textsc{sum contrast}. This contrast compares each tested group not against a baseline / control condition, but instead to the average response across all groups. Consider an example where three different priming conditions are compared to each other, such as orthographic, phonological, and semantic priming. The question of interest may be whether two of the priming conditions (e.g., orthographic and phonological priming) elicit stronger responses than the average response across all conditions. This could be done \textcolor{black}{with} \textsc{sum contrasts}: The first contrast would compare orthographic priming with the average response, and the second contrast would compare phonological priming with the average response. \textsc{Sum contrasts} also have an important role in factors with two levels, where they simply test the difference between those two factor levels (e.g., the difference between phonological versus orthographic priming).

A third contrast coding is \textsc{repeated contrasts}. These are probably less often used in empirical studies, but are arguably the contrast of highest relevance for research in psychology and cognitive science. \textsc{Repeated contrasts} successively test neighboring factor levels against each other. For example, a study may manipulate the frequency of some target words into three categories of \enquote{low frequency}, \enquote{medium frequency}, and \enquote{high frequency}. What may be of interest in the study is whether low frequency words differ from medium frequency words, and whether medium frequency words differ from high frequency words. \textsc{Repeated contrasts} test exactly these differences between neighoring factor levels.

A fourth type of contrast is \textsc{polynomial contrasts}. These are useful when trends are of interest that span across multiple factor levels. In the example with different levels of word frequency, a simple hypothesis may state that the response increases with increasing levels of word frequency. That is, one may expect that a response increases from low to medium frequency words \textcolor{black}{by the same magnitude} as it increases from medium to high frequency words. That is, a linear trend is expected. Here, it is possible to test a quadratic trend of word frequency; for example, when the effect is expected to be \textcolor{black}{larger or smaller between medium and high frequency words compared to the effect between low and medium frequency words.}

One additional option for contrast coding is provided by \textsc{Helmert contrasts}. In an example with three factor levels, for \textsc{Helmert contrasts} the first contrast codes the difference between the first two factor levels, and the second contrast codes the difference between the mean of the first two levels and the third level. (In cases of a four-level factor, the third contrast tests the difference between (i) the average of the first three levels and (ii) the fourth level.) An example for the use of \textsc{Helmert contrasts} is a priming paradigm with the three experimental conditions \enquote{valid prime}, \enquote{invalid prime 1}, and \enquote{invalid prime 2}. The first contrast may here test the difference between conditions \enquote{invalid prime 1} and \enquote{invalid prime 2}. The second contrast could then test the difference between valid versus invalid conditions. This coding would provide maximal power to test the difference between valid and invalid conditions, as it pools across both invalid conditions for the comparison.

\hypertarget{basic-concepts-illustrated-using-a-two-level-factor}{%
\section{Basic concepts illustrated using a two-level factor}\label{basic-concepts-illustrated-using-a-two-level-factor}}

Consider the simplest case: suppose we want to compare the means of a dependent variable (DV) such as response times between two groups of subjects. R can be used to simulate data for such an example using the function \texttt{mixedDesign()} \textcolor{black}{(for details regarding this function, see Appendix}~\ref{app:mixedDesign}). The simulations assume longer response times in condition F1 (\(\mu_1 = 0.8\) sec) than F2 (\(\mu_2 = 0.4\) sec). The data from the \(10\) simulated subjects \textcolor{black}{are aggregated} and summary statistics \textcolor{black}{are computed} for the two groups.

\begin{Shaded}
\begin{Highlighting}[]
\KeywordTok{library}\NormalTok{(dplyr)}
\CommentTok{# load mixedDesign function for simulating data}
\KeywordTok{source}\NormalTok{(}\StringTok{"functions/mixedDesign.v0.6.3.R"}\NormalTok{)}
\NormalTok{M <-}\StringTok{ }\KeywordTok{matrix}\NormalTok{(}\KeywordTok{c}\NormalTok{(}\FloatTok{0.8}\NormalTok{, }\FloatTok{0.4}\NormalTok{), }\DataTypeTok{nrow=}\DecValTok{2}\NormalTok{, }\DataTypeTok{ncol=}\DecValTok{1}\NormalTok{, }\DataTypeTok{byrow=}\OtherTok{FALSE}\NormalTok{)}
\KeywordTok{set.seed}\NormalTok{(}\DecValTok{1}\NormalTok{) }\CommentTok{# set seed of random number generator for replicability}
\NormalTok{simdat <-}\StringTok{ }\KeywordTok{mixedDesign}\NormalTok{(}\DataTypeTok{B=}\DecValTok{2}\NormalTok{, }\DataTypeTok{W=}\OtherTok{NULL}\NormalTok{, }\DataTypeTok{n=}\DecValTok{5}\NormalTok{, }\DataTypeTok{M=}\NormalTok{M,  }\DataTypeTok{SD=}\NormalTok{.}\DecValTok{20}\NormalTok{, }\DataTypeTok{long =} \OtherTok{TRUE}\NormalTok{) }
\KeywordTok{names}\NormalTok{(simdat)[}\DecValTok{1}\NormalTok{] <-}\StringTok{ "F"}  \CommentTok{# Rename B_A to F(actor)}
\KeywordTok{levels}\NormalTok{(simdat}\OperatorTok{$}\NormalTok{F) <-}\StringTok{ }\KeywordTok{c}\NormalTok{(}\StringTok{"F1"}\NormalTok{, }\StringTok{"F2"}\NormalTok{)}
\NormalTok{simdat}
\end{Highlighting}
\end{Shaded}

\begin{verbatim}
##     F id    DV
## 1  F1  1 0.997
## 2  F1  2 0.847
## 3  F1  3 0.712
## 4  F1  4 0.499
## 5  F1  5 0.945
## 6  F2  6 0.183
## 7  F2  7 0.195
## 8  F2  8 0.608
## 9  F2  9 0.556
## 10 F2 10 0.458
\end{verbatim}

\begin{Shaded}
\begin{Highlighting}[]
\KeywordTok{str}\NormalTok{(simdat)}
\end{Highlighting}
\end{Shaded}

\begin{verbatim}
## 'data.frame':    10 obs. of  3 variables:
##  $ F : Factor w/ 2 levels "F1","F2": 1 1 1 1 1 2 2 2 2 2
##  $ id: Factor w/ 10 levels "1","2","3","4",..: 1 2 3 4 5 6 7 8 9 10
##  $ DV: num  0.997 0.847 0.712 0.499 0.945 ...
\end{verbatim}

\begin{Shaded}
\begin{Highlighting}[]
\NormalTok{table1 <-}\StringTok{ }\NormalTok{simdat }\OperatorTok{%>%}\StringTok{ }\KeywordTok{group_by}\NormalTok{(F) }\OperatorTok{%>%}\StringTok{ }\CommentTok{# Table for main effect F}
\StringTok{   }\KeywordTok{summarize}\NormalTok{(}\DataTypeTok{N=}\KeywordTok{n}\NormalTok{(), }\DataTypeTok{M=}\KeywordTok{mean}\NormalTok{(DV), }\DataTypeTok{SD=}\KeywordTok{sd}\NormalTok{(DV), }\DataTypeTok{SE=}\NormalTok{SD}\OperatorTok{/}\KeywordTok{sqrt}\NormalTok{(N) )}
\NormalTok{(GM <-}\StringTok{  }\KeywordTok{mean}\NormalTok{(table1}\OperatorTok{$}\NormalTok{M)) }\CommentTok{# Grand Mean}
\end{Highlighting}
\end{Shaded}

\begin{verbatim}
## [1] 0.6
\end{verbatim}

\begin{figure}

{\centering \includegraphics{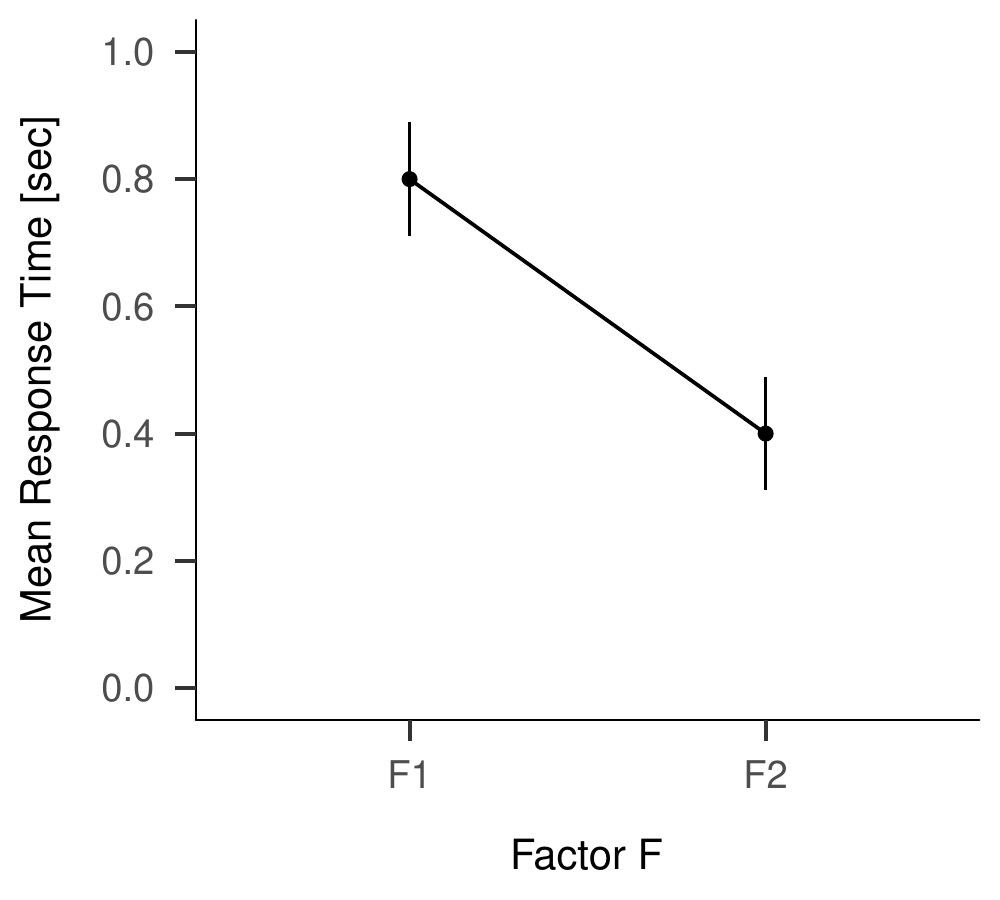} 

}

\caption{Means and standard errors of the simulated dependent variable (e.g., response times in seconds) in two conditions F1 and F2.}\label{fig:Fig1Means}
\end{figure}

\begin{table}[b]
\begin{center}
\begin{threeparttable}
\caption{\label{tab:table1}Summary statistics per condition for the simulated data.}
\begin{tabular}{lllll}
\toprule
Factor F & \multicolumn{1}{c}{N data points} & \multicolumn{1}{c}{Estimated means} & \multicolumn{1}{c}{Standard deviations} & \multicolumn{1}{c}{Standard errors}\\
\midrule
F1 & 5 & 0.8 & 0.2 & 0.1\\
F2 & 5 & 0.4 & 0.2 & 0.1\\
\bottomrule
\end{tabular}
\end{threeparttable}
\end{center}
\end{table}

The results, displayed in Table~\ref{tab:table1} and in Figure~\ref{fig:Fig1Means}, show that the assumed true condition means are exactly realized with the simulated data. The numbers are exact because the \texttt{mixedDesign()} function ensures that the data are generated so as to have the true means for each level. In real data-sets, of course, the sample means will vary from experiment to experiment.

A simple regression of \texttt{DV} on \texttt{F} yields a straightforward test of the difference between the group means. Part of the output of the \texttt{summary} function \textcolor{black}{is presented below, and the same results are also displayed in} Table~\ref{tab:table2}:

\begin{Shaded}
\begin{Highlighting}[]
\NormalTok{m_F <-}\StringTok{ }\KeywordTok{lm}\NormalTok{(DV }\OperatorTok{~}\StringTok{ }\DecValTok{1} \OperatorTok{+}\StringTok{ }\NormalTok{F, simdat)}
\KeywordTok{round}\NormalTok{(}\KeywordTok{summary}\NormalTok{(m_F)}\OperatorTok{$}\NormalTok{coef,}\DecValTok{3}\NormalTok{)}
\end{Highlighting}
\end{Shaded}

\begin{verbatim}
##             Estimate Std. Error t value Pr(>|t|)
## (Intercept)      0.8      0.089    8.94    0.000
## FF2             -0.4      0.126   -3.16    0.013
\end{verbatim}

\begin{table}[h]
\begin{center}
\begin{threeparttable}
\caption{\label{tab:table2}Estimated regression model. Confidence intervals are obtained in R, e.g., using the function confint().}
\begin{tabular}{lllll}
\toprule
Predictor & \multicolumn{1}{c}{$Estimate$} & \multicolumn{1}{c}{95\% CI} & \multicolumn{1}{c}{$t(8)$} & \multicolumn{1}{c}{$p$}\\
\midrule
Intercept & 0.8 & $[0.6$, $1.0]$ & 8.94 & < .001\\
FF2 & -0.4 & $[-0.7$, $-0.1]$ & -3.16 & .013\\
\bottomrule
\end{tabular}
\end{threeparttable}
\end{center}
\end{table}

Comparing the means for each condition with the coefficients (\emph{Estimates}) reveals that (i) the intercept \textcolor{black}{($0.8$)} is the mean for condition F1, \(\hat\mu_1\); and (ii) the slope (\texttt{FF2}\textcolor{black}{: $-0.4$)} is the difference between the true means for the two groups, \(\hat\mu_2 - \hat\mu_1\) (Bolker, 2018):

\begin{equation}
\begin{array}{lcl}
\text{Intercept} = & \hat{\mu}_1 & = \text{estimated mean for \texttt{F1}} \\
\text{Slope (\texttt{FF2})} = & \hat{\mu}_2 - \hat{\mu}_1 & = \text{estim. mean for \texttt{F2}} - \text{estim. mean for \texttt{F1}} 
\end{array}
\label{def:beta}
\end{equation}

The new information is confidence intervals associated with the regression coefficients. The t-test suggests that response times in group F2 are lower than in group F1.

\hypertarget{treatmentcontrasts}{%
\subsection{Default contrast coding: Treatment contrasts}\label{treatmentcontrasts}}

How does R arrive at these particular values for the intercept and slope? That is, why does the intercept assess the mean of condition \texttt{F1} and how do we know the slope measures the difference in means between \texttt{F2}\(-\)\texttt{F1}? This result is a consequence of the default contrast coding of the factor \texttt{F}. R assigns \textsc{treatment contrasts} to factors and orders their levels alphabetically. The first factor level (here: \texttt{F1}) is coded as \(0\) and the second level (here: \texttt{F2}) is coded as \(1\). This is visible when inspecting the current contrast attribute of the factor using the \texttt{contrasts} command:

\begin{Shaded}
\begin{Highlighting}[]
\KeywordTok{contrasts}\NormalTok{(simdat}\OperatorTok{$}\NormalTok{F)}
\end{Highlighting}
\end{Shaded}

\begin{verbatim}
##    F2
## F1  0
## F2  1
\end{verbatim}

Why does this contrast coding yield these particular regression coefficients? Let's take a look at the regression equation.
Let \(\beta_0\) represent the intercept, and \(\beta_1\) the slope. Then, the simple regression above expresses the belief that the expected response time \(y\) is a linear function of the factor \texttt{F}. In a more general formulation, this is written as follows: \(y\) is a linear function of some predictor \(x\) with regression coefficients for the intercept, \(\beta_0\), and for the factor, \(\beta_1\):

\begin{equation}
y = \beta_0 + \beta_1x
\label{eq:lm1}
\end{equation}

So, if \(x = 0\) \textcolor{black}{(condition} \texttt{F1}), \(y\) is \(\beta_0 + \beta_1 \cdot 0 = \beta_0\); and if \(x = 1\) \textcolor{black}{(condition} \texttt{F2}), \(y\) is \(\beta_0 + \beta_1 \cdot 1 = \beta_0 + \beta_1\).

Expressing the above in terms of the estimated coefficients:

\begin{equation}
\begin{array}{lccll}
\text{estim. value for \texttt{F1}} = & \hat{\mu}_1 = & \hat{\beta}_0 = & \text{Intercept} \\
\text{estim. value for \texttt{F2}} = & \hat{\mu}_2 = & \hat{\beta}_0 + \hat{\beta}_1 = & \text{Intercept} + \text{Slope (\texttt{FF2})}
\end{array}
\label{eq:predVal}
\end{equation}

\textcolor{black}{It is useful to think of such unstandardized regression coefficients as difference scores; they express the increase in the dependent variable $y$ associated with a change in the independent variable $x$ of $1$ unit, such as going from $0$ to $1$ in this example. The difference between condition means is $0.4 - 0.8 = -0.4$, which is exactly the estimated regression coefficient $\hat{\beta}_1$. The sign of the slope is negative because we have chosen to subtract the larger mean \texttt{F1} score from the smaller mean \texttt{F2} score.}

\hypertarget{inverseMatrix}{%
\subsection{Defining hypotheses}\label{inverseMatrix}}

The analysis of the \textcolor{black}{regression equation} demonstrates that in the \textsc{treatment contrast} the intercept assesses the average response in the baseline condition, whereas the slope tests the difference between condition means. \textcolor{black}{However, these are just verbal descriptions of what each coefficient assesses. Is it also possible to formally write down the null hypotheses that are tested by each of these two coefficients?} From the perspective of formal hypothesis tests, the slope represents the main test of interest, so we consider this first. The \textsc{treatment contrast} expresses the null hypothesis that the difference in means between the two levels of the factor F is \(0\); formally, the null hypothesis \(H_0\) is that \(H_0: \; \beta_1 = 0\):

\begin{equation}
H_0: - 1 \cdot \mu_{F1} + 1 \cdot \mu_{F2} = 0
\end{equation}

or equivalently:

\begin{equation} \label{eq:f2minusf1}
H_0: \mu_{F2} - \mu_{F1} = 0
\end{equation}

The \(\pm 1\) weights in the null hypothesis statement \textcolor{black}{directly express which means are compared by the treatment contrast.}

\textcolor{black}{The intercept in the \textsc{treatment contrast} expresses a null hypothesis that is usually of no interest: that the mean in condition F1 of the factor F is $0$.}
Formally, the null hypothesis is \(H_0: \; \beta_0 = 0\):

\begin{equation} \label{eq:trmtcontrfirstmention}
H_0: 1 \cdot \mu_{F1} + 0 \cdot \mu_{F2} = 0
\end{equation}

\noindent
or equivalently:

\begin{equation}
H_0: \mu_{F1} = 0 .
\end{equation}

\noindent
The fact that the intercept term formally tests the null hypothesis that the mean of condition \texttt{F1} is zero is in line with our previous derivation (see equation \ref{def:beta}).

In R, factor levels are ordered alphabetically and by default the first level is used as the baseline in \textsc{treatment contrasts}. Obviously, this default mapping will only be correct for a given data-set if the levels' alphabetical ordering matches the desired contrast coding. When it does not, it is possible to re-order the levels. Here is one way of re-ordering the levels in R:

\begin{Shaded}
\begin{Highlighting}[]
\NormalTok{simdat}\OperatorTok{$}\NormalTok{Fb <-}\StringTok{ }\KeywordTok{factor}\NormalTok{(simdat}\OperatorTok{$}\NormalTok{F,  }\DataTypeTok{levels =} \KeywordTok{c}\NormalTok{(}\StringTok{"F2"}\NormalTok{,}\StringTok{"F1"}\NormalTok{))}
\KeywordTok{contrasts}\NormalTok{(simdat}\OperatorTok{$}\NormalTok{Fb)}
\end{Highlighting}
\end{Shaded}

\begin{verbatim}
##    F1
## F2  0
## F1  1
\end{verbatim}

\noindent
\textcolor{black}{This re-ordering} did not change any data associated with the factor, only one of its attributes. With this new contrast attribute the simple regression yields the following result (see Table \ref{tab:table4}).

\begin{Shaded}
\begin{Highlighting}[]
\NormalTok{m1_mr <-}\StringTok{ }\KeywordTok{lm}\NormalTok{(DV }\OperatorTok{~}\StringTok{ }\DecValTok{1} \OperatorTok{+}\StringTok{ }\NormalTok{Fb, simdat)}
\end{Highlighting}
\end{Shaded}

\begin{table}[h]
\begin{center}
\begin{threeparttable}
\caption{\label{tab:table4}Reordering factor levels}
\begin{tabular}{lllll}
\toprule
Predictor & \multicolumn{1}{c}{$Estimate$} & \multicolumn{1}{c}{95\% CI} & \multicolumn{1}{c}{$t(8)$} & \multicolumn{1}{c}{$p$}\\
\midrule
Intercept & 0.4 & $[0.2$, $0.6]$ & 4.47 & .002\\
FbF1 & 0.4 & $[0.1$, $0.7]$ & 3.16 & .013\\
\bottomrule
\end{tabular}
\end{threeparttable}
\end{center}
\end{table}

\textcolor{black}{The model now tests different hypotheses. The intercept now codes the mean of condition} \texttt{F2}\textcolor{black}{, and the slope measures the difference in means between} \texttt{F1} \textcolor{black}{minus} \texttt{F2}\textcolor{black}{. This represents an alternative coding of the treatment contrast.}

\hypertarget{effectcoding}{%
\subsection{Sum contrasts}\label{effectcoding}}

Treatment contrasts are only one of many options. It is also possible to use \textsc{sum contrasts}, which code one of the conditions as \(-1\) and the other as \(+1\), effectively `centering' the effects at the grand mean (GM, i.e., the mean of the two group means). Here, we rescale the contrast to values of \(-0.5\) and \(+0.5\), which makes the estimated treatment effect the same as for treatment coding and easier to interpret.

\textcolor{black}{To} use this contrast in a linear regression, use the \texttt{contrasts} function (for results see Table \ref{tab:table5}):

\begin{Shaded}
\begin{Highlighting}[]
\NormalTok{(}\KeywordTok{contrasts}\NormalTok{(simdat}\OperatorTok{$}\NormalTok{F) <-}\StringTok{ }\KeywordTok{c}\NormalTok{(}\OperatorTok{-}\FloatTok{0.5}\NormalTok{,}\OperatorTok{+}\FloatTok{0.5}\NormalTok{))}
\end{Highlighting}
\end{Shaded}

\begin{verbatim}
## [1] -0.5  0.5
\end{verbatim}

\begin{Shaded}
\begin{Highlighting}[]
\NormalTok{m1_mr <-}\StringTok{ }\KeywordTok{lm}\NormalTok{(DV }\OperatorTok{~}\StringTok{ }\DecValTok{1} \OperatorTok{+}\StringTok{ }\NormalTok{F, simdat)}
\end{Highlighting}
\end{Shaded}

\begin{table}[h]
\begin{center}
\begin{threeparttable}
\caption{\label{tab:table5}Estimated regression model}
\begin{tabular}{lllll}
\toprule
Predictor & \multicolumn{1}{c}{$Estimate$} & \multicolumn{1}{c}{95\% CI} & \multicolumn{1}{c}{$t(8)$} & \multicolumn{1}{c}{$p$}\\
\midrule
Intercept & 0.6 & $[0.5$, $0.7]$ & 9.49 & < .001\\
F1 & -0.4 & $[-0.7$, $-0.1]$ & -3.16 & .013\\
\bottomrule
\end{tabular}
\end{threeparttable}
\end{center}
\end{table}

Here, the slope (\texttt{F1}) again codes the difference of the groups associated with the first and second factor levels. \textcolor{black}{It has the same value as in the \textsc{treatment contrast}.}
However, the intercept now represents the estimate of the average of condition means for F1 and F2, that is, the GM. \textcolor{black}{This differs from the \textsc{treatment contrast}. For the scaled \textsc{sum contrast}:}

\begin{equation}
\begin{array}{lcl}
\text{Intercept} = & (\hat{\mu}_1 + \hat{\mu}_2)/2 & = \text{estimated mean of \texttt{F1} and \texttt{F2}} \\
\text{Slope (\texttt{F1})} = & \hat{\mu}_2 - \hat{\mu}_1 & = \text{estim. mean for \texttt{F2}} - \text{estim. mean for \texttt{F1}} 
\end{array}
\label{def:beta2}
\end{equation}

\textcolor{black}{How does R arrive at these values for the intercept and the slope? Why does the intercept assess the GM and why does the slope test the group-difference? This is the result of rescaling the \textsc{sum contrast}. The first factor level (}\texttt{F1}\textcolor{black}{) was coded as $-0.5$, and the second factor level (}\texttt{F1}\textcolor{black}{) as $+0.5$:}

\begin{Shaded}
\begin{Highlighting}[]
\KeywordTok{contrasts}\NormalTok{(simdat}\OperatorTok{$}\NormalTok{F)}
\end{Highlighting}
\end{Shaded}

\begin{verbatim}
##    [,1]
## F1 -0.5
## F2  0.5
\end{verbatim}

\textcolor{black}{Let's again look at the regression equation to better understand what computations are performed. Again, $\beta_0$ represents the intercept, $\beta_1$ represents the slope, and the predictor variable $x$ represents the factor} \texttt{F}\textcolor{black}{. The regression equation is written as:}

\begin{equation}
y = \beta_0 + \beta_1x
\label{eq:lm2}
\end{equation}

The group of \texttt{F1} subjects is \textcolor{black}{then coded as $-0.5$, and t}he response time for the group of \texttt{F1} subjects is \(\beta_0 + \beta_1 \cdot x_1 = 0.6 + (-0.4) \cdot (-0.5) = 0.8\). \textcolor{black}{The \texttt{F2} group, to the contrary, is coded as $+0.5$.} By implication, the mean of the \texttt{F2} group must be \(\beta_0 + \beta_1 \cdot x_1 = 0.6 + (-0.4) \cdot 0.5 = 0.4\).
\textcolor{black}{Expressed in terms of the estimated coefficients:}

\begin{equation}
\begin{array}{lccll}
\text{estim. value for \texttt{F1}} = & \hat{\mu}_1 = & \hat{\beta}_0 - 0.5 \cdot \hat{\beta}_1 = & \text{Intercept} - 0.5 \cdot \text{Slope (\texttt{F1})}\\
\text{estim. value for \texttt{F2}} = & \hat{\mu}_2 = & \hat{\beta}_0 + 0.5 \cdot \hat{\beta}_1 = & \text{Intercept} + 0.5 \cdot \text{Slope (\texttt{F1})}
\end{array}
\label{eq:predVal2}
\end{equation}

\textcolor{black}{The unstandardized regression coefficient is a difference score: Taking a step of one unit on the predictor variable $x$, e.g., from $-0.5$ to $+0.5$, reflecting a step from condition \texttt{F1} to \texttt{F2}, changes the dependent variable from $0.8$ (for condition \texttt{F1}) to $0.4$ (condition \texttt{F2}), reflecting a difference of $0.4 - 0.8 = -0.4$; and this is again exactly the estimated regression coefficient $\hat{\beta}_1$.}
\textcolor{black}{Moreover, a}s mentioned above, the intercept now assesses the GM of conditions F1 and F2\textcolor{black}{: it is exactly in the middle between condition means for F1 and F2.}

\textcolor{black}{So far we gave verbal statements about what is tested by the intercept and the slope in the case of the scaled \textsc{sum contrast}. It is possible to write these statements as formal null hypotheses that are tested by each regression coefficient: }
Sum contrasts express the null hypothesis that the difference in means between the two levels of factor F is 0; formally, the null hypothesis \(H_0\) is that

\begin{equation}
H_0: -1 \cdot \mu_{F1} + 1 \cdot \mu_{F2} = 0
\end{equation}

\noindent
\textcolor{black}{This is the same hypothesis that was also tested by the slope in the treatment contrast.}
The intercept, however, now expresses a different hypothesis about the data: it expresses the null hypothesis that the average of the two conditions F1 and F2 is 0:

\begin{equation}
H_0: 1/2 \cdot \mu_{F1} + 1/2 \cdot \mu_{F2} = \frac{\mu_{F1} + \mu_{F2}}{2} = 0
\end{equation}

\noindent
In balanced data, i.e., in data-sets where there are no missing data points, the average of the two conditions F1 and F2 is the GM. In unbalanced data-sets, where there are missing values, this average is the \textcolor{black}{weighted GM}.
To illustrate this point, consider an example with fully balanced data and two equal group sizes of \(5\) subjects for each group F1 and F2. Here, the GM is also the mean across all subjects. Next, consider a highly simplified unbalanced data-set, where \textcolor{black}{in condition} \texttt{F1} two observations of the dependent variable \textcolor{black}{are available} with values of \(2\) and \(3\), and where \textcolor{black}{in condition} \texttt{F2} only one observation of the dependent variables \textcolor{black}{is available} with a value of \(4\). In this data-set, the mean across all subjects is \(\frac{2 + 3 + 4}{3} = \frac{9}{3} = 3\). However, the \textcolor{black}{(weighted)} GM as assessed in the intercept in a model using \textcolor{black}{sum contrasts} for factor \texttt{F} \textcolor{black}{would first compute the mean for each group separately (i.e., $\frac{2 + 3}{2} = 2.5$, and $4$), and then compute the mean across conditions} \(\frac{2.5 + 4}{2} = \frac{6.5}{2} = 3.25\). \textcolor{black}{The GM of $3.25$ is different from} the mean across subjects of \(3\).

To summarize, \textsc{treatment contrasts} and \textsc{sum contrasts} are two possible ways to parameterize the difference between two groups; they test different hypotheses (there are cases, however, where the hypotheses are equivalent). \textsc{Treatment contrasts} compare one or more means against a baseline condition, whereas \textsc{sum contrasts} allow us to determine whether we can reject the null hypothesis that a condition's mean is the same as the GM (which in the two-group case also implies a hypothesis test that the two group means are the same). \textcolor{black}{One question that comes up here, is how one knows or formally derives what hypotheses a given set of contrasts tests. This question will be discussed in detail below for the general case of any arbitrary contrasts.}

\hypertarget{cell-means-parameterization}{%
\subsection{Cell means parameterization}\label{cell-means-parameterization}}

\textcolor{black}{One alternative option is use the so-called cell means parameterization. In this approach, one does not estimate an intercept term, and then differences between factor levels. Instead, each degree of freedom in a design is used to simply estimate the mean of one of the factor levels. As a consequence, no comparisons between condition means are tested, but it is simply tested for each factor level whether the associated condition mean differs from zero. Cell means parameterization is specified by explicitly removing the intercept term (which is added automatically) by adding a $-1$ in the regression formula:}

\begin{Shaded}
\begin{Highlighting}[]
\NormalTok{m2_mr <-}\StringTok{ }\KeywordTok{lm}\NormalTok{(DV }\OperatorTok{~}\StringTok{ }\DecValTok{-1} \OperatorTok{+}\StringTok{ }\NormalTok{F, simdat)}
\end{Highlighting}
\end{Shaded}

\begin{table}[h]
\begin{center}
\begin{threeparttable}
\caption{\label{tab:table2a}Estimated regression model}
\begin{tabular}{lllll}
\toprule
Predictor & \multicolumn{1}{c}{$Estimate$} & \multicolumn{1}{c}{95\% CI} & \multicolumn{1}{c}{$t(8)$} & \multicolumn{1}{c}{$p$}\\
\midrule
FF1 & 0.8 & $[0.6$, $1.0]$ & 8.94 & < .001\\
FF2 & 0.4 & $[0.2$, $0.6]$ & 4.47 & .002\\
\bottomrule
\end{tabular}
\end{threeparttable}
\end{center}
\end{table}

\textcolor{black}{Now, the regression coefficients (see the column labeled `Estimate') estimate the mean of the first factor level ($0.8$) and the mean of the second factor level ($0.4$). Each of these means is compared to zero in the statistical tests, and each of these means is significantly larger than zero. This cell means parameterization usually does not allow a test of the hypotheses of interest, as these hypotheses usually relate to differences between conditions rather than to whether each condition differs from zero.}

\hypertarget{examples-of-different-default-contrast-types}{%
\section{Examples of different default contrast types}\label{examples-of-different-default-contrast-types}}

\textcolor{black}{Above, we introduced conceptual explanations of different default contrasts by discussing example applications. Moreover, the preceding section on basic concepts mentioned that contrasts are implemented by numerical contrast coefficients. These contrast coefficients are usually represented in matrices of coefficients. Each of the discussed default contrasts is implemented using a different contrast matrix. 
These default contrast matrices are available in the R System for Statistical Computing in the basic distribution of R in the stats package.
Here, we provide a quick overview of the different contrast matrices specifically for the example applications discussed above.}

\textcolor{black}{For the \textsc{treatment contrasts}, we discussed an example where a psychotherapy group and a pharmacotherapy group are each compared to the same control or baseline group; the latter could be a group that receives no treatment. The corresponding contrast matrix is obtained using the following function call:}

\begin{Shaded}
\begin{Highlighting}[]
\KeywordTok{contr.treatment}\NormalTok{(}\DecValTok{3}\NormalTok{)}
\end{Highlighting}
\end{Shaded}

\begin{verbatim}
##   2 3
## 1 0 0
## 2 1 0
## 3 0 1
\end{verbatim}

\textcolor{black}{The number of rows specifies the number of factor levels. The three rows indicate the three levels of the factor. The first row codes the baseline or control condition (the baseline always only contains $0$s as contrast coefficients), the second row codes the psychotherapy group, and the third row codes the pharmacotherapy group. The two columns reflect the two comparisons that are being tested by the contrasts: the first column tests the second group (i.e., psychotherapy) against the baseline / control group, and the second column tests the third group (i.e., pharmacotherapy) against the baseline / control group.}

\textcolor{black}{For the \textsc{sum contrast}, our example involved conditions for orthographic priming, phonological priming, and semantic priming. The orthographic and phonological priming conditions were compared to the average priming effect across all three groups. In R, there is again a standard function call for the \textsc{sum contrast}:}

\begin{Shaded}
\begin{Highlighting}[]
\KeywordTok{contr.sum}\NormalTok{(}\DecValTok{3}\NormalTok{)}
\end{Highlighting}
\end{Shaded}

\begin{verbatim}
##   [,1] [,2]
## 1    1    0
## 2    0    1
## 3   -1   -1
\end{verbatim}

\textcolor{black}{Again, the three rows indicate three groups, and the two columns reflect the two comparisons. The first row codes orthographic priming, the second row phonological priming, and the last row semantic priming. Now, the first column codes a contrast that compares the response in orthographic priming against the average response, and the second column codes a contrast comparing phonological priming against the average response. Why these contrasts test these hypotheses is not transparent here. We will return to this issue below.}

\textcolor{black}{For \textsc{repeated contrasts}, our example compared response times in low frequency words vs.  medium frequency words, and medium frequency words vs.  high frequency words. In R, the corresponding contrast matrix is available in the MASS package} (Venables \& Ripley, 2002):

\begin{Shaded}
\begin{Highlighting}[]
\KeywordTok{library}\NormalTok{(MASS)}
\KeywordTok{contr.sdif}\NormalTok{(}\DecValTok{3}\NormalTok{)}
\end{Highlighting}
\end{Shaded}

\begin{verbatim}
##      2-1    3-2
## 1 -0.667 -0.333
## 2  0.333 -0.333
## 3  0.333  0.667
\end{verbatim}

\textcolor{black}{The first row represents low frequency words, the second row medium frequency words, and the last row high frequency words. Now the first contrast (column) tests the difference between the second minus the first row, i.e., the response to medium frequency words minus response to low frequency words. The second contrast (column) tests the difference between the third minus the second row, i.e., the difference in the response to high frequency words minus the response to medium frequency words. Why the \textsc{repeated contrast} tests exactly these differences is not transparent either.}

\textcolor{black}{Below, we will explain how these and other contrasts are generated from a careful definition of the hypotheses that one wishes to test for a given data-set. We will introduce a basic workflow for how to create one's own custom contrasts.}

\textcolor{black}{We discussed that for the example with three different word frequency levels, it is possible to test the hypothesis of a linear (or quadratic) trend across all levels of word frequency. Such \textsc{polynomial contrasts} are specified in R using the following command:}

\begin{Shaded}
\begin{Highlighting}[]
\KeywordTok{contr.poly}\NormalTok{(}\DecValTok{3}\NormalTok{)}
\end{Highlighting}
\end{Shaded}

\begin{verbatim}
##             .L     .Q
## [1,] -7.07e-01  0.408
## [2,] -7.85e-17 -0.816
## [3,]  7.07e-01  0.408
\end{verbatim}

\textcolor{black}{As in the other contrasts mentioned above, it is not clear from this contrast matrix what hypotheses are being tested. As before, the three rows represent three levels of word frequency. The first column codes a linear increase with word frequency levels. The second column codes a quadratic trend. The} \texttt{contr.poly()} \textcolor{black}{function tests orthogonalized trends - a concept that we will explain below.}

\textcolor{black}{We had also discussed \textsc{Helmert contrasts} above, and had given the example that one might want to compare two "invalid" priming conditions to each other, and then compare both "invalid" priming conditions to one "valid" prime condition. \textsc{Helmert contrasts} are specified in R using the following command:}

\begin{Shaded}
\begin{Highlighting}[]
\KeywordTok{contr.helmert}\NormalTok{(}\DecValTok{3}\NormalTok{)}
\end{Highlighting}
\end{Shaded}

\begin{verbatim}
##   [,1] [,2]
## 1   -1   -1
## 2    1   -1
## 3    0    2
\end{verbatim}

\textcolor{black}{The first row represents "invalid prime 1", the second row "invalid prime 2", and the third row the "valid prime" condition. The first column tests the difference between the two "invalid" prime conditions. The coefficients in the second column test both "invalid prime" conditions against the "valid prime" condition.}

\textcolor{black}{How can one make use of these contrast matrices for a specific regression analysis? As discussed above for the case of two factor levels, one needs to tell R to use one of these contrast coding schemes for a factor of interest in a linear model (LM)/regression analysis. Let's assume a data-frame called} \texttt{dat} \textcolor{black}{with a dependent variable} \texttt{dat\$DV} \textcolor{black}{and a three-level factor} \texttt{dat\$WordFrequency} \textcolor{black}{with levels low, medium, and high frequency words. One chooses one of the above contrast matrices and `assigns' this contrast to the factor. Here, we choose the \textsc{repeated contrast}:}

\begin{Shaded}
\begin{Highlighting}[]
\KeywordTok{contrasts}\NormalTok{(dat}\OperatorTok{$}\NormalTok{WordFrequency) <-}\StringTok{ }\KeywordTok{contr.sdif}\NormalTok{(}\DecValTok{3}\NormalTok{)}
\end{Highlighting}
\end{Shaded}

\textcolor{black}{This way, when running a linear model using this factor, R will automatically use the contrast matrix assigned to the factor. This is done in R with the simple call of a linear model, where} \texttt{dat} \textcolor{black}{is specified as the data-frame to use, where the numeric variable} \texttt{DV} \textcolor{black}{is defined as the dependent variable, and where the factor} \texttt{WordFrequency} \textcolor{black}{is added as predictor in the analysis:}

\begin{Shaded}
\begin{Highlighting}[]
\KeywordTok{lm}\NormalTok{(DV }\OperatorTok{~}\StringTok{ }\DecValTok{1} \OperatorTok{+}\StringTok{ }\NormalTok{WordFrequency, }\DataTypeTok{data=}\NormalTok{dat)}
\end{Highlighting}
\end{Shaded}

\textcolor{black}{Assuming that there are three levels to the WordFrequency factor, 
the lm function will estimate three regression coefficients: one intercept and two slopes. What these regression coefficients test will depend on which contrast coding is specified. Given that we have used \textsc{repeated contrasts}, the resulting regression coefficients will now test the difference between medium and low frequency words (first slope) and will test the difference between high and medium frequency words (second slope). Examples of output from regression models for concrete sitations will be shown below. Moreover, it will be shown how contrast matrices are generated for whatever hypotheses one wants to test in a given data-set.}

\hypertarget{the-hypothesis-matrix-illustrated-with-a-three-level-factor}{%
\section{The hypothesis matrix illustrated with a three-level factor}\label{the-hypothesis-matrix-illustrated-with-a-three-level-factor}}

Consider again the example with the three low, medium, and high frequency conditions. The \texttt{mixedDesign} function \textcolor{black}{can be used} to simulate data from a lexical decision task with response times as dependent variable. The research question is: do response times differ as a function of the between-subject factor word frequency with three levels: low, medium, and high? We assume that lower word frequency results in longer response times. Here, we specify word frequency as a between-subject factor. In cognitive science experiments, frequency will usually vary within subjects and between items. However, the within- or between-subjects status of an effect is independent of its contrast coding; we assume the manipulation to be between subjects for ease of exposition. \textcolor{black}{The concepts presented here extend to repeated measures designs that are usually analyzed using linear mixed models.}

\textcolor{black}{The following R code} simulates the data and computes the table of means and standard deviations for the three frequency categories:

\begin{Shaded}
\begin{Highlighting}[]
\NormalTok{M <-}\StringTok{ }\KeywordTok{matrix}\NormalTok{(}\KeywordTok{c}\NormalTok{(}\DecValTok{500}\NormalTok{, }\DecValTok{450}\NormalTok{, }\DecValTok{400}\NormalTok{), }\DataTypeTok{nrow=}\DecValTok{3}\NormalTok{, }\DataTypeTok{ncol=}\DecValTok{1}\NormalTok{, }\DataTypeTok{byrow=}\OtherTok{FALSE}\NormalTok{)}
\KeywordTok{set.seed}\NormalTok{(}\DecValTok{1}\NormalTok{)}
\NormalTok{simdat2 <-}\StringTok{ }\KeywordTok{mixedDesign}\NormalTok{(}\DataTypeTok{B=}\DecValTok{3}\NormalTok{, }\DataTypeTok{W=}\OtherTok{NULL}\NormalTok{, }\DataTypeTok{n=}\DecValTok{4}\NormalTok{, }\DataTypeTok{M=}\NormalTok{M,  }\DataTypeTok{SD=}\DecValTok{20}\NormalTok{, }\DataTypeTok{long =} \OtherTok{TRUE}\NormalTok{) }
\KeywordTok{names}\NormalTok{(simdat2)[}\DecValTok{1}\NormalTok{] <-}\StringTok{ "F"}  \CommentTok{# Rename B_A to F(actor)/F(requency)}
\KeywordTok{levels}\NormalTok{(simdat2}\OperatorTok{$}\NormalTok{F) <-}\StringTok{ }\KeywordTok{c}\NormalTok{(}\StringTok{"low"}\NormalTok{, }\StringTok{"medium"}\NormalTok{, }\StringTok{"high"}\NormalTok{)}
\NormalTok{simdat2}\OperatorTok{$}\NormalTok{DV <-}\StringTok{ }\KeywordTok{round}\NormalTok{(simdat2}\OperatorTok{$}\NormalTok{DV)}
\KeywordTok{head}\NormalTok{(simdat2)}
\end{Highlighting}
\end{Shaded}

\begin{verbatim}
##        F id  DV
## 1    low  1 497
## 2    low  2 474
## 3    low  3 523
## 4    low  4 506
## 5 medium  5 422
## 6 medium  6 467
\end{verbatim}

\begin{Shaded}
\begin{Highlighting}[]
\NormalTok{table.word <-}\StringTok{ }\NormalTok{simdat2 }\OperatorTok{%>%}\StringTok{ }\KeywordTok{group_by}\NormalTok{(F) }\OperatorTok{%>%}\StringTok{ }
\StringTok{  }\KeywordTok{summarise}\NormalTok{(}\DataTypeTok{N =} \KeywordTok{length}\NormalTok{(DV), }\DataTypeTok{M =} \KeywordTok{mean}\NormalTok{(DV), }\DataTypeTok{SD =} \KeywordTok{sd}\NormalTok{(DV), }\DataTypeTok{SE =} \KeywordTok{sd}\NormalTok{(DV)}\OperatorTok{/}\KeywordTok{sqrt}\NormalTok{(N))}
\end{Highlighting}
\end{Shaded}

\begin{table}[h]
\begin{center}
\begin{threeparttable}
\caption{\label{tab:table6}Summary statistics of the simulated lexical decision data per frequency level.}
\begin{tabular}{lllll}
\toprule
Factor F & \multicolumn{1}{c}{N data points} & \multicolumn{1}{c}{Estimated means} & \multicolumn{1}{c}{Standard deviations} & \multicolumn{1}{c}{Standard errors}\\
\midrule
low & 4 & 500 & 20 & 10\\
medium & 4 & 450 & 20 & 10\\
high & 4 & 400 & 20 & 10\\
\bottomrule
\end{tabular}
\end{threeparttable}
\end{center}
\end{table}

As shown in Table \ref{tab:table6}, the estimated means reflect our assumptions about the true means in the data simulation: Response times decrease with increasing word frequency.
The effect is significant in an ANOVA (see Table \ref{tab:table7}).

\begin{Shaded}
\begin{Highlighting}[]
\NormalTok{aovF <-}\StringTok{ }\KeywordTok{aov}\NormalTok{(DV }\OperatorTok{~}\StringTok{ }\DecValTok{1} \OperatorTok{+}\StringTok{ }\NormalTok{F }\OperatorTok{+}\StringTok{ }\KeywordTok{Error}\NormalTok{(id), }\DataTypeTok{data=}\NormalTok{simdat2)}
\end{Highlighting}
\end{Shaded}

\begin{table}[h]
\begin{center}
\begin{threeparttable}
\caption{\label{tab:table7}ANOVA results. The effect F stands for the word frequency factor.}
\begin{tabular}{lllllll}
\toprule
Effect & \multicolumn{1}{c}{$F$} & \multicolumn{1}{c}{$\mathit{df}_1$} & \multicolumn{1}{c}{$\mathit{df}_2$} & \multicolumn{1}{c}{$\mathit{MSE}$} & \multicolumn{1}{c}{$p$} & \multicolumn{1}{c}{$\hat{\eta}^2_G$}\\
\midrule
F & 24.93 & 2 & 9 & 403.19 & < .001 & .847\\
\bottomrule
\end{tabular}
\end{threeparttable}
\end{center}
\end{table}

The ANOVA, however, does not tell us the source of the difference. In the following sections, we use this and an additional data-set to illustrate \protect\hyperlink{sumcontrasts}{\textsc{sum}}, \protect\hyperlink{repeatedcontrasts}{\textsc{repeated}}, \protect\hyperlink{polynomialContrasts}{\textsc{polynomial}}, and \protect\hyperlink{customContrasts}{\textsc{custom}} contrasts. In practice, usually only one set of contrasts is selected when the expected pattern of means is formulated during the design of the experiment. The decision about which contrasts to use is made before the pattern of means is known.

\hypertarget{sumcontrasts}{%
\subsection{Sum contrasts}\label{sumcontrasts}}

For didactic purposes, \textcolor{black}{the next sections describe} \textsc{sum contrasts}. Suppose \textcolor{black}{that the expectation is} that \textcolor{black}{low-frequency words are responded to slower and} medium-frequency words are responded to faster than the GM response time. Then, \textcolor{black}{the} research question could be: \textcolor{black}{Do low-frequency words differ from the GM and do} medium-frequency words differ from the GM? And if so, are they above or below the GM? We want to test the following two hypotheses:

\begin{equation}
H_{0_1}: \mu_1 = \frac{\mu_1+\mu_2+\mu_3}{3} = GM
\end{equation}

\noindent
and

\begin{equation}
H_{0_2}: \mu_2 = \frac{\mu_1+\mu_2+\mu_3}{3} = GM
\end{equation}

\(H_{0_1}\) \textcolor{black}{can also be written} as:

\begin{align} \label{h01}
& \mu_1 =\frac{\mu_1+\mu_2+\mu_3}{3}\\
\Leftrightarrow & \mu_1 - \frac{\mu_1+\mu_2+\mu_3}{3} = 0\\
\Leftrightarrow & \frac{2}{3} \mu_1 - \frac{1}{3}\mu_2 - \frac{1}{3}\mu_3 = 0
\end{align}

Here, the weights \(2/3, -1/3, -1/3\) \textcolor{black}{are informative about} how to combine the condition means to define the null hypothesis.

\(H_{0_2}\) \textcolor{black}{is also rewritten} as:

\begin{align}\label{h02}
&  \mu_2 = \frac{\mu_1+\mu_2+\mu_3}{3}\\
\Leftrightarrow & \mu_2 - \frac{\mu_1+\mu_2+\mu_3}{3} = 0 \\
\Leftrightarrow & -\frac{1}{3}\mu_1 + \frac{2}{3} \mu_2 - \frac{1}{3} \mu_3 = 0
\end{align}

Here, the weights are \(-1/3, 2/3, -1/3\), and they again \textcolor{black}{indicate} how to combine the condition means for defining the null hypothesis.

\hypertarget{the-hypothesis-matrix}{%
\subsection{The hypothesis matrix}\label{the-hypothesis-matrix}}

\textcolor{black}{The weights of the condition means are not only useful to define hypotheses. They also provide the starting step in a very powerful method which allows the researcher to generate the contrasts that are needed to test these hypotheses in a linear model. That is, what we did so far is to explain some kinds of different contrast codings that exist and what the hypotheses are that they test. That is, if a \textcolor{black}{certain} data-set \textcolor{black}{is given} and certain hypotheses \textcolor{black}{exist} that \textcolor{black}{need to be tested} in this data-set, then \textcolor{black}{the procedure would be} to check whether any of the contrasts that we encountered above happen to test exactly the hypotheses \textcolor{black}{of interest. Sometimes it suffices to use one of these existing contrasts.} However, at other times, \textcolor{black}{our research} hypotheses \textcolor{black}{may} not correspond exactly to any of the contrasts in the default set of standard contrasts provided in R. For these cases, or simply for more complex designs, it is very useful to know how contrast matrices are created. Indeed, a relatively simple procedure exists in which we write our hypotheses formally, extract the weights of the condition means from the hypotheses, and then automatically generate the correct contrast matrix that we need in order to test these hypotheses in a linear model. Using this powerful method, \textcolor{black}{it is not necessary} to find a match to a contrast matrix provided by the family of functions in R starting with the prefix contr. Instead, \textcolor{black}{it is possible to} simply define the hypotheses that one wants to test, and to obtain the correct contrast matrix for these in an automatic procedure. Here, for pedagogical reasons, we show some examples of how to apply this procedure in cases where the hypotheses} \emph{do} \textcolor{black}{correspond to some of the existing contrasts.}

\textcolor{black}{Defining a custom contrast matrix involves four steps:}

\begin{enumerate}
\def\labelenumi{\arabic{enumi}.}
\tightlist
\item
  \textcolor{black}{Write down the hypotheses}
\item
  \textcolor{black}{Extract the weights and write them into what we will call a} \emph{hypothesis matrix}
\item
  \textcolor{black}{Apply the} \emph{generalized matrix inverse} \textcolor{black}{to the hypothesis matrix to create the contrast matrix}
\item
  \textcolor{black}{Assign the contrast matrix to the factor and run the linear model}
\end{enumerate}

\textcolor{black}{Let us apply this four-step procedure to our example of the \textsc{sum contrast}. The first step, writing down the hypotheses, is shown above.  The second step involves writing down the weights that each hypothesis gives to condition means. The weights for the first null hypothesis are} \texttt{wH01=c(+2/3,\ -1/3,\ -1/3)}, and the weights for the second null hypothesis are \texttt{wH02=c(-1/3,\ +2/3,\ -1/3)}.

\textcolor{black}{Before writing these into a} hypothesis matrix, we also define a null hypothesis for the intercept term. For the intercept, \textcolor{black}{the} hypothesis is that the mean across all conditions is zero:

\begin{align}
H_{0_0}: &\frac{\mu_1 + \mu_2 + \mu_3}{3} = 0 \\
H_{0_0}: &\frac{1}{3} \mu_1 + \frac{1}{3}\mu_2 + \frac{1}{3}\mu_3 = 0
\end{align}

This null hypothesis has weights of \(1/3\) for all condition means.
The weights from all three hypotheses that \textcolor{black}{were defined are now combined} and written into a matrix that we refer to as the \emph{hypothesis matrix} (\texttt{Hc}):

\begin{Shaded}
\begin{Highlighting}[]
\NormalTok{HcSum <-}\StringTok{ }\KeywordTok{rbind}\NormalTok{(}\DataTypeTok{cH00=}\KeywordTok{c}\NormalTok{(}\DataTypeTok{low=} \DecValTok{1}\OperatorTok{/}\DecValTok{3}\NormalTok{, }\DataTypeTok{med=} \DecValTok{1}\OperatorTok{/}\DecValTok{3}\NormalTok{, }\DataTypeTok{hi=} \DecValTok{1}\OperatorTok{/}\DecValTok{3}\NormalTok{), }
               \DataTypeTok{cH01=}\KeywordTok{c}\NormalTok{(}\DataTypeTok{low=}\OperatorTok{+}\DecValTok{2}\OperatorTok{/}\DecValTok{3}\NormalTok{, }\DataTypeTok{med=}\OperatorTok{-}\DecValTok{1}\OperatorTok{/}\DecValTok{3}\NormalTok{, }\DataTypeTok{hi=}\OperatorTok{-}\DecValTok{1}\OperatorTok{/}\DecValTok{3}\NormalTok{), }
               \DataTypeTok{cH02=}\KeywordTok{c}\NormalTok{(}\DataTypeTok{low=}\OperatorTok{-}\DecValTok{1}\OperatorTok{/}\DecValTok{3}\NormalTok{, }\DataTypeTok{med=}\OperatorTok{+}\DecValTok{2}\OperatorTok{/}\DecValTok{3}\NormalTok{, }\DataTypeTok{hi=}\OperatorTok{-}\DecValTok{1}\OperatorTok{/}\DecValTok{3}\NormalTok{))}
\KeywordTok{fractions}\NormalTok{(}\KeywordTok{t}\NormalTok{(HcSum))}
\end{Highlighting}
\end{Shaded}

\begin{verbatim}
##     cH00 cH01 cH02
## low  1/3  2/3 -1/3
## med  1/3 -1/3  2/3
## hi   1/3 -1/3 -1/3
\end{verbatim}

Each set of weights \textcolor{black}{is first entered} as a row into the matrix (command \texttt{rbind()}\textcolor{black}{). This has mathematical reasons that we discuss below. However, we then switch rows and columns of the matrix for easier readability using the command} \texttt{t()} \textcolor{black}{(this transposes the matrix, i.e., switches rows and columns).}\footnote{Matrix transpose changes the arrangement of the columns and rows of a matrix, but leaves the content of the matrix unchanged. For example, for the matrix \(X\) with three rows and two columns \(X = \left(\begin{array}{cc} a & b \\ c & d \\ e & f \end{array} \right)\), the transpose yields a matrix with two rows and three columns, where the rows and columns are flipped: \(X^T = \left(\begin{array}{ccc} a & c & e \\ b & d & f \end{array} \right)\).} The command \texttt{fractions()} turns the decimals into fractions to improve readability.

Now that the condition weights from the hypotheses \textcolor{black}{have been written} into the hypothesis matrix, the third step of the procedure \textcolor{black}{is implemented}: a matrix operation called the \enquote{generalized matrix inverse}\footnote{At this point, there is no need to understand in detail what this means. We refer the interested reader to Appendix \ref{app:LinearAlgebra}. For a quick overview, we recommend a vignette explaining the generalized inverse in the \href{https://cran.r-project.org/web/packages/matlib/vignettes/ginv.html}{matlib package} (Friendly, Fox, \& Chalmers, 2018).} is used to obtain the contrast matrix that \textcolor{black}{is needed} to test these hypotheses in a linear model.
In R this next step is done using the function \texttt{ginv()} \textcolor{black}{from the} \texttt{MASS} \textcolor{black}{package}. We here define a function \texttt{ginv2()} for nicer formatting of the output.\footnote{The function \texttt{fractions()} from the \texttt{MASS} package \textcolor{black}{is used} to make the output more easily readable, and \textcolor{black}{the function} \texttt{provideDimnames()} \textcolor{black}{is used} to keep row and column names.}

\begin{Shaded}
\begin{Highlighting}[]
\NormalTok{ginv2 <-}\StringTok{ }\ControlFlowTok{function}\NormalTok{(x) }\CommentTok{# define a function to make the output nicer}
  \KeywordTok{fractions}\NormalTok{(}\KeywordTok{provideDimnames}\NormalTok{(}\KeywordTok{ginv}\NormalTok{(x),}\DataTypeTok{base=}\KeywordTok{dimnames}\NormalTok{(x)[}\DecValTok{2}\OperatorTok{:}\DecValTok{1}\NormalTok{]))}
\end{Highlighting}
\end{Shaded}

Applying the generalized inverse to the hypothesis matrix results in the new matrix \texttt{XcSum}. This is the contrast matrix \(X_c\) that tests exactly those hypotheses that were specified earlier:

\begin{Shaded}
\begin{Highlighting}[]
\NormalTok{(XcSum <-}\StringTok{ }\KeywordTok{ginv2}\NormalTok{(HcSum))}
\end{Highlighting}
\end{Shaded}

\begin{verbatim}
##     cH00 cH01 cH02
## low  1    1    0  
## med  1    0    1  
## hi   1   -1   -1
\end{verbatim}

This contrast matrix corresponds exactly to the \textsc{sum contrasts} \textcolor{black}{described} above. In the case of the \textsc{sum contrast}, the contrast matrix looks very different from the hypothesis matrix. The contrast matrix in \textsc{sum contrasts} codes with \(+1\) the condition that is to be compared to the GM. The condition that is never compared to the GM is coded as \(-1\). Without knowing the \textcolor{black}{relationship between the hypothesis matrix and the contrast matrix}, the meaning of the coefficients is completely opaque.

To verify this custom-made contrast matrix, it \textcolor{black}{is compared} to the \textsc{sum contrast} matrix as generated by the R function \texttt{contr.sum()} in the \texttt{stats} package. The resulting contrast matrix is identical to \textcolor{black}{the} result when adding the intercept term, a column of ones, to the contrast matrix:

\begin{Shaded}
\begin{Highlighting}[]
\KeywordTok{fractions}\NormalTok{(}\KeywordTok{cbind}\NormalTok{(}\DecValTok{1}\NormalTok{,}\KeywordTok{contr.sum}\NormalTok{(}\DecValTok{3}\NormalTok{)))}
\end{Highlighting}
\end{Shaded}

\begin{verbatim}
##   [,1] [,2] [,3]
## 1  1    1    0  
## 2  1    0    1  
## 3  1   -1   -1
\end{verbatim}

In order to test the hypotheses, step four in our procedure \textcolor{black}{involves} assigning \textsc{sum contrasts} to the factor \texttt{F} \textcolor{black}{in our example data, and running a linear model.}\footnote{Alternative ways to specify \textcolor{black}{default} contrasts in R are to set contrasts globally using \texttt{options(contrasts="contr.sum")} or to set contrasts locally only for a specific analysis, by including a named list specifying contrasts for each factor in a linear model: \texttt{lm(DV\ \textasciitilde{}\ 1\ +\ F,\ contrasts=list(F="contr.sum"))}.} \textcolor{black}{This allows estimating the regression coefficients associated with each contrast. We compare these to the data in Table}~\ref{tab:table6} to test whether the regression coefficients actually correspond to the differences of condition means, as intended. To define the contrast, \textcolor{black}{it is necessary to} remove the intercept term, as this is automatically added by the linear model function \texttt{lm()} in R.

\begin{Shaded}
\begin{Highlighting}[]
\KeywordTok{contrasts}\NormalTok{(simdat2}\OperatorTok{$}\NormalTok{F) <-}\StringTok{ }\NormalTok{XcSum[,}\DecValTok{2}\OperatorTok{:}\DecValTok{3}\NormalTok{]}
\NormalTok{m1_mr <-}\StringTok{ }\KeywordTok{lm}\NormalTok{(DV }\OperatorTok{~}\StringTok{ }\DecValTok{1} \OperatorTok{+}\StringTok{ }\NormalTok{F, }\DataTypeTok{data=}\NormalTok{simdat2)}
\end{Highlighting}
\end{Shaded}

\begin{table}[h]
\begin{center}
\begin{threeparttable}
\caption{\label{tab:table8}Regression model using the sum contrast.}
\begin{tabular}{lllll}
\toprule
Predictor & \multicolumn{1}{c}{$Estimate$} & \multicolumn{1}{c}{95\% CI} & \multicolumn{1}{c}{$t(9)$} & \multicolumn{1}{c}{$p$}\\
\midrule
Intercept & 450 & $[437$, $463]$ & 77.62 & < .001\\
FcH01 & 50 & $[32$, $69]$ & 6.11 & < .001\\
FcH02 & 0 & $[-18$, $19]$ & 0.01 & .992\\
\bottomrule
\end{tabular}
\end{threeparttable}
\end{center}
\end{table}

The linear model regression coefficients (see Table \ref{tab:table8}) show the GM response time of \(450\) ms in the intercept. \textcolor{black}{Remember that the first regression coefficient} \texttt{FcH01} \textcolor{black}{was designed to test our first hypothesis that low frequency words are responded to slower than the GM. The regression coefficient} \texttt{FcH01} (\enquote{Estimate}) of \(50\) exactly reflects the difference between low frequency words (\(500\) ms) and the GM of \(450\) ms. The second hypothesis was that response times for medium frequency words differ from the GM. The fact that the second regression coefficient \texttt{FcH02} is exactly \(0\) indicates that response times for medium frequency words (\(450\) ms) are identical with the GM of \(450\) ms. Although there is evidence against the null hypothesis that low-frequency words have the same reading time as the GM, there is no evidence against the null hypothesis that medium frequency words have the same reading times as the GM.

\textcolor{black}{We have now not only derived contrasts and hypothesis tests for the sum contrast, we have also used a powerful and highly general procedure that is used to generate contrasts for many kinds of different hypotheses and experimental designs.}

\hypertarget{further-examples-of-contrasts-illustrated-with-a-factor-with-four-levels}{%
\section{Further examples of contrasts illustrated with a factor with four levels}\label{further-examples-of-contrasts-illustrated-with-a-factor-with-four-levels}}

\begin{figure}

{\centering \includegraphics{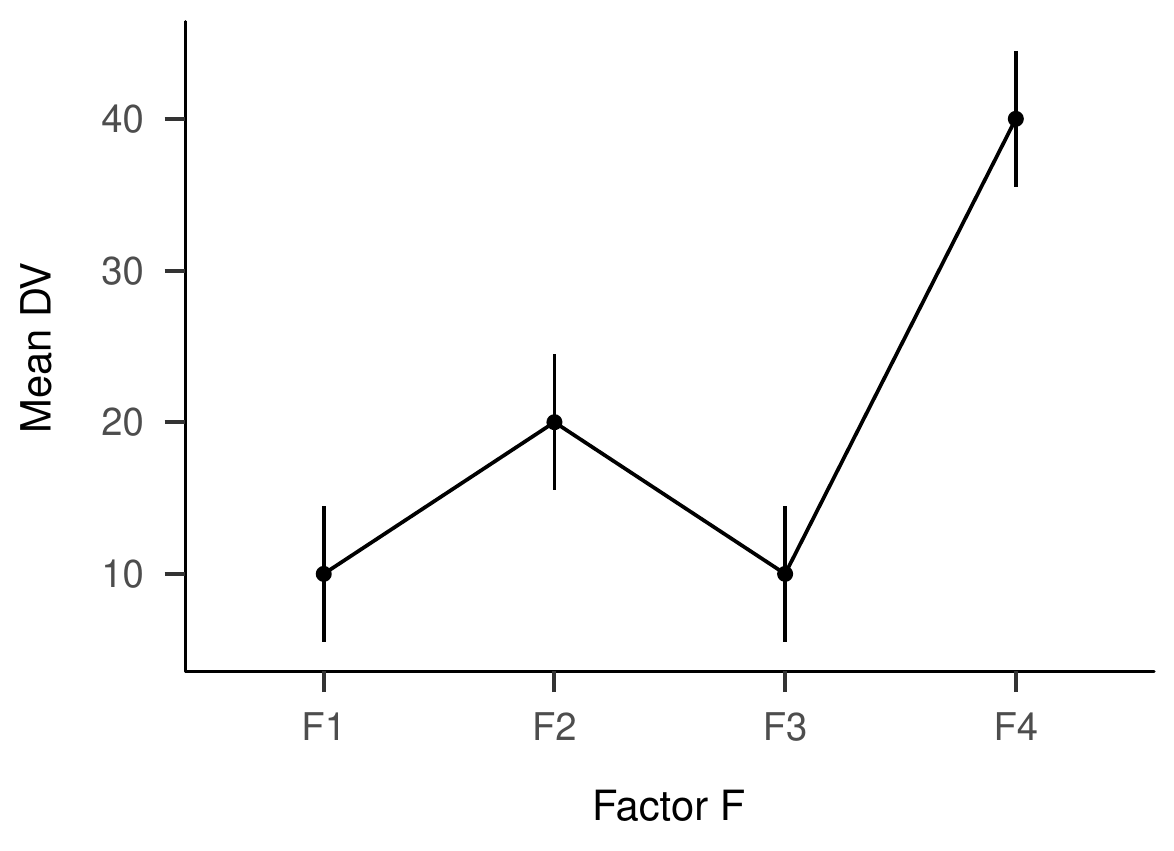} 

}

\caption{Means and error bars (showing standard errors) for a simulated data-set with one between-subjects factor with four levels.}\label{fig:helmertsimdatFig}
\end{figure}

In order to understand \textcolor{black}{\textsc{repeated difference} and \textsc{polynomial contrasts}}, it may be instructive to consider an experiment with one between-subject factor with four levels.
We simulate such a data-set using the function \texttt{mixedDesign}.
The sample sizes for each level and the means and standard errors are shown in Table \ref{tab:helmertsimdatTab}, and the means and standard errors are also shown graphically in Figure \ref{fig:helmertsimdatFig}.

We assume that the four factor levels \texttt{F1} to \texttt{F4} reflect levels of word frequency, including the levels \texttt{low}, \texttt{medium-low}, \texttt{medium-high}, and \texttt{high} frequency words, and that the dependent variable reflects some response time.\footnote{Qualitatively, the simulated pattern of results is actually empirically observed for word frequency effects on single fixation durations (Heister, Würzner, \& Kliegl, 2012).}

\begin{Shaded}
\begin{Highlighting}[]
\CommentTok{# Data, means, and figure}
\NormalTok{M <-}\StringTok{ }\KeywordTok{matrix}\NormalTok{(}\KeywordTok{c}\NormalTok{(}\DecValTok{10}\NormalTok{, }\DecValTok{20}\NormalTok{, }\DecValTok{10}\NormalTok{, }\DecValTok{40}\NormalTok{), }\DataTypeTok{nrow=}\DecValTok{4}\NormalTok{, }\DataTypeTok{ncol=}\DecValTok{1}\NormalTok{, }\DataTypeTok{byrow=}\OtherTok{FALSE}\NormalTok{)}
\KeywordTok{set.seed}\NormalTok{(}\DecValTok{1}\NormalTok{)}
\NormalTok{simdat3 <-}\StringTok{ }\KeywordTok{mixedDesign}\NormalTok{(}\DataTypeTok{B=}\DecValTok{4}\NormalTok{, }\DataTypeTok{W=}\OtherTok{NULL}\NormalTok{, }\DataTypeTok{n=}\DecValTok{5}\NormalTok{, }\DataTypeTok{M=}\NormalTok{M,  }\DataTypeTok{SD=}\DecValTok{10}\NormalTok{, }\DataTypeTok{long =} \OtherTok{TRUE}\NormalTok{) }
\KeywordTok{names}\NormalTok{(simdat3)[}\DecValTok{1}\NormalTok{] <-}\StringTok{ "F"}  \CommentTok{# Rename B_A to F(actor)}
\KeywordTok{levels}\NormalTok{(simdat3}\OperatorTok{$}\NormalTok{F) <-}\StringTok{ }\KeywordTok{c}\NormalTok{(}\StringTok{"F1"}\NormalTok{, }\StringTok{"F2"}\NormalTok{, }\StringTok{"F3"}\NormalTok{, }\StringTok{"F4"}\NormalTok{)}
\NormalTok{table3 <-}\StringTok{ }\NormalTok{simdat3 }\OperatorTok{%>%}\StringTok{ }\KeywordTok{group_by}\NormalTok{(F) }\OperatorTok{%>%}\StringTok{ }
\StringTok{   }\KeywordTok{summarize}\NormalTok{(}\DataTypeTok{N=}\KeywordTok{length}\NormalTok{(DV), }\DataTypeTok{M=}\KeywordTok{mean}\NormalTok{(DV), }\DataTypeTok{SD=}\KeywordTok{sd}\NormalTok{(DV), }\DataTypeTok{SE=}\NormalTok{SD}\OperatorTok{/}\KeywordTok{sqrt}\NormalTok{(N) )}
\NormalTok{(GM <-}\StringTok{  }\KeywordTok{mean}\NormalTok{(table3}\OperatorTok{$}\NormalTok{M)) }\CommentTok{# Grand Mean}
\end{Highlighting}
\end{Shaded}

\begin{verbatim}
## [1] 20
\end{verbatim}

\begin{table}[h]
\begin{center}
\begin{threeparttable}
\caption{\label{tab:helmertsimdatTab}Summary statistics for simulated data with one between-subjects factor with four levels.}
\begin{tabular}{lllll}
\toprule
Factor F & \multicolumn{1}{c}{N data points} & \multicolumn{1}{c}{Estimated means} & \multicolumn{1}{c}{Standard deviations} & \multicolumn{1}{c}{Standard errors}\\
\midrule
F1 & 5 & 10.0 & 10.0 & 4.5\\
F2 & 5 & 20.0 & 10.0 & 4.5\\
F3 & 5 & 10.0 & 10.0 & 4.5\\
F4 & 5 & 40.0 & 10.0 & 4.5\\
\bottomrule
\end{tabular}
\end{threeparttable}
\end{center}
\end{table}

\hypertarget{repeatedcontrasts}{%
\subsection{Repeated contrasts}\label{repeatedcontrasts}}

Arguably, the most popular contrast psychologists and psycholinguists are interested in is the comparison between neighboring levels of a factor. This type of contrast is called \textsc{repeated contrast}. In our example, our research question might be whether the frequency level \texttt{low} leads to slower response times than frequency level \texttt{medium-low}, whether frequency level \texttt{medium-low} leads to slower response times than frequency level \texttt{medium-high}, and whether frequency level \texttt{medium-high} leads to slower response times than frequency level \texttt{high}.

\textcolor{black}{\textsc{Repeated contrasts} are used to implement these comparisons. Consider first how to derive the contrast matrix for \textsc{repeated contrasts}, starting out by specifying the hypotheses that \textcolor{black}{are to be tested} about the data. Importantly, this again applies the general strategy of how to translate (any) hypotheses about differences between groups or conditions into a set of contrasts, yielding a powerful tool of great value in many research settings. We follow the four-step procedure outlined above.}

The first step is to specify our hypotheses, and to write them down in a way such that their weights can be extracted easily. For a four-level factor, the three hypotheses are:

\begin{equation}
H_{0_{2-1}}: -1 \cdot \mu_1 + 1 \cdot \mu_2 + 0 \cdot \mu_3 + 0 \cdot \mu_4 = 0
\end{equation}

\begin{equation}
H_{0_{3-2}}: 0 \cdot \mu_1 - 1 \cdot \mu_2 + 1 \cdot \mu_3 + 0 \cdot \mu_4 = 0
\end{equation}

\begin{equation}
H_{0_{4-3}}: 0 \cdot \mu_1 + 0 \cdot \mu_2 - 1 \cdot \mu_3 + 1 \cdot \mu_4 = 0
\end{equation}

Here, the \(\mu_x\) are the mean response times in condition \(x\). Each hypothesis gives weights to the different condition means. The first hypothesis (\(H_{0_{2-1}}\)) tests the difference between condition mean for \texttt{F2} (\(\mu_2\)) minus the condition mean for \texttt{F1} (\(\mu_1\)), but ignores condition means for \texttt{F3} and \texttt{F4} (\(\mu_3\), \(\mu_4\)). \(\mu_1\) has a weight of \(-1\), \(\mu_2\) has a weight of \(+1\), and \(\mu_3\) and \(\mu_4\) have weights of \(0\). As \textcolor{black}{the} second step, the vector of weights for the first hypothesis \textcolor{black}{is extracted} as \texttt{c2vs1\ \textless{}-\ c(F1=-1,F2=+1,F3=0,F4=0)}. Next, the same thing \textcolor{black}{is done} for the other hypotheses - the weights for all hypotheses \textcolor{black}{are extracted} and coded into a \emph{hypothesis matrix} in R:

\begin{Shaded}
\begin{Highlighting}[]
\KeywordTok{t}\NormalTok{(HcRE <-}\StringTok{ }\KeywordTok{rbind}\NormalTok{(}\DataTypeTok{c2vs1=}\KeywordTok{c}\NormalTok{(}\DataTypeTok{F1=}\OperatorTok{-}\DecValTok{1}\NormalTok{,}\DataTypeTok{F2=}\OperatorTok{+}\DecValTok{1}\NormalTok{,}\DataTypeTok{F3=} \DecValTok{0}\NormalTok{,}\DataTypeTok{F4=} \DecValTok{0}\NormalTok{),}
                \DataTypeTok{c3vs2=}\KeywordTok{c}\NormalTok{(}\DataTypeTok{F1=} \DecValTok{0}\NormalTok{,}\DataTypeTok{F2=}\OperatorTok{-}\DecValTok{1}\NormalTok{,}\DataTypeTok{F3=}\OperatorTok{+}\DecValTok{1}\NormalTok{,}\DataTypeTok{F4=} \DecValTok{0}\NormalTok{),}
                \DataTypeTok{c4vs3=}\KeywordTok{c}\NormalTok{(}\DataTypeTok{F1=} \DecValTok{0}\NormalTok{,}\DataTypeTok{F2=} \DecValTok{0}\NormalTok{,}\DataTypeTok{F3=}\OperatorTok{-}\DecValTok{1}\NormalTok{,}\DataTypeTok{F4=}\OperatorTok{+}\DecValTok{1}\NormalTok{)))}
\end{Highlighting}
\end{Shaded}

\begin{verbatim}
##    c2vs1 c3vs2 c4vs3
## F1    -1     0     0
## F2     1    -1     0
## F3     0     1    -1
## F4     0     0     1
\end{verbatim}

\textcolor{black}{Again, we show the transposed version of the hypothesis matrix (switching rows and columns), but now we leave out the hypothesis for the intercept (we discuss below when this can be neglected).}

Next, the new contrast matrix \texttt{XcRE} \textcolor{black}{is obtained}. This is the contrast matrix \(X_c\) that exactly tests the hypotheses written down above:

\begin{Shaded}
\begin{Highlighting}[]
\NormalTok{(XcRE <-}\StringTok{ }\KeywordTok{ginv2}\NormalTok{(HcRE))}
\end{Highlighting}
\end{Shaded}

\begin{verbatim}
##    c2vs1 c3vs2 c4vs3
## F1 -3/4  -1/2  -1/4 
## F2  1/4  -1/2  -1/4 
## F3  1/4   1/2  -1/4 
## F4  1/4   1/2   3/4
\end{verbatim}

\textcolor{black}{In the case of the \textsc{repeated contrast}, the contrast matrix again looks very different from the hypothesis matrix.} In this case, the contrast matrix looks a lot less intuitive than the hypothesis matrix, and if one did not know the associated hypothesis matrix, it seems unclear what the contrast matrix would actually test. \textcolor{black}{To verify this custom-made contrast matrix, we compare it to the \textsc{repeated contrast} matrix as generated by the R function} \texttt{contr.sdif()} \textcolor{black}{in the \texttt{MASS} package} (Venables \& Ripley, 2002)\textcolor{black}{. The resulting contrast matrix is identical to our result:}

\begin{Shaded}
\begin{Highlighting}[]
\KeywordTok{fractions}\NormalTok{(}\KeywordTok{contr.sdif}\NormalTok{(}\DecValTok{4}\NormalTok{))}
\end{Highlighting}
\end{Shaded}

\begin{verbatim}
##   2-1  3-2  4-3 
## 1 -3/4 -1/2 -1/4
## 2  1/4 -1/2 -1/4
## 3  1/4  1/2 -1/4
## 4  1/4  1/2  3/4
\end{verbatim}

Step four in the procedure is to apply \textsc{repeated contrasts} to the factor \texttt{F} in the example data, and to run a linear model. This allows us to estimate the regression coefficients associated with each contrast. \textcolor{black}{These are compared} to the data in Figure~\ref{fig:helmertsimdatFig} \textcolor{black}{to test whether the regression coefficients actually correspond to the differences between successive condition means, as intended.}

\begin{Shaded}
\begin{Highlighting}[]
\KeywordTok{contrasts}\NormalTok{(simdat3}\OperatorTok{$}\NormalTok{F) <-}\StringTok{ }\NormalTok{XcRE}
\NormalTok{m2_mr <-}\StringTok{ }\KeywordTok{lm}\NormalTok{(DV }\OperatorTok{~}\StringTok{ }\DecValTok{1} \OperatorTok{+}\StringTok{ }\NormalTok{F, }\DataTypeTok{data=}\NormalTok{simdat3)}
\end{Highlighting}
\end{Shaded}

\begin{table}[h]
\begin{center}
\begin{threeparttable}
\caption{\label{tab:table0}Repeated contrasts. \label{tab:table0}}
\begin{tabular}{lllll}
\toprule
Predictor & \multicolumn{1}{c}{$Estimate$} & \multicolumn{1}{c}{95\% CI} & \multicolumn{1}{c}{$t(16)$} & \multicolumn{1}{c}{$p$}\\
\midrule
Intercept & 20 & $[15$, $25]$ & 8.94 & < .001\\
Fc2vs1 & 10 & $[-3$, $23]$ & 1.58 & .133\\
Fc3vs2 & -10 & $[-23$, $3]$ & -1.58 & .133\\
Fc4vs3 & 30 & $[17$, $43]$ & 4.74 & < .001\\
\bottomrule
\end{tabular}
\end{threeparttable}
\end{center}
\end{table}

The results (see Table~\ref{tab:table0}) show that as expected, the regression coefficients reflect exactly the differences \textcolor{black}{that were of interest}: the regression coefficient (\enquote{Estimate}) \texttt{Fc2vs1} has a value of \(10\), which exactly corresponds to the difference between condition mean for \texttt{F2} (\(20\)) minus condition mean for \texttt{F1} (\(10\)), i.e., \(20 - 10 = 10\). Likewise, the regression coefficient \texttt{Fc3vs2} has a value of \(-10\), which corresponds to the difference between condition mean for \texttt{F3} (\(10\)) minus condition mean for \texttt{F2} (\(20\)), i.e., \(10 - 20 = -10\). Finally, the regression coefficient \texttt{Fc4vs3} has a value of \(30\), which reflects the difference between condition \texttt{F4} (\(40\)) minus condition \texttt{F3} (\(10\)), i.e., \(40 - 10 = 30\). Thus, the regression coefficients reflect differences between successive or neighboring condition means, and test the corresponding null hypotheses.

To sum up, formally writing down the hypotheses, extracting the weights into a hypothesis matrix, and applying the generalized matrix inverse operation yields a set of contrast coefficients that provide the desired estimates. This procedure is very general: it allows us to derive the contrast matrix corresponding to any set of hypotheses that one may want to test. The four-step procedure described above allows us to construct contrast matrices that are among the standard set of contrasts in R (\textsc{repeated contrasts} or \textsc{sum contrasts}, etc.), and also allows us to construct non-standard custom contrasts that are specifically tailored to the particular hypotheses one wants to test. The hypothesis matrix and the contrast matrix are linked by the generalized inverse; understanding this link is the key ingredient to understanding contrasts in diverse settings.

\hypertarget{contrasts-in-linear-regression-analysis-the-design-or-model-matrix}{%
\subsection{Contrasts in linear regression analysis: The design or model matrix}\label{contrasts-in-linear-regression-analysis-the-design-or-model-matrix}}

We have now discussed how different contrasts are created from the hypothesis matrix. However, we have not treated in detail how exactly contrasts are used in a linear model. Here, we will see that the contrasts for a factor in a linear model are just the same thing as continuous numeric predictors (i.e., covariates) in a linear/multiple regression analysis. That is, contrasts are \textcolor{black}{the} way to encode discrete factor levels into numeric predictor variables to use in linear/multiple regression analysis, by encoding which differences between factor levels are tested.
The contrast matrix \(X_c\) that we have looked at so far has one entry (row) for each experimental condition. For use in a linear model, however, the contrast matrix is coded into a design or model matrix \(X\), where each individual data point has one row. The design matrix \(X\) can be extracted using the function \texttt{model.matrix()}:

\begin{Shaded}
\begin{Highlighting}[]
\NormalTok{(}\KeywordTok{contrasts}\NormalTok{(simdat3}\OperatorTok{$}\NormalTok{F) <-}\StringTok{ }\NormalTok{XcRE) }\CommentTok{# contrast matrix}
\end{Highlighting}
\end{Shaded}

\begin{verbatim}
##    c2vs1 c3vs2 c4vs3
## F1 -3/4  -1/2  -1/4 
## F2  1/4  -1/2  -1/4 
## F3  1/4   1/2  -1/4 
## F4  1/4   1/2   3/4
\end{verbatim}

\begin{Shaded}
\begin{Highlighting}[]
\NormalTok{(covars <-}\StringTok{ }\KeywordTok{as.data.frame}\NormalTok{(}\KeywordTok{model.matrix}\NormalTok{(}\OperatorTok{~}\StringTok{ }\DecValTok{1} \OperatorTok{+}\StringTok{ }\NormalTok{F, simdat3))) }\CommentTok{# design matrix}
\end{Highlighting}
\end{Shaded}

\begin{verbatim}
##    (Intercept) Fc2vs1 Fc3vs2 Fc4vs3
## 1            1  -0.75   -0.5  -0.25
## 2            1  -0.75   -0.5  -0.25
## 3            1  -0.75   -0.5  -0.25
## 4            1  -0.75   -0.5  -0.25
## 5            1  -0.75   -0.5  -0.25
## 6            1   0.25   -0.5  -0.25
## 7            1   0.25   -0.5  -0.25
## 8            1   0.25   -0.5  -0.25
## 9            1   0.25   -0.5  -0.25
## 10           1   0.25   -0.5  -0.25
## 11           1   0.25    0.5  -0.25
## 12           1   0.25    0.5  -0.25
## 13           1   0.25    0.5  -0.25
## 14           1   0.25    0.5  -0.25
## 15           1   0.25    0.5  -0.25
## 16           1   0.25    0.5   0.75
## 17           1   0.25    0.5   0.75
## 18           1   0.25    0.5   0.75
## 19           1   0.25    0.5   0.75
## 20           1   0.25    0.5   0.75
\end{verbatim}

\textcolor{black}{For each of the $20$ subjects, four numbers are stored in this model matrix. They represent the three values of three predictor variables used to predict response times in the task. Indeed, this matrix is exactly the design matrix $X$ commonly used in multiple regression analysis, where each column represents one numeric predictor variable (covariate), and the first column codes the intercept term.}

To further illustrate this, the covariates \textcolor{black}{are extracted} from this design matrix and stored separately as numeric predictor variables in the data-frame:

\begin{Shaded}
\begin{Highlighting}[]
\NormalTok{simdat3[,}\KeywordTok{c}\NormalTok{(}\StringTok{"Fc2vs1"}\NormalTok{,}\StringTok{"Fc3vs2"}\NormalTok{,}\StringTok{"Fc4vs3"}\NormalTok{)] <-}\StringTok{ }\NormalTok{covars[,}\DecValTok{2}\OperatorTok{:}\DecValTok{4}\NormalTok{]}
\end{Highlighting}
\end{Shaded}

They are now used as numeric predictor variables in a multiple regression analysis:

\begin{Shaded}
\begin{Highlighting}[]
\NormalTok{m3_mr <-}\StringTok{ }\KeywordTok{lm}\NormalTok{(DV }\OperatorTok{~}\StringTok{ }\DecValTok{1} \OperatorTok{+}\StringTok{ }\NormalTok{Fc2vs1 }\OperatorTok{+}\StringTok{ }\NormalTok{Fc3vs2 }\OperatorTok{+}\StringTok{ }\NormalTok{Fc4vs3, }\DataTypeTok{data=}\NormalTok{simdat3)}
\end{Highlighting}
\end{Shaded}

\begin{table}[h]
\begin{center}
\begin{threeparttable}
\caption{\label{tab:table0.3}Repeated contrasts as linear regression.}
\begin{tabular}{lllll}
\toprule
Predictor & \multicolumn{1}{c}{$Estimate$} & \multicolumn{1}{c}{95\% CI} & \multicolumn{1}{c}{$t(16)$} & \multicolumn{1}{c}{$p$}\\
\midrule
Intercept & 20 & $[15$, $25]$ & 8.94 & < .001\\
Fc2vs1 & 10 & $[-3$, $23]$ & 1.58 & .133\\
Fc3vs2 & -10 & $[-23$, $3]$ & -1.58 & .133\\
Fc4vs3 & 30 & $[17$, $43]$ & 4.74 & < .001\\
\bottomrule
\end{tabular}
\end{threeparttable}
\end{center}
\end{table}

The results show that the regression coefficients are exactly the same as in the contrast-based analysis shown in the previous section (Table \ref{tab:table0}). This demonstrates that contrasts serve to code discrete factor levels into a linear/multiple regression analysis by numerically encoding comparisons between specific condition means.

\begin{figure}

{\centering \includegraphics{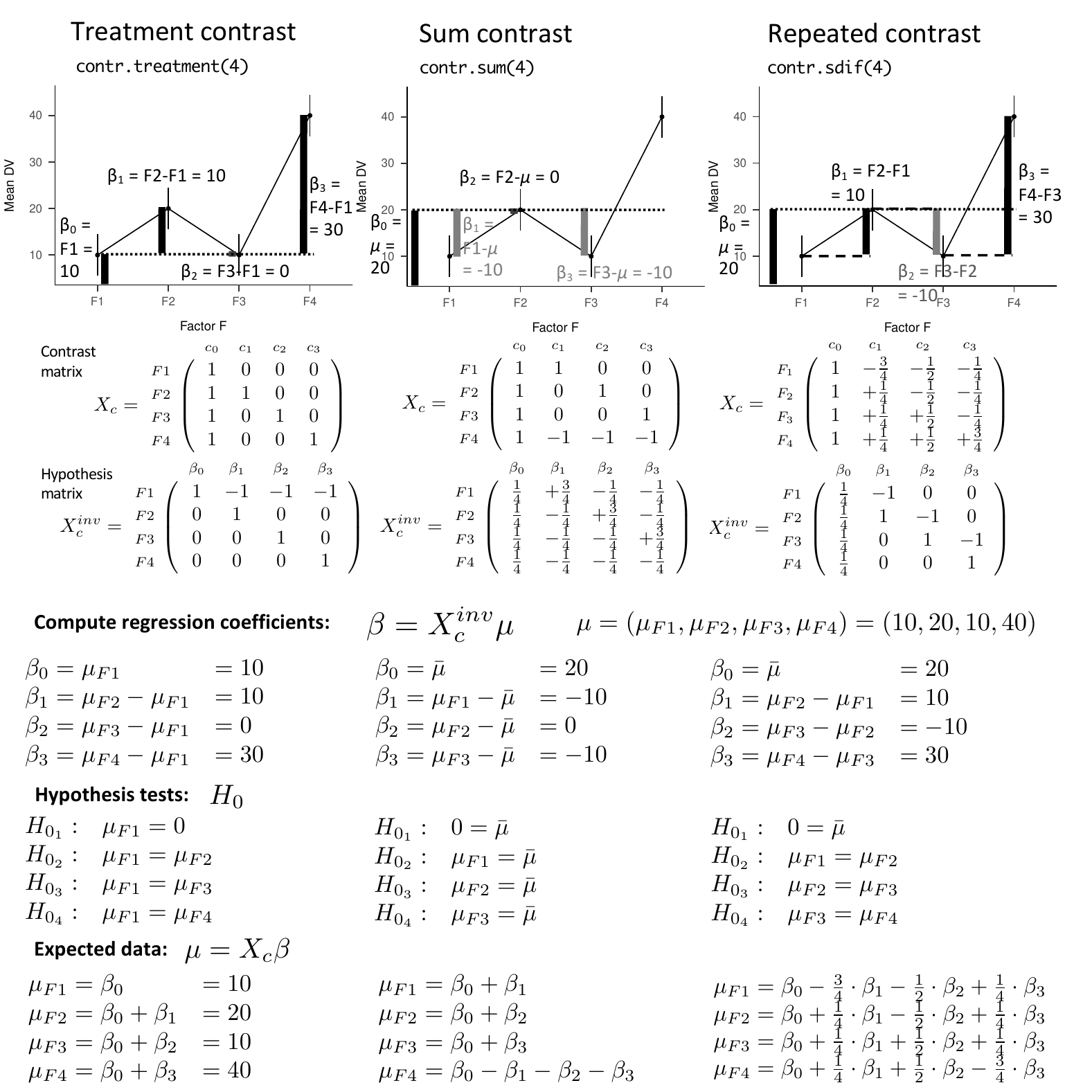} 

}

\caption{Overview of contrasts including treatment, sum, and repeated contrasts. From top to bottom panels, we illustrate the computation of regression coefficients, show the contrast and hypothesis matrices, formulas for computing regression coefficients, the null hypotheses tested by each coefficient, and formulas for estimated data.}\label{fig:Overview}
\end{figure}

Figure \ref{fig:Overview} provides an overview of the introduced contrasts.

\FloatBarrier

\hypertarget{polynomialContrasts}{%
\subsection{Polynomial contrasts}\label{polynomialContrasts}}

\textsc{Polynomial contrasts} are another option for analyzing factors. Suppose that we expect a linear trend across conditions, where the response increases by a constant magnitude with each successive factor level. This could be \textcolor{black}{the} expectation when four levels of a factor reflect decreasing levels of word frequency (i.e., four factor levels: high, medium-high, medium-low, and low word frequency), where \textcolor{black}{one} expects the lowest response for high frequency words, and successively higher responses for \textcolor{black}{lower} word frequencies. The effect for each individual level of a factor may \textcolor{black}{not be strong enough for detecting} it in the statistical model. Specifying a linear trend in a polynomial constrast allows us to pool the whole increase into a single coefficient for the linear trend, increasing statistical power to detect the increase. Such a specification constrains the hypothesis to one interpretable degree of freedom, e.g., a linear increase across factor levels. The larger the number of factor levels, the more parsimonious are \textsc{polynomial contrasts} compared to contrast-based specifications as introduced in the previous sections or compared to an omnibus F-test. Going beyond a linear trend, one may also have expectations about quadratic trends. For example, one may expect an increase only among very low frequency words, but no difference between high and medium-high frequency words.

\begin{Shaded}
\begin{Highlighting}[]
\NormalTok{Xpol <-}\StringTok{ }\KeywordTok{contr.poly}\NormalTok{(}\DecValTok{4}\NormalTok{)}
\NormalTok{(}\KeywordTok{contrasts}\NormalTok{(simdat3}\OperatorTok{$}\NormalTok{F) <-}\StringTok{ }\NormalTok{Xpol)}
\end{Highlighting}
\end{Shaded}

\begin{verbatim}
##          .L   .Q     .C
## [1,] -0.671  0.5 -0.224
## [2,] -0.224 -0.5  0.671
## [3,]  0.224 -0.5 -0.671
## [4,]  0.671  0.5  0.224
\end{verbatim}

\begin{Shaded}
\begin{Highlighting}[]
\NormalTok{m1_mr.Xpol <-}\StringTok{ }\KeywordTok{lm}\NormalTok{(DV }\OperatorTok{~}\StringTok{ }\DecValTok{1} \OperatorTok{+}\StringTok{ }\NormalTok{F, }\DataTypeTok{data=}\NormalTok{simdat3)}
\end{Highlighting}
\end{Shaded}

\begin{table}[h]
\begin{center}
\begin{threeparttable}
\caption{\label{tab:table14polA}Polynomial contrasts.}
\begin{tabular}{lllll}
\toprule
Predictor & \multicolumn{1}{c}{$Estimate$} & \multicolumn{1}{c}{95\% CI} & \multicolumn{1}{c}{$t(16)$} & \multicolumn{1}{c}{$p$}\\
\midrule
Intercept & 20 & $[15$, $25]$ & 8.94 & < .001\\
F L & 18 & $[8$, $27]$ & 4.00 & .001\\
F Q & 10 & $[1$, $19]$ & 2.24 & .040\\
F C & 13 & $[4$, $23]$ & 3.00 & .008\\
\bottomrule
\end{tabular}
\end{threeparttable}
\end{center}
\end{table}

In this example (see Table \ref{tab:table14polA}), condition means increase across factor levels in a linear fashion, but the quadratic and cubic trends are also significant.

\hypertarget{customContrasts}{%
\subsection{Custom contrasts}\label{customContrasts}}

Sometimes, a hypothesis about a pattern of means takes a form that cannot be expressed by the standard sets of contrasts available in R. For example, a theory or model may make quantitative predictions about the expected pattern of means. Alternatively, prior empirical research findings or logical reasoning may suggest a specific qualitative pattern. Such predictions could be quantitatively constrained when they come from a computational or mathematical model, but when a theory only predicts a qualitative pattern, these predictions can be represented by choosing some plausible values for the means (Baguley, 2012). For example, assume that a theory predicts for the pattern of means presented in Figure \ref{fig:helmertsimdatFig} that the first two means (for \texttt{F1} and \texttt{F2}) are identical, but that means for levels \texttt{F3} and \texttt{F4} increase linearly. One starts approximating a contrast by giving a potential expected outcome of means, such as \texttt{M\ =\ c(10,\ 10,\ 20,\ 30)}. \textcolor{black}{It is possible to} turn these predicted means into a contrast by centering them (i.e., subtracting the mean of \(17.5\)): \texttt{M\ =\ c(-7.5,\ -7.5,\ 2.5,\ 12.5)}. This already works as a contrast. \textcolor{black}{It is possible to} further simplify this by dividing by \(2.5\), which yields \texttt{M\ =\ c(-3,\ -3,\ 1,\ 5)}. We will use this contrast in a regression model. \textcolor{black}{Notice that if you have $I$ conditions, you can specify $I-1$ contrasts. However, it is also possible to specify less contrasts than this maximum number of $I-1$. The example below illustrates this point:}

\begin{Shaded}
\begin{Highlighting}[]
\NormalTok{(}\KeywordTok{contrasts}\NormalTok{(simdat3}\OperatorTok{$}\NormalTok{F) <-}\StringTok{ }\KeywordTok{cbind}\NormalTok{(}\KeywordTok{c}\NormalTok{(}\OperatorTok{-}\DecValTok{3}\NormalTok{, }\DecValTok{-3}\NormalTok{, }\DecValTok{1}\NormalTok{, }\DecValTok{5}\NormalTok{)))}
\end{Highlighting}
\end{Shaded}

\begin{verbatim}
##      [,1]
## [1,]   -3
## [2,]   -3
## [3,]    1
## [4,]    5
\end{verbatim}

\begin{Shaded}
\begin{Highlighting}[]
\NormalTok{C <-}\StringTok{  }\KeywordTok{model.matrix}\NormalTok{(}\OperatorTok{~}\StringTok{ }\DecValTok{1} \OperatorTok{+}\StringTok{ }\NormalTok{F, }\DataTypeTok{data=}\NormalTok{simdat3)}
\NormalTok{simdat3}\OperatorTok{$}\NormalTok{Fcust <-}\StringTok{ }\NormalTok{C[,}\StringTok{"F1"}\NormalTok{]}
\NormalTok{m1_mr.Xcust <-}\StringTok{ }\KeywordTok{lm}\NormalTok{(DV }\OperatorTok{~}\StringTok{ }\DecValTok{1} \OperatorTok{+}\StringTok{ }\NormalTok{Fcust, }\DataTypeTok{data=}\NormalTok{simdat3)}
\end{Highlighting}
\end{Shaded}

\begin{table}[h]
\begin{center}
\begin{threeparttable}
\caption{\label{tab:table14cust}Custom contrasts.}
\begin{tabular}{lllll}
\toprule
Predictor & \multicolumn{1}{c}{$Estimate$} & \multicolumn{1}{c}{95\% CI} & \multicolumn{1}{c}{$t(18)$} & \multicolumn{1}{c}{$p$}\\
\midrule
Intercept & 20 & $[14$, $26]$ & 6.97 & < .001\\
Fcust & 3 & $[1$, $5]$ & 3.15 & .006\\
\bottomrule
\end{tabular}
\end{threeparttable}
\end{center}
\end{table}

For cases where a qualitative pattern of means can be realized with more than one set of quantitative values, Baguley (2012) notes that often the precise numbers may not be decisive. He also suggests selecting the simplest set of integer numbers that matches the desired pattern.

\hypertarget{nonOrthogonal}{%
\section{What makes a good set of contrasts?}\label{nonOrthogonal}}

Contrasts decompose ANOVA omnibus F tests into several component comparisons (Baguley, 2012). Orthogonal contrasts decompose the sum of squares of the F test into additive independent subcomponents, which allows for clarity in interpreting each effect.
As mentioned earlier, for a factor with \(I\) levels one can make \(I-1\) comparisons. For example, in a design with one factor with two levels, only one comparison is possible (between the two factor levels). \textcolor{black}{More generally, if we have a factor with $I_1$ and another factor with $I_2$ levels, then the total number of conditions is $I_1\times I_2 = \nu$ (not $I_1 + I_2$!), which implies a maximum of $\nu-1$  contrasts.}

For example, in a design with one factor with three levels, A, B, and C, in principle one could make three comparisons (A vs.~B, A vs.~C, B vs.~C).
However, after defining an intercept, only two means can be compared. Therefore, for a factor with three levels, we define two comparisons within one statistical model. F tests are nothing but combinations, or bundles of contrasts. F tests are less specific and they lack focus, but they are useful when the hypothesis in question is vague. However, a significant F test leaves unclear what effects the data actually show. Contrasts are very useful to test specific effects in the data.

\textcolor{black}{One critical precondition for contrasts is that they implement different hypotheses that are not collinear, that is, that none of the contrasts can be generated from the other contrasts by linear combination. For example, the contrast} \texttt{c1\ =\ c(1,2,3)} \textcolor{black}{can be generated from the contrast} \texttt{c2\ =\ c(3,4,5)} \textcolor{black}{simply by computing} \texttt{c2\ -\ 2}\textcolor{black}{. Therefore, contrasts c1 and c2 cannot be used simultaneously. That is, each contrast needs to encode some independent information about the data. Otherwise, the model cannot be estimated, and the lm() function gives an error, indicating that the design matrix is "rank deficient".}

There are (at least) two criteria to decide what a good contrast is. First, \textit{orthogonal contrasts} have advantages as they test mutually independent hypotheses about the data (see Dobson \& Barnett, 2011, sec. 6.2.5, p.~91 for a detailed explanation of orthogonality). \textcolor{black}{Second, it is} crucial that contrasts are defined in a way such that they answer the research questions. \textcolor{black}{This second point is crucial. One way to accomplish this, is to use the hypothesis matrix to generate contrasts, as this ensures that one uses contrasts that exactly test the hypotheses of interest in a given study.}

\hypertarget{centered-contrasts}{%
\subsection{Centered contrasts}\label{centered-contrasts}}

Contrasts are often constrained to be centered, such that the individual contrast coefficients \(c_i\) \textcolor{black}{for different factor levels $i$ sum to $0$: $\sum_{i=1}^I c_i = 0$. This has advantages when testing interactions with other factors or covariates (we discuss interactions between factors below).
All contrasts discussed here are centered except for the \textsc{treatment contrast}, in which the contrast coefficients for each contrast do not sum to zero:}

\begin{Shaded}
\begin{Highlighting}[]
\KeywordTok{colSums}\NormalTok{(}\KeywordTok{contr.treatment}\NormalTok{(}\DecValTok{4}\NormalTok{))}
\end{Highlighting}
\end{Shaded}

\begin{verbatim}
## 2 3 4 
## 1 1 1
\end{verbatim}

\textcolor{black}{Other contrasts, such as \textsc{repeated contrasts}, are centered and the contrast coefficients for each contrast sum to $0$:}

\begin{Shaded}
\begin{Highlighting}[]
\KeywordTok{colSums}\NormalTok{(}\KeywordTok{contr.sdif}\NormalTok{(}\DecValTok{4}\NormalTok{))}
\end{Highlighting}
\end{Shaded}

\begin{verbatim}
## 2-1 3-2 4-3 
##   0   0   0
\end{verbatim}

The contrast coefficients mentioned above appear in the contrast matrix. By contrast, the weights in the hypothesis matrix are always centered. This is also true for the \textsc{treatment contrast}. The reason is that they code hypotheses, which always relate to comparisons between conditions or bundles of conditions.
The only exception are the weights for the intercept, which always sum to \(1\) in the hypothesis matrix. This is done to ensure that when applying the generalized matrix inverse, the intercept results in a constant term with values of \(1\) in the contrast matrix. That the intercept is coded by a column of \(1\)s in the contrast matrix accords to convention as it provides a scaling of the intercept coefficient that is simple to interpret.
\textcolor{black}{An important question concerns whether (or when) the intercept needs to be considered in the generalized matrix inversion, and whether (or when) it can be ignored. This question is closely related to the concept of orthogonal contrasts, a concept we turn to below.}

\hypertarget{orthogonal-contrasts}{%
\subsection{Orthogonal contrasts}\label{orthogonal-contrasts}}

Two centered contrasts \(c_1\) and \(c_2\) are orthogonal to each other if the following condition applies. Here, \(i\) is the \(i\)-th cell of the vector representing the contrast.

\begin{equation}
\sum_{i=1}^I c_{1,i} \cdot c_{2,i} = 0
\end{equation}

Orthogonality can be determined easily in R by computing the correlation between two contrasts. Orthogonal contrasts have a correlation of \(0\). Contrasts are therefore just a special case for the general case of predictors in regression models, where two numeric predictor variables are orthogonal if they are un-correlated.

For example, coding two factors in a \(2 \times 2\) design (we return to this case in a section on ANOVA below) using \textsc{sum contrasts}, these sum contrasts and their interaction are orthogonal to each other:

\begin{Shaded}
\begin{Highlighting}[]
\NormalTok{(Xsum <-}\StringTok{ }\KeywordTok{cbind}\NormalTok{(}\DataTypeTok{F1=}\KeywordTok{c}\NormalTok{(}\DecValTok{1}\NormalTok{,}\DecValTok{1}\NormalTok{,}\OperatorTok{-}\DecValTok{1}\NormalTok{,}\OperatorTok{-}\DecValTok{1}\NormalTok{), }\DataTypeTok{F2=}\KeywordTok{c}\NormalTok{(}\DecValTok{1}\NormalTok{,}\OperatorTok{-}\DecValTok{1}\NormalTok{,}\DecValTok{1}\NormalTok{,}\OperatorTok{-}\DecValTok{1}\NormalTok{), }\DataTypeTok{F1xF2=}\KeywordTok{c}\NormalTok{(}\DecValTok{1}\NormalTok{,}\OperatorTok{-}\DecValTok{1}\NormalTok{,}\OperatorTok{-}\DecValTok{1}\NormalTok{,}\DecValTok{1}\NormalTok{)))}
\end{Highlighting}
\end{Shaded}

\begin{verbatim}
##      F1 F2 F1xF2
## [1,]  1  1     1
## [2,]  1 -1    -1
## [3,] -1  1    -1
## [4,] -1 -1     1
\end{verbatim}

\begin{Shaded}
\begin{Highlighting}[]
\KeywordTok{cor}\NormalTok{(Xsum)}
\end{Highlighting}
\end{Shaded}

\begin{verbatim}
##       F1 F2 F1xF2
## F1     1  0     0
## F2     0  1     0
## F1xF2  0  0     1
\end{verbatim}

\noindent
Notice that the correlations between the different contrasts (i.e., the off-diagonals) are exactly \(0\). \textsc{Sum contrasts} coding one multi-level factor, however, are not orthogonal to each other:

\begin{Shaded}
\begin{Highlighting}[]
\KeywordTok{cor}\NormalTok{(}\KeywordTok{contr.sum}\NormalTok{(}\DecValTok{4}\NormalTok{))}
\end{Highlighting}
\end{Shaded}

\begin{verbatim}
##      [,1] [,2] [,3]
## [1,]  1.0  0.5  0.5
## [2,]  0.5  1.0  0.5
## [3,]  0.5  0.5  1.0
\end{verbatim}

\noindent
Here, the correlations between individual contrasts, which appear in the off-diagonals, deviate from \(0\), indicating non-orthogonality. The same is also true for \textsc{treatment} and \textsc{repeated contrasts}:

\begin{Shaded}
\begin{Highlighting}[]
\KeywordTok{cor}\NormalTok{(}\KeywordTok{contr.sdif}\NormalTok{(}\DecValTok{4}\NormalTok{))}
\end{Highlighting}
\end{Shaded}

\begin{verbatim}
##       2-1   3-2   4-3
## 2-1 1.000 0.577 0.333
## 3-2 0.577 1.000 0.577
## 4-3 0.333 0.577 1.000
\end{verbatim}

\begin{Shaded}
\begin{Highlighting}[]
\KeywordTok{cor}\NormalTok{(}\KeywordTok{contr.treatment}\NormalTok{(}\DecValTok{4}\NormalTok{))}
\end{Highlighting}
\end{Shaded}

\begin{verbatim}
##        2      3      4
## 2  1.000 -0.333 -0.333
## 3 -0.333  1.000 -0.333
## 4 -0.333 -0.333  1.000
\end{verbatim}

Orthogonality of contrasts plays a critical role when computing the generalized inverse. In the inversion operation, orthogonal contrasts are converted independently from each other. That is, the presence or absence of another orthogonal contrast does not change the resulting weights. \textcolor{black}{In fact, for orthogonal contrasts, applying the generalized matrix inverse to the hypothesis matrix simply produces a scaled version of the hypothesis matrix into the contrast matrix 
(for mathematical details see Appendix}~\ref{app:InverseOperation}).

\textcolor{black}{The crucial point here is the following.  As long as contrasts are fully orthogonal, and as long as one does not care about the scaling of predictors, it is not necessary to use the generalized matrix inverse, and one can code the contrast matrix directly. However, when scaling is of interest, or when non-orthogonal or non-centered contrasts are involved, then the generalized inverse formulation of the hypothesis matrix is needed to specify contrasts correctly.}

\hypertarget{the-role-of-the-intercept-in-non-centered-contrasts}{%
\subsection{The role of the intercept in non-centered contrasts}\label{the-role-of-the-intercept-in-non-centered-contrasts}}

A related question concerns whether the intercept needs to be considered when computing the generalized inverse for a contrast. It turns out that considering the intercept is necessary for contrasts that are not centered. This is the case for \textsc{treatment contrasts} which are not centered; e.g., the treatment contrast for two factor levels \texttt{c1vs0\ =\ c(0,1)}: \(\sum_i c_i = 0 + 1 = 1\). One can actually show that the formula to determine whether contrasts are centered (i.e., \(\sum_i c_i = 0\)) is the same formula as the formula to test whether a contrast is \enquote{orthogonal to the intercept}. Remember that for the intercept, all contrast coefficients are equal to one: \(c_{1,i} = 1\) \textcolor{black}{(here, $c_{1,i}$ indicates the vector of contrast coefficients associated with the intercept)}. We enter these contrast coefficient values into the formula testing whether a contrast is orthogonal to the intercept \textcolor{black}{(here, $c_{2,i}$ indicates the vector of contrast coefficients associated with some contrast for which we want to test whether it is "orthogonal to the intercept")}: \(\sum_i c_{1,i} \cdot c_{2,i} = \sum_i 1 \cdot c_{2,i} = \sum_i c_{2,i} = 0\). The resulting formula is: \(\sum_i c_{2,i} = 0\), which is exactly the formula for whether a contrast is centered. Because of this analogy, \textsc{treatment contrasts} can be viewed to be `not orthogonal to the intercept'. \textcolor{black}{This means that the intercept needs to be considered when computing the generalized inverse for treatment contrasts. As we have discussed above, when the intercept is included in the hypothesis matrix, the weights for this intercept term should sum to one, as this yields a column of ones for the intercept term in the contrast matrix.}

\hypertarget{a-closer-look-at-hypothesis-and-contrast-matrices}{%
\section{A closer look at hypothesis and contrast matrices}\label{a-closer-look-at-hypothesis-and-contrast-matrices}}

\hypertarget{inverting-the-procedure-from-a-contrast-matrix-to-the-associated-hypothesis-matrix}{%
\subsection{Inverting the procedure: From a contrast matrix to the associated hypothesis matrix}\label{inverting-the-procedure-from-a-contrast-matrix-to-the-associated-hypothesis-matrix}}

\textcolor{black}{One important point to appreciate about the generalized inverse matrix operation is that applying the inverse twice yields back the original matrix. 
It follows that applying the inverse operation twice to the hypothesis matrix $H_c$ yields back the original hypothesis matrix: $(H_c^{inv})^{inv} = H_c$. For example, let us look at the hypothesis matrix of a \textsc{repeated contrast}:}

\begin{Shaded}
\begin{Highlighting}[]
\KeywordTok{t}\NormalTok{(HcRE <-}\StringTok{ }\KeywordTok{rbind}\NormalTok{(}\DataTypeTok{c2vs1=}\KeywordTok{c}\NormalTok{(}\DataTypeTok{F1=}\OperatorTok{-}\DecValTok{1}\NormalTok{,}\DataTypeTok{F2=} \DecValTok{1}\NormalTok{,}\DataTypeTok{F3=} \DecValTok{0}\NormalTok{), }
                \DataTypeTok{c3vs1=}\KeywordTok{c}\NormalTok{(    }\DecValTok{0}\NormalTok{,   }\DecValTok{-1}\NormalTok{,    }\DecValTok{1}\NormalTok{)))}
\end{Highlighting}
\end{Shaded}

\begin{verbatim}
##    c2vs1 c3vs1
## F1    -1     0
## F2     1    -1
## F3     0     1
\end{verbatim}

\begin{Shaded}
\begin{Highlighting}[]
\KeywordTok{t}\NormalTok{(}\KeywordTok{ginv2}\NormalTok{(}\KeywordTok{ginv2}\NormalTok{(HcRE)))}
\end{Highlighting}
\end{Shaded}

\begin{verbatim}
##    c2vs1 c3vs1
## F1 -1     0   
## F2  1    -1   
## F3  0     1
\end{verbatim}

It is clear that applying the generalized inverse twice to the hypothesis matrix yields back the same matrix.
This also implies that taking the contrast matrix \(X_c\) (i.e., \(X_c = H_c^{inv}\)), and applying the generalized inverse operation, gets back the hypothesis matrix \(X_c^{inv} = H_c\).

Why is this of interest? This means that if one has a given contrast matrix, e.g., one that is provided by standard software packages, or one that is described in a research paper, then one can apply the generalized inverse operation to obtain the hypothesis matrix. This will tell us exactly which hypotheses were tested by the given contrast matrix.

\textcolor{black}{As an example, let us take a closer look at this using the \textsc{treatment contrast}. Let's start with a \textsc{treatment contrast} for a factor with three levels F1, F2, and F3. Adding a column of $1$s adds the intercept (}\texttt{int}\textcolor{black}{):}

\begin{Shaded}
\begin{Highlighting}[]
\NormalTok{(XcTr <-}\StringTok{ }\KeywordTok{cbind}\NormalTok{(}\DataTypeTok{int=}\DecValTok{1}\NormalTok{,}\KeywordTok{contr.treatment}\NormalTok{(}\DecValTok{3}\NormalTok{)))}
\end{Highlighting}
\end{Shaded}

\begin{verbatim}
##   int 2 3
## 1   1 0 0
## 2   1 1 0
## 3   1 0 1
\end{verbatim}

\textcolor{black}{The next step is to} apply the generalized inverse operation:

\begin{Shaded}
\begin{Highlighting}[]
\KeywordTok{t}\NormalTok{(}\KeywordTok{ginv2}\NormalTok{(XcTr))}
\end{Highlighting}
\end{Shaded}

\begin{verbatim}
##   int 2  3 
## 1  1  -1 -1
## 2  0   1  0
## 3  0   0  1
\end{verbatim}

\textcolor{black}{This shows} the hypotheses that the \textsc{treatment contrasts} test, by extracting the weights from the hypothesis matrix. The first contrast (\texttt{int}) has weights \texttt{cH00\ \textless{}-\ c(1,\ 0,\ 0)}. \textcolor{black}{Writing} this down as a formal hypothesis test \textcolor{black}{yields}:

\begin{equation}
H_{0_0}: 1 \cdot \mu_1 + 0 \cdot \mu_2 + 0 \cdot \mu_3 = 0
\end{equation}

That is, the first contrast tests the hypothesis \(H_{0_0}: \mu_1 = 0\) that the mean of the first factor level \(\mu_1\) is zero. As the factor level F1 was defined as the baseline condition in the treatment contrast, this means that for treatment contrasts, the intercept captures the condition mean of the baseline condition. This is the exact same result that \textcolor{black}{was shown} at the beginning of this paper, when first introducing treatment contrasts (see equation \ref{eq:trmtcontrfirstmention}).

\textcolor{black}{We also extract the weights for the other contrasts from the hypothesis matrix. The weights for the second contrast are} \texttt{cH01\ \textless{}-\ c(-1,\ 1,\ 0)}. \textcolor{black}{This is written} as a formal hypothesis test:

\begin{equation}
H_{0_1}: -1 \cdot \mu_1 + 1 \cdot \mu_2 + 0 \cdot \mu_3 = 0
\end{equation}

\textcolor{black}{The second contrast tests the difference in condition means between the first and the second factor level, i.e., it tests the null hypothesis that the difference in condition means of the second minus the first factor levels is zero $H_{0_1}: \mu_2 - \mu_1 = 0$.}

\textcolor{black}{We also extract the weights for the last contrast, which are} \texttt{cH02\ \textless{}-\ c(-1,\ 0,\ 1)}\textcolor{black}{, and write them as a formal hypothesis test:}

\begin{equation}
H_{0_2}: -1 \cdot \mu_1 + 0 \cdot \mu_2 + 1 \cdot \mu_3 = 0
\end{equation}

\textcolor{black}{This contrast tests the difference between the third (F3) and the first (F1) condition means, and tests the null hypothesis that the difference is zero: $H_{0_2}: \mu_3 - \mu_1 = 0$. These results correspond to what we know about treatment contrasts, i.e., that treatment contrasts test the difference of each group to the baseline condition. They demonstrate that it is possible to use the generalized inverse to learn about the hypotheses that a given set of contrasts tests.}

\hypertarget{the-importance-of-the-intercept-when-transforming-between-the-contrast-matrix-and-the-hypothesis-matrix-in-non-centered-contrasts}{%
\subsection{The importance of the intercept when transforming between the contrast matrix and the hypothesis matrix in non-centered contrasts}\label{the-importance-of-the-intercept-when-transforming-between-the-contrast-matrix-and-the-hypothesis-matrix-in-non-centered-contrasts}}

The above example of the treatment contrast also demonstrates that it is vital to consider the intercept when doing the transformation between the contrast matrix and the hypothesis matrix. Let us have a look at what the hypothesis matrix looks like when the intercept is ignored in the inversion:

\begin{Shaded}
\begin{Highlighting}[]
\NormalTok{(XcTr <-}\StringTok{ }\KeywordTok{contr.treatment}\NormalTok{(}\DecValTok{3}\NormalTok{))}
\end{Highlighting}
\end{Shaded}

\begin{verbatim}
##   2 3
## 1 0 0
## 2 1 0
## 3 0 1
\end{verbatim}

\begin{Shaded}
\begin{Highlighting}[]
\KeywordTok{t}\NormalTok{(Hc <-}\StringTok{ }\KeywordTok{ginv2}\NormalTok{(XcTr))}
\end{Highlighting}
\end{Shaded}

\begin{verbatim}
##   2 3
## 1 0 0
## 2 1 0
## 3 0 1
\end{verbatim}

\textcolor{black}{Now, the hypothesis matrix looks very different. In fact it looks just the same as the contrast matrix. However, the hypothesis matrix does not code any reasonable hypotheses or comparisons any more: The first contrast now tests the hypothesis that the condition mean for F2 is zero, $H_{0_1}: \mu_2 = 0$. The second contrast now tests the hypothesis that the condition mean for F3 is zero, $H_{0_2}: \mu_3 = 0$. However, we know that these are the wrong hypotheses for the  \textsc{treatment contrast} when the intercept is included in the model. This demonstrates that it is important to consider the intercept in the generalized inverse. As explained earlier in the section on non-/orthogonal contrasts, this is important for contrasts that are not centered. For centered contrasts, such as the \textsc{sum contrast} or the \textsc{repeated contrast}, including or excluding the intercept does not change the results.}

\hypertarget{the-hypothesis-matrix-and-contrast-matrix-in-matrix-form}{%
\subsection{The hypothesis matrix and contrast matrix in matrix form}\label{the-hypothesis-matrix-and-contrast-matrix-in-matrix-form}}

\hypertarget{matrix-notation-for-the-contrast-matrix}{%
\subsubsection{Matrix notation for the contrast matrix}\label{matrix-notation-for-the-contrast-matrix}}

\textcolor{black}{We have discussed above the relation of contrasts to linear/multiple regression analysis, i.e., that contrasts encode numeric predictor variables (covariates) for testing comparisons between discrete conditions in a linear/multiple regression model. 
The introduction of \textsc{treatment contrasts} had shown that a contrast can be used as the predictior $x$ in the linear regression equation $y = \beta_0 + \beta_1 \cdot x$. To repeat: in the treatment contrast, if $x$ is $0$ for the baseline condition, the predicted data is $\beta_0 + \beta_1 \cdot 0 = \beta_0$, indicating the intercept $\beta_0$ is the prediction for the mean of the baseline factor level (}\texttt{F1}\textcolor{black}{). If $x$ is $1$ (}\texttt{F2}\textcolor{black}{), then the predicted data is $\beta_0 + \beta_1 \cdot 1 = \beta_0 + \beta_1$. Both of these predictions for conditions $x = 0$ and $x = 1$ are summarized in a single equation using matrix notation (also see equation} \eqref{eq:lm1} in the introduction; cf.~Bolker, 2018).
Here, the different possible values of \(x\) \textcolor{black}{are represented} in the contrast matrix \(X_c\).

\begin{equation}
X_c = \left(\begin{array}{cc}
1 & 0 \\
1 & 1 
\end{array}
\right)
\end{equation}

\textcolor{black}{This matrix has one row for each condition/group of the study, i.e., here, it has 2 rows. The matrix has two columns. 
The second column (i.e., on the right-hand side) contains the treatment contrast with $x_1 = 0$ and $x_2 = 1$. 
The first column of $X_c$ contains a column of $1$s, which indicate that the intercept $\beta_0$ is added in each condition.}

\textcolor{black}{Multiplying}\footnote{Matrix multiplication is defined as follows. Consider a matrix \(X\) with three rows and two columns \(X = \left(\begin{array}{cc} x_{1,1} & x_{1,2} \\ x_{2,1} & x_{2,2} \\ x_{3,1} & x_{3,3} \end{array} \right)\), and a vector with two entries \(\beta = \left(\begin{array}{cc} \beta_0 \\ \beta_1 \end{array} \right)\).
  The matrix \(X\) can be multiplied with the vector \(\beta\) as follows:
  \(X \beta = \left(\begin{array}{cc} x_{1,1} & x_{1,2} \\ x_{2,1} & x_{2,2} \\ x_{3,1} & x_{3,2} \end{array} \right) \left(\begin{array}{cc} \beta_0 \\ \beta_1 \end{array} \right) = \left(\begin{array}{cc} x_{1,1} \cdot \beta_0 + x_{1,2} \cdot \beta_1 \\ x_{2,1} \cdot \beta_0 + x_{2,2} \cdot \beta_1 \\ x_{3,1} \cdot \beta_0 + x_{3,2} \cdot \beta_1 \end{array} \right) = \left(\begin{array}{cc} y_1 \\ y_2 \\ y_3 \end{array} \right) = y_{3,1}\)
  Multiplying an \(n\times p\) matrix with another \(p\times m\) matrix will yield an \(n\times m\) matrix.
  If the number of columns of the first matrix is not the same as the number of rows of the second matrix, matrix multiplication is undefined.} \textcolor{black}{this contrast matrix $X_c$ with the vector of regression coefficients $\beta=(\beta_0, \beta_1)$ containing the intercept $\beta_0$ and the effect of the factor $x$, $\beta_1$ (i.e., the slope), yields the expected response times for conditions} \texttt{F1} and \texttt{F2}\textcolor{black}{, $y_1$ and $y_2$, which here correspond to the condition means, $\mu_1$ and $\mu_2$:}

\begin{equation}
X_c \beta = \left(\begin{array}{cc}
1 & 0 \\
1 & 1 
\end{array}
\right)
\left(\begin{array}{c}
\beta_0 \\ \beta_1
\end{array}
\right) =
\left(\begin{array}{c}
1 \cdot \beta_0 + 0 \cdot \beta_1 \\ 
1 \cdot \beta_0 + 1 \cdot \beta_1
\end{array}
\right) =
\left(\begin{array}{c}
\beta_0 \\ 
\beta_0 + \beta_1
\end{array}
\right) =
\left(\begin{array}{c}
\mu_1 \\ \mu_2
\end{array}\right) = \mu \quad
\label{cdef}
\end{equation}

More compactly:

\begin{equation}
\mu = X_c \beta
\label{eq:lm0}
\end{equation}

This matrix formulation can be implemented in R. Consider again the simulated data displayed in Figure~\ref{fig:Fig1Means} \textcolor{black}{with two factor levels} \texttt{F1} and \texttt{F2}\textcolor{black}{, and condition means of $\mu_1 = 0.8$ and $\mu_2 = 0.4$. The  \textsc{treatment contrast} codes condition} \texttt{F1} \textcolor{black}{as the baseline condition with $x = 0$, and condition} \texttt{F2} \textcolor{black}{as $x = 1$. As shown in Table}~\ref{tab:table2}, the estimated regression coefficients were \(\beta_0 = 0.8\) and \(\beta_1 = -0.4\). The contrast matrix \(X_c\) can now be constructed as follows:

\begin{Shaded}
\begin{Highlighting}[]
\NormalTok{(XcTr <-}\StringTok{ }\KeywordTok{cbind}\NormalTok{(}\DataTypeTok{int=}\KeywordTok{c}\NormalTok{(}\DataTypeTok{F1=}\DecValTok{1}\NormalTok{,}\DataTypeTok{F2=}\DecValTok{1}\NormalTok{), }\DataTypeTok{c2vs1=}\KeywordTok{c}\NormalTok{(}\DecValTok{0}\NormalTok{,}\DecValTok{1}\NormalTok{)))}
\end{Highlighting}
\end{Shaded}

\begin{verbatim}
##    int c2vs1
## F1   1     0
## F2   1     1
\end{verbatim}

The regression coefficients can be written as:

\begin{Shaded}
\begin{Highlighting}[]
\NormalTok{(beta <-}\StringTok{ }\KeywordTok{c}\NormalTok{(}\FloatTok{0.8}\NormalTok{,}\OperatorTok{-}\FloatTok{0.4}\NormalTok{))}
\end{Highlighting}
\end{Shaded}

\begin{verbatim}
## [1]  0.8 -0.4
\end{verbatim}

\begin{Shaded}
\begin{Highlighting}[]
\CommentTok{## convert to a 2x1 vector:}
\NormalTok{beta <-}\StringTok{ }\KeywordTok{matrix}\NormalTok{(beta,}\DataTypeTok{ncol =} \DecValTok{1}\NormalTok{)}
\end{Highlighting}
\end{Shaded}

Multiplying the \(2\times 2\) contrast matrix with the estimated regression coefficients, a \(2\times 1\) vector, gives predictions of the condition means:

\begin{Shaded}
\begin{Highlighting}[]
\NormalTok{XcTr }\OperatorTok{%*%}\StringTok{ }\NormalTok{beta }\CommentTok{# The symbol %*% indicates matrix multiplication}
\end{Highlighting}
\end{Shaded}

\begin{verbatim}
##    [,1]
## F1  0.8
## F2  0.4
\end{verbatim}

As expected, the matrix multiplication yields the condition means.

\hypertarget{using-the-generalized-inverse-to-estimate-regression-coefficients}{%
\subsubsection{Using the generalized inverse to estimate regression coefficients}\label{using-the-generalized-inverse-to-estimate-regression-coefficients}}

One key question remains unanswered by this representation of the linear model: Although in our example we know the contrast matrix \(X_c\) and the condition means \(\mu\), both the condition means \(\mu\) and the regression coefficients \(\beta\) are unknown, and need to be estimated from the data \(y\) (this is what the command \texttt{lm()} does).

\textcolor{black}{Can we use the matrix notation $\mu = X_c \beta$ (cf. equation} \ref{eq:lm0}) to estimate the regression coefficients \(\beta\)? That is, can we re-formulate the equation, by writing the regression coefficients \(\beta\) on one side of the equation, such that we can compute \(\beta\)? Intuitively, what \textcolor{black}{needs to be done} for this, is to \enquote{divide by \(X_c\)}. This would yield \(1 \cdot \beta = \beta\) on the right side of the equation, and \enquote{one divided by \(X_c\)} times \(\mu\) on the left hand side. That is, this would allow us to solve for the regression coefficients \(\beta\) and would provide a formula to compute them. Indeed, the generalized matrix inverse operation does for matrices exactly what we intuitively refer to as \enquote{divide by \(X_c\)}, that is, it computes the inverse of a matrix such that pre-multiplying the inverse of a matrix with the matrix yields \(1\): \(X_c^{inv} \cdot X_c = 1\) (where \(1\) is the identity matrix, with \(1\)s on the diagonal and off-diagonal \(0\)s). For example, we take the contrast matrix of a \textsc{treatment contrast} for a factor with two levels, and we compute the related hypothesis matrix using the generalized inverse: Pre-multiplying the contrast matrix with its inverse yields the identity matrix, that is, a matrix where the diagonals are all \(1\) and the off-diagonals are all \(0\):

\begin{Shaded}
\begin{Highlighting}[]
\NormalTok{XcTr}
\end{Highlighting}
\end{Shaded}

\begin{verbatim}
##    int c2vs1
## F1   1     0
## F2   1     1
\end{verbatim}

\begin{Shaded}
\begin{Highlighting}[]
\KeywordTok{ginv2}\NormalTok{(XcTr)}
\end{Highlighting}
\end{Shaded}

\begin{verbatim}
##       F1 F2
## int    1  0
## c2vs1 -1  1
\end{verbatim}

\begin{Shaded}
\begin{Highlighting}[]
\KeywordTok{fractions}\NormalTok{( }\KeywordTok{ginv2}\NormalTok{(XcTr) }\OperatorTok{%*%}\StringTok{ }\NormalTok{XcTr )}
\end{Highlighting}
\end{Shaded}

\begin{verbatim}
##       int c2vs1
## int   1   0    
## c2vs1 0   1
\end{verbatim}

\textcolor{black}{That is, multiplying equation} \ref{eq:lm0} \textcolor{black}{by $X_c^{inv}$, yields (for details see Appendix}~\ref{app:LinearAlgebra}):

\begin{align}
X_c^{inv} \mu &= X_c^{inv} X_c \beta \\
X_c^{inv} \mu &= \beta
\end{align}

\textcolor{black}{This shows that the generalized matrix inverse actually allows to estimate regression coefficients from the data. This is done by (matrix) multiplying the hypothesis matrix (i.e., the inverse contrast matrix) with the condition means: $\hat{\beta} = X_c^{inv} \mu$. Importantly, this derivation ignored residual errors in the regression equation. For a full derivation see the Appendix}~\ref{app:LinearAlgebra}.

Consider again the simple example of a \textsc{treatment contrast} for a factor with two levels.

\begin{Shaded}
\begin{Highlighting}[]
\NormalTok{XcTr}
\end{Highlighting}
\end{Shaded}

\begin{verbatim}
##    int c2vs1
## F1   1     0
## F2   1     1
\end{verbatim}

Inverting the contrast matrix yields the hypothesis matrix.

\begin{Shaded}
\begin{Highlighting}[]
\NormalTok{(HcTr <-}\StringTok{ }\KeywordTok{ginv2}\NormalTok{(XcTr))}
\end{Highlighting}
\end{Shaded}

\begin{verbatim}
##       F1 F2
## int    1  0
## c2vs1 -1  1
\end{verbatim}

\textcolor{black}{So far, we always displayed the resulting hypothesis matrix with rows and columns switched using matrix transpose (R function} \texttt{t()}\textcolor{black}{) to make it more easily readable. That is, each column represented one contrast, and each row represented one condition. However, in the present paragraph we discuss the matrix notation of the linear model, and we therefore show the hypothesis matrix in its original, untransposed form, where each row represents one contrast/hypothesis and each column represents one factor level. That is, the first row of} \texttt{HcTr} (\texttt{int\ =\ c(1,0)}) \textcolor{black}{encodes the intercept and the null hypothesis that the condition mean} \texttt{F1} \textcolor{black}{is zero. The second row} \texttt{cH01\ =\ c(-1,1)} \textcolor{black}{encodes the null hypothesis that the condition mean of} \texttt{F2} \textcolor{black}{is identical to the condition mean} \texttt{F1}.

\textcolor{black}{When we multiply this hypothesis matrix $H_c$ with the observed condition means $\mu$, this yields the regression coefficients. For illustration we use the condition means from our first example of the treatment contrast (see Figure}~\ref{fig:Fig1Means}), \textcolor{black}{encoded in the data frame} \texttt{table1} \textcolor{black}{in variable} \texttt{table1\$M}. \textcolor{black}{The resulting regression coefficients are the same values as in the \texttt{lm} command presented in Table}~\ref{tab:table2}:

\begin{Shaded}
\begin{Highlighting}[]
\NormalTok{mu <-}\StringTok{ }\NormalTok{table1}\OperatorTok{$}\NormalTok{M}
\NormalTok{HcTr }\OperatorTok{%*%}\StringTok{ }\NormalTok{mu }\CommentTok{# %*% indicates matrix multiplication.}
\end{Highlighting}
\end{Shaded}

\begin{verbatim}
##       [,1]
## int    0.8
## c2vs1 -0.4
\end{verbatim}

\hypertarget{estimating-regression-coefficients-and-testing-hypotheses-using-condition-means}{%
\subsubsection{Estimating regression coefficients and testing hypotheses using condition means}\label{estimating-regression-coefficients-and-testing-hypotheses-using-condition-means}}

We explained above that the hypothesis matrix contains weights for the condition means to define the hypothesis that a given contrast tests. For example, the hypothesis matrix from a treatment contrast \textcolor{black}{was used} for a factor with two levels:

\begin{Shaded}
\begin{Highlighting}[]
\NormalTok{HcTr}
\end{Highlighting}
\end{Shaded}

\begin{verbatim}
##       F1 F2
## int    1  0
## c2vs1 -1  1
\end{verbatim}

\textcolor{black}{The first row codes the weights for the intercept. This encodes the following hypothesis:}

\begin{equation}
H_{0_0}: 1 \cdot \mu_1 + 0 \cdot \mu_2 = \mu_1 = 0
\end{equation}

In fact, writing this down as a hypothesis involves a short-cut: What the hypothesis matrix actually encodes are weights for how to combine condition means to compute regression coefficients. Our null hypotheses are then tested as \(H_{0_x}: \beta_x = 0\).
For the present example, \textcolor{black}{estimates for} the regression coefficient for the intercept \textcolor{black}{are}:

\begin{equation}
\hat{\beta_0} = 1 \cdot \mu_1 + 0 \cdot \mu_2 = \mu_1
\end{equation}

\textcolor{black}{For the example data-set:}

\begin{equation}
\hat{\beta_0} = 1 \cdot 0.8 + 0 \cdot 0.4 = 0.8
\end{equation}

\textcolor{black}{This is exactly the same value that the} \texttt{lm()} \textcolor{black}{command showed.
A second step tests the null hypothesis that the regression coefficient is zero, i.e., $H_{0_0}: \beta_0 = 0$.}

The same analysis \textcolor{black}{is done} for the slope, which is coded in the second row of the hypothesis matrix. The hypothesis is expressed as:

\begin{equation}
H_{0_1}: -1 \cdot \mu_1 + 1 \cdot \mu_2 = \mu_2 - \mu_1 = 0
\end{equation}

\textcolor{black}{This involves first computing the regression coefficient for the slope:}

\begin{equation}
\hat{\beta_0} = -1 \cdot \mu_1 + 1 \cdot \mu_2 = \mu_2 - \mu_1
\end{equation}

\textcolor{black}{For the example data-set, this yields:}

\begin{equation}
\hat{\beta_1} = -1 \cdot 0.8 + 1 \cdot 0.4 = - 0.4
\end{equation}

\textcolor{black}{Again, this is the same value for the slope as given by the command} \texttt{lm()}\textcolor{black}{.
A second step tests the hypothesis that the slope is zero: $H_{0_1}: \beta_1 = 0$.}

To summarize, we write the formulas for both regression coefficients in a single equation.
\textcolor{black}{A first step writes down} the hypothesis matrix:

\begin{equation}
H_c = X_c^{inv} = 
\left(\begin{array}{rr}
1 & 0 \\
-1 & 1 
\end{array}
\right)
\end{equation}

\textcolor{black}{Here, the first row contains the weights for the intercept (}\texttt{int\ =\ c(1,0)}\textcolor{black}{), and the second row contains the weights for the slope (}\texttt{c2vs1\ =\ c(-1,1)}\textcolor{black}{).
Multiplying this hypothesis matrix with the average response times in conditions} \texttt{F1} and \texttt{F2}\textcolor{black}{, $\mu_1$ and $\mu_2$, yields estimates of the regression coefficients:  $\hat{\beta} = \left(\begin{array}{c} \hat{\beta_0} \\ \hat{\beta_1} \end{array}\right)$. They show that the intercept $\beta_0$ is equal to the average value of} \texttt{F1}\textcolor{black}{, $\mu_1$, and the slope $\beta_1$ is equal to the difference in average values between} \texttt{F2} \textcolor{black}{minus} \texttt{F1}\textcolor{black}{, $\mu_2 - \mu_1$:}

\begin{equation}
X_c^{inv} \mu = 
\left(\begin{array}{rr}
1 & 0 \\
-1 & 1 
\end{array}
\right)
\left(\begin{array}{c}
\mu_1 \\ \mu_2
\end{array}
\right) = 
\left(\begin{array}{rr}
1 \cdot \mu_1 + 0 \cdot \mu_2 \\
-1 \cdot \mu_1 + 1 \cdot \mu_2
\end{array}
\right) = 
\left(\begin{array}{c}
\mu_1 \\
\mu_2 - \mu_1
\end{array}
\right) =
\left(\begin{array}{c}
\hat{\beta_0} \\ \hat{\beta_1}
\end{array}\right) = \hat{\beta}
\label{cinvdef}
\end{equation}

\textcolor{black}{Or in short (see equation}~\ref{eq:beta}):

\begin{equation}
\hat{\beta} = X_c^{inv} \mu
\end{equation}

\textcolor{black}{These analyses show the important result that the hypothesis matrix is used to compute regression coefficients from the condition means. This important result is derived from the matrix formulation of the linear model (for details see Appendix}~\ref{app:LinearAlgebra}). \textcolor{black}{This is one reason why it is important to understand the matrix formulation of the linear model.}

\hypertarget{the-design-or-model-matrix-and-its-generalized-inverse}{%
\subsubsection{The design or model matrix and its generalized inverse}\label{the-design-or-model-matrix-and-its-generalized-inverse}}

\textcolor{black}{Until now we have been applying the generalized inverse to the hypothesis matrix to obtain the contrast matrix $X_c$. The contrast matrix contains one row per condition. When performing a linear model analysis, the contrast matrix is translated into a so-called \textit{design or model matrix} $X$. This matrix contains one row for every data point in the vector of data $y$. As a consequence, the same condition will appear more than once in the design matrix $X$.}

\textcolor{black}{The key point to note here is that} the generalized inverse operation can not only be applied to the contrast matrix \(X_c\) but also to the design matrix \(X\), which contains one row per data point. The important result is that taking the generalized inverse of this design matrix \(X\) and multiplying it with the raw data (i.e., the dependent variable) also yields the regression coefficients:

\begin{Shaded}
\begin{Highlighting}[]
\KeywordTok{contrasts}\NormalTok{(simdat}\OperatorTok{$}\NormalTok{F) <-}\StringTok{ }\KeywordTok{c}\NormalTok{(}\DecValTok{0}\NormalTok{,}\DecValTok{1}\NormalTok{)}
\KeywordTok{data.frame}\NormalTok{(X <-}\StringTok{ }\KeywordTok{model.matrix}\NormalTok{( }\OperatorTok{~}\StringTok{ }\DecValTok{1} \OperatorTok{+}\StringTok{ }\NormalTok{F,simdat)) }\CommentTok{# obtain design matrix X}
\end{Highlighting}
\end{Shaded}

\begin{verbatim}
##    X.Intercept. F1
## 1             1  0
## 2             1  0
## 3             1  0
## 4             1  0
## 5             1  0
## 6             1  1
## 7             1  1
## 8             1  1
## 9             1  1
## 10            1  1
\end{verbatim}

\begin{Shaded}
\begin{Highlighting}[]
\NormalTok{(Xinv <-}\StringTok{ }\KeywordTok{ginv2}\NormalTok{(X)) }\CommentTok{# take generalized inverse of X}
\end{Highlighting}
\end{Shaded}

\begin{verbatim}
##             1    2    3    4    5    6    7    8    9    10  
## (Intercept)  1/5  1/5  1/5  1/5  1/5    0    0    0    0    0
## F1          -1/5 -1/5 -1/5 -1/5 -1/5  1/5  1/5  1/5  1/5  1/5
\end{verbatim}

\begin{Shaded}
\begin{Highlighting}[]
\NormalTok{(y <-}\StringTok{ }\NormalTok{simdat}\OperatorTok{$}\NormalTok{DV) }\CommentTok{# raw data}
\end{Highlighting}
\end{Shaded}

\begin{verbatim}
##  [1] 0.997 0.847 0.712 0.499 0.945 0.183 0.195 0.608 0.556 0.458
\end{verbatim}

\begin{Shaded}
\begin{Highlighting}[]
\NormalTok{Xinv }\OperatorTok{%*%}\StringTok{ }\NormalTok{y }\CommentTok{# matrix multiplication}
\end{Highlighting}
\end{Shaded}

\begin{verbatim}
##             [,1]
## (Intercept)  0.8
## F1          -0.4
\end{verbatim}

\textcolor{black}{The generalized inverse automatically generates covariates} that perform the averaging across the individual data points per condition. E.g., the estimate of the intercept \(\hat{\beta}_0\) is computed from the \(i\) observations of the dependent variable \(y_i\) with the formula:

\begin{equation}
\hat{\beta}_0 = \frac{1}{5} \cdot y_1 + \frac{1}{5} \cdot y_2 + \frac{1}{5} \cdot y_3 + \frac{1}{5} \cdot y_4 + \frac{1}{5} \cdot y_5 + 0 \cdot y_6 + 0 \cdot y_7 + 0 \cdot y_8 + 0 \cdot y_9 + 0 \cdot y_{10} = \frac{1}{5} \cdot \sum_{i=1}^{5} y_i
\end{equation}

\noindent
which expresses the formula for estimating \(\mu_1\), the mean response time in condition F1.

\hypertarget{effectSize}{%
\section{The variance explained by each predictor and effect size statistics for linear contrasts}\label{effectSize}}

\textcolor{black}{Should one include all predictors in a model? And what is the effect size of a contrast? To answer these questions, it is important to understand the concept of variance explained. This is explained next.}

\hypertarget{sum-of-squares-and-r2_alerting-as-measures-of-variability-and-effect-size}{%
\subsection{\texorpdfstring{Sum of squares and \(r^2_{alerting}\) as measures of variability and effect size}{Sum of squares and r\^{}2\_\{alerting\} as measures of variability and effect size}}\label{sum-of-squares-and-r2_alerting-as-measures-of-variability-and-effect-size}}

\textcolor{black}{The sum of squares is a measure of the variability in the data. The total sum of squares in a data set equals the sum of squared deviations of individual data points $y_i$ from their mean $\bar{y}$: $SS_{total} = \sum_i (y_i - \bar{y})^2$. This can be partitioned into different components, which add up to the total sum of squares. One component is the sum of squares associated with the residuals, that is the sum of squared deviations of $i$ individual observed data points $y_i$ from the value predicted by a linear model $y_{pred}$: $SS_{residuals} = \sum_i (y_i - y_{pred})^2$; another component, that we are interested in here, is the sum of squares associated with a certain factor, which is the squared deviation of $j$ condition means $\mu_j$ from their mean $\bar{\mu}$, where each squared deviation is multiplied by the number of data points $n_j$ that go into its computation: $SS_{effect} = \sum_j n_j \cdot (\mu_j - \bar{\mu})^2$.}

\textcolor{black}{The sum of squares for a factor ($SS_{effect}$) can be further partitioned into the contributions of $k$ different linear contrasts: $SS_{effect} = \sum_k SS_{contrast \;k}$.}\footnote{\textcolor{black}{The $SS_{contrast}$ can be computed as follows: $SS_{contrast} = \frac{(\sum_j c_j \mu_j)^2}{\sum_j {c_j}^2 / n_j}$, where $j$ is the number of factor levels, $c_j$ are the contrast coefficients for each factor level, $\mu_j$ are the condition means, and $n_j$ is the number of data points per factor level $j$.}} A measure for the effect size of a linear contrast is the proportion of variance that it explains (i.e., sum of squares \emph{contrast}) from the total variance explained by the factor (i.e., sum of squares \emph{effect}), that is, \(r^2_{alerting} = \text{SS}_{contrast} / \text{SS}_{effect}\) (Baguley, 2012; Rosenthal, Rosnow, \& Rubin, 2000). This is computed by entering each contrast as an individual predictor into a linear model, and by then extracting the corresponding sum of squares from the output of the \texttt{anova()} function. This is illustrated with the example of \textsc{polynomial contrasts}.

First, we assign \textsc{polynomial contrasts} to the factor \texttt{F}, extract the numeric predictor variables using the R function \texttt{model.matrix()}, and use these as covariates in a linear model analysis.

\begin{Shaded}
\begin{Highlighting}[]
\NormalTok{(}\KeywordTok{contrasts}\NormalTok{(simdat3}\OperatorTok{$}\NormalTok{F) <-}\StringTok{ }\KeywordTok{contr.poly}\NormalTok{(}\DecValTok{4}\NormalTok{))}
\end{Highlighting}
\end{Shaded}

\begin{verbatim}
##          .L   .Q     .C
## [1,] -0.671  0.5 -0.224
## [2,] -0.224 -0.5  0.671
## [3,]  0.224 -0.5 -0.671
## [4,]  0.671  0.5  0.224
\end{verbatim}

\begin{Shaded}
\begin{Highlighting}[]
\NormalTok{simdat3X <-}\StringTok{ }\KeywordTok{model.matrix}\NormalTok{(}\OperatorTok{~}\StringTok{ }\DecValTok{1} \OperatorTok{+}\StringTok{ }\NormalTok{F, }\DataTypeTok{data=}\NormalTok{simdat3)}
\NormalTok{simdat3[,}\KeywordTok{c}\NormalTok{(}\StringTok{"cLinear"}\NormalTok{,}\StringTok{"cQuadratic"}\NormalTok{,}\StringTok{"cCubic"}\NormalTok{)] <-}\StringTok{ }\NormalTok{simdat3X[,}\DecValTok{2}\OperatorTok{:}\DecValTok{4}\NormalTok{]}
\NormalTok{m1_mr.Xpol2 <-}\StringTok{ }\KeywordTok{lm}\NormalTok{(DV }\OperatorTok{~}\StringTok{ }\DecValTok{1} \OperatorTok{+}\StringTok{ }\NormalTok{cLinear }\OperatorTok{+}\StringTok{ }\NormalTok{cQuadratic }\OperatorTok{+}\StringTok{ }\NormalTok{cCubic, }\DataTypeTok{data=}\NormalTok{simdat3)}
\end{Highlighting}
\end{Shaded}

Next, this model \textcolor{black}{is analyzed} using the R function \texttt{anova()}. \textcolor{black}{This yields the sum of squares (Sum Sq) explained by each of the covariates.}

\begin{Shaded}
\begin{Highlighting}[]
\NormalTok{(aovModel <-}\StringTok{ }\KeywordTok{anova}\NormalTok{(m1_mr.Xpol2))}
\end{Highlighting}
\end{Shaded}

\begin{verbatim}
## Analysis of Variance Table
## 
## Response: DV
##            Df Sum Sq Mean Sq F value Pr(>F)   
## cLinear     1   1600    1600      16 0.0010 **
## cQuadratic  1    500     500       5 0.0399 * 
## cCubic      1    900     900       9 0.0085 **
## Residuals  16   1600     100                  
## ---
## Signif. codes:  0 '***' 0.001 '**' 0.01 '*' 0.05 '.' 0.1 ' ' 1
\end{verbatim}

\begin{Shaded}
\begin{Highlighting}[]
\CommentTok{# SumSq contrast}
\NormalTok{SumSq <-}\StringTok{ }\NormalTok{aovModel[}\DecValTok{1}\OperatorTok{:}\DecValTok{3}\NormalTok{,}\StringTok{"Sum Sq"}\NormalTok{]}
\KeywordTok{names}\NormalTok{(SumSq) <-}\StringTok{ }\KeywordTok{c}\NormalTok{(}\StringTok{"cLinear"}\NormalTok{,}\StringTok{"cQuadratic"}\NormalTok{,}\StringTok{"cCubic"}\NormalTok{)}
\NormalTok{SumSq}
\end{Highlighting}
\end{Shaded}

\begin{verbatim}
##    cLinear cQuadratic     cCubic 
##       1600        500        900
\end{verbatim}

Summing across the three contrasts that encode the factor \texttt{F} allows us to compute the total effect sum of squares associated with it. Now, \textcolor{black}{everything is available that is needed} to compute the \(r^2_{alerting}\) summary statistic: dividing the individual sum of squares by the total effect size of squares yields \(r^2_{alerting}\).

\begin{Shaded}
\begin{Highlighting}[]
\CommentTok{# SumSq effect}
\KeywordTok{sum}\NormalTok{(SumSq)}
\end{Highlighting}
\end{Shaded}

\begin{verbatim}
## [1] 3000
\end{verbatim}

\begin{Shaded}
\begin{Highlighting}[]
\CommentTok{# r2 alerting}
\KeywordTok{round}\NormalTok{(SumSq }\OperatorTok{/}\StringTok{ }\KeywordTok{sum}\NormalTok{(SumSq), }\DecValTok{2}\NormalTok{)}
\end{Highlighting}
\end{Shaded}

\begin{verbatim}
##    cLinear cQuadratic     cCubic 
##       0.53       0.17       0.30
\end{verbatim}

\textcolor{black}{The results show that the expected linear trend explains} 53\(\%\) \textcolor{black}{of the variance in condition means of factor} \texttt{F}\textcolor{black}{. Based on the statistical test shown in the anova output, the linear trend has a significant effect on the dependent variable. However, the effect size analysis shows it does not explain the full pattern of results, as nearly half the variance associated with factor} \texttt{F} \textcolor{black}{remains unexplained by it, whereas other effects, namely non-linear trends, seem to contribute to explaining the effect of factor} \texttt{F}\textcolor{black}{. This situation is a so-called} \emph{ecumenical} \textcolor{black}{outcome} (Abelson \& Prentice, 1997)\textcolor{black}{, where the a priori contrast (linear trend) is only one of several contrasts explaining the factor's effect. The fact that the a priori linear contrast is significant but  has a $r^2_{alerting}$ clearly smaller than $1$, suggests that other contrasts seem to contribute to the effect of factor} \texttt{F}\textcolor{black}{. In a} so-called \emph{canonical} \textcolor{black}{outcome, to the contrary, $r^2_{alerting}$ for a contrast would approach $1$, such that no other additional contrasts are needed to explain the effect of factor} \texttt{F}.

\(r^2_{alerting}\) is useful for comparing the relative importance of different contrasts for a given data-set. However, it is not a general measure of effect size, such as \(\eta^2\), which is computed using the function call \texttt{etasq(Model)} from the package \texttt{heplots} (\(\eta^2_{partial}\)) (Friendly, 2010).

\hypertarget{adding-all-contrasts-associated-with-a-factor}{%
\subsection{Adding all contrasts associated with a factor}\label{adding-all-contrasts-associated-with-a-factor}}

\textcolor{black}{The above discussion shows that the total variability associated with a factor can be composed into contributions from individual contrasts. This implies, that even if only one of the contrasts associated with a factor is of interest in an analysis, it still makes sense to include the other contrasts associated with the factor. For example, when using polynomial contrasts, e}ven if only the linear trend is of interest in an analysis, it still makes sense to include the contrasts for the quadratic and cubic trends.
This is because there is a total amount of variance associated with each factor. One can capture all of this variance by using polynomial contrasts up \textcolor{black}{to} \(I-1\) degrees. That is, for a two-level factor, only a linear trend is tested. For a three-level factor, we test a linear and a quadratic trend. For a four-level factor, we additionally test a cubic trend. Specifying all these \(I-1\) polynomial trends allows us to capture all the variance associated with the factor. In case one is not interested in the cubic trend, one can simply leave this out of the model. However, this would also mean that some of the variance associated with the factor could be left unexplained. This unexplained variance would be added to the residual variance, and would impair our ability to detect effects in the linear (or in the quadratic) trend.

This shows that if the variance associated with a factor is not explained by contrasts, then this unexplained variance will increase the residual variance, and reduce the statistical power to detect the effects of interest. It is therefore good practice to code contrasts that capture all the variance associated with a factor.

\hypertarget{MR_ANOVA}{%
\section{Examples of contrast coding in a factorial design with two factors}\label{MR_ANOVA}}

\begin{figure}

{\centering \includegraphics{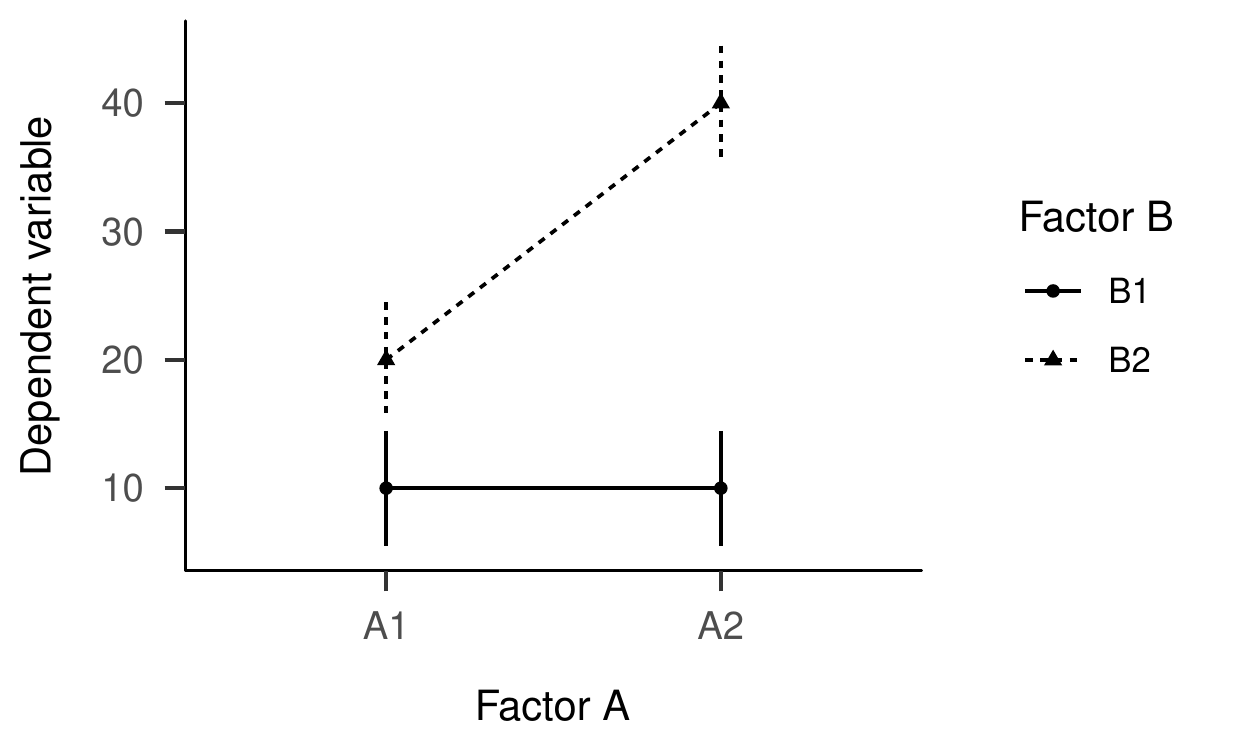} 

}

\caption{Means and error bars (showing standard errors) for a simulated data-set with a two-by-two  between-subjects factorial design.}\label{fig:twobytwosimdatFig}
\end{figure}

Let us assume that the exact same four means that we have simulated above actually come from an \(A(2) \times B(2)\) between-subject-factor design rather than an F(4) between-subject-factor design. We simulate the data as shown below in Table \ref{tab:twobytwosimdatTab} and Figure \ref{fig:twobytwosimdatFig}. The means and standard deviations are exactly the same as in Figure \ref{fig:helmertsimdatFig}.

\begin{table}[h]
\begin{center}
\begin{threeparttable}
\caption{\label{tab:twobytwosimdatTab}Summary statistics for a two-by-two between-subjects factorial design.}
\begin{tabular}{llllll}
\toprule
Factor A & \multicolumn{1}{c}{Factor B} & \multicolumn{1}{c}{N data} & \multicolumn{1}{c}{Means} & \multicolumn{1}{c}{Std. dev.} & \multicolumn{1}{c}{Std. errors}\\
\midrule
A1 & B1 & 5 & 10.0 & 10.0 & 4.5\\
A1 & B2 & 5 & 20.0 & 10.0 & 4.5\\
A2 & B1 & 5 & 10.0 & 10.0 & 4.5\\
A2 & B2 & 5 & 40.0 & 10.0 & 4.5\\
\bottomrule
\end{tabular}
\end{threeparttable}
\end{center}
\end{table}

\clearpage

\begin{Shaded}
\begin{Highlighting}[]
\CommentTok{# generate 2 times 2 between subjects data:}
\NormalTok{simdat4 <-}\StringTok{ }\KeywordTok{mixedDesign}\NormalTok{(}\DataTypeTok{B=}\KeywordTok{c}\NormalTok{(}\DecValTok{2}\NormalTok{,}\DecValTok{2}\NormalTok{), }\DataTypeTok{W=}\OtherTok{NULL}\NormalTok{, }\DataTypeTok{n=}\DecValTok{5}\NormalTok{, }\DataTypeTok{M=}\NormalTok{M,  }\DataTypeTok{SD=}\DecValTok{10}\NormalTok{, }\DataTypeTok{long =} \OtherTok{TRUE}\NormalTok{) }
\KeywordTok{names}\NormalTok{(simdat4)[}\DecValTok{1}\OperatorTok{:}\DecValTok{2}\NormalTok{] <-}\StringTok{ }\KeywordTok{c}\NormalTok{(}\StringTok{"A"}\NormalTok{,}\StringTok{"B"}\NormalTok{)}
\KeywordTok{head}\NormalTok{(simdat4)}
\end{Highlighting}
\end{Shaded}

\begin{verbatim}
##    A  B id     DV
## 1 A1 B1  1 26.195
## 2 A1 B1  2  5.758
## 3 A1 B1  3 11.862
## 4 A1 B1  4  6.321
## 5 A1 B1  5 -0.136
## 6 A1 B2  6 18.380
\end{verbatim}

\begin{Shaded}
\begin{Highlighting}[]
\NormalTok{table4 <-}\StringTok{ }\NormalTok{simdat4 }\OperatorTok{%>%}\StringTok{ }\KeywordTok{group_by}\NormalTok{(A, B) }\OperatorTok{%>%}\StringTok{ }\CommentTok{# plot interaction}
\StringTok{    }\KeywordTok{summarize}\NormalTok{(}\DataTypeTok{N=}\KeywordTok{length}\NormalTok{(DV), }\DataTypeTok{M=}\KeywordTok{mean}\NormalTok{(DV), }\DataTypeTok{SD=}\KeywordTok{sd}\NormalTok{(DV), }\DataTypeTok{SE=}\NormalTok{SD}\OperatorTok{/}\KeywordTok{sqrt}\NormalTok{(N))}
\NormalTok{GM <-}\StringTok{  }\KeywordTok{mean}\NormalTok{(table4}\OperatorTok{$}\NormalTok{M) }\CommentTok{# Grand Mean}
\end{Highlighting}
\end{Shaded}

\hypertarget{the-difference-between-an-anova-and-a-multiple-regression}{%
\subsection{The difference between an ANOVA and a multiple regression}\label{the-difference-between-an-anova-and-a-multiple-regression}}

Let's compare the traditional ANOVA with a multiple regression (i.e., using contrasts as covariates) for analyzing these data.

\begin{Shaded}
\begin{Highlighting}[]
\CommentTok{# ANOVA: B_A(2) times B_B(2)}
\NormalTok{m2_aov <-}\StringTok{ }\KeywordTok{aov}\NormalTok{(DV }\OperatorTok{~}\StringTok{ }\NormalTok{A}\OperatorTok{*}\NormalTok{B }\OperatorTok{+}\StringTok{ }\KeywordTok{Error}\NormalTok{(id), }\DataTypeTok{data=}\NormalTok{simdat4)}

\CommentTok{# MR: B_A(2) times B_B(2)}
\NormalTok{m2_mr <-}\StringTok{ }\KeywordTok{lm}\NormalTok{(DV }\OperatorTok{~}\StringTok{ }\DecValTok{1} \OperatorTok{+}\StringTok{ }\NormalTok{A}\OperatorTok{*}\NormalTok{B, }\DataTypeTok{data=}\NormalTok{simdat4)}
\end{Highlighting}
\end{Shaded}

\begin{table}[!htbp]
\begin{center}
\begin{threeparttable}
\caption{\label{tab:table16}Estimated ANOVA model.}
\begin{tabular}{lllllll}
\toprule
Effect & \multicolumn{1}{c}{$F$} & \multicolumn{1}{c}{$\mathit{df}_1$} & \multicolumn{1}{c}{$\mathit{df}_2$} & \multicolumn{1}{c}{$\mathit{MSE}$} & \multicolumn{1}{c}{$p$} & \multicolumn{1}{c}{$\hat{\eta}^2_G$}\\
\midrule
A & 5.00 & 1 & 16 & 100.00 & .040 & .238\\
B & 20.00 & 1 & 16 & 100.00 & < .001 & .556\\
A $\times$ B & 5.00 & 1 & 16 & 100.00 & .040 & .238\\
\bottomrule
\end{tabular}
\end{threeparttable}
\end{center}
\end{table}

\begin{table}[!htbp]
\begin{center}
\begin{threeparttable}
\caption{\label{tab:table17}Estimated regression model.}
\begin{tabular}{lllll}
\toprule
Predictor & \multicolumn{1}{c}{$Estimate$} & \multicolumn{1}{c}{95\% CI} & \multicolumn{1}{c}{$t(16)$} & \multicolumn{1}{c}{$p$}\\
\midrule
Intercept & 10 & $[1$, $19]$ & 2.24 & .040\\
AA2 & 0 & $[-13$, $13]$ & 0.00 & > .999\\
BB2 & 10 & $[-3$, $23]$ & 1.58 & .133\\
AA2 $\times$ BB2 & 20 & $[1$, $39]$ & 2.24 & .040\\
\bottomrule
\end{tabular}
\end{threeparttable}
\end{center}
\end{table}

The results from the two analyses, shown in Tables \ref{tab:table16} and \ref{tab:table17}, are very different.
How do we see these are different? Notice that \textcolor{black}{it is possible to} compute F-values from t-values from the fact that \(F(1,df) = t(df)^2\) (Snedecor \& Cochran, 1967) \textcolor{black}{(where $df$ indicates degrees of freedom)}. When applying this to the above multiple regression model, the F-value for factor \(A\) (i.e., \(AA2\)) is \(0.00^2 = 0\). This is obviously not the same as in the ANOVA, where the F-value for factor \(A\) is \(5\). Likewise, in the multiple regression factor \(B\) (i.e., \(BB2\)) has an F-value of \(1.58^2 = 2.5\), which also does not correspond to the F-value for factor \(B\) in the ANOVA of \(20\). Interestingly, however, the F-value for the interaction is identical in both models, as \(2.24^2 = 5\).

The reason that the two results are different is that one needs \textsc{sum contrasts} in the linear model to \textcolor{black}{get the conventional tests from an ANOVA model. (This is true for factors with two levels, but does not generalize to factors with more levels.)}

\begin{Shaded}
\begin{Highlighting}[]
\CommentTok{# define sum contrasts:}
\KeywordTok{contrasts}\NormalTok{(simdat4}\OperatorTok{$}\NormalTok{A) <-}\StringTok{ }\KeywordTok{contr.sum}\NormalTok{(}\DecValTok{2}\NormalTok{)}
\KeywordTok{contrasts}\NormalTok{(simdat4}\OperatorTok{$}\NormalTok{B) <-}\StringTok{ }\KeywordTok{contr.sum}\NormalTok{(}\DecValTok{2}\NormalTok{)}
\NormalTok{m2_mr.sum <-}\StringTok{ }\KeywordTok{lm}\NormalTok{(DV }\OperatorTok{~}\StringTok{ }\DecValTok{1} \OperatorTok{+}\StringTok{ }\NormalTok{A}\OperatorTok{*}\NormalTok{B, }\DataTypeTok{data=}\NormalTok{simdat4)}

\CommentTok{# Alternative using covariates}
\NormalTok{mat_myC <-}\StringTok{ }\KeywordTok{model.matrix}\NormalTok{(}\OperatorTok{~}\StringTok{ }\DecValTok{1} \OperatorTok{+}\StringTok{ }\NormalTok{A}\OperatorTok{*}\NormalTok{B, simdat4)}
\NormalTok{simdat4[, }\KeywordTok{c}\NormalTok{(}\StringTok{"GM"}\NormalTok{, }\StringTok{"FA"}\NormalTok{, }\StringTok{"FB"}\NormalTok{, }\StringTok{"FAxB"}\NormalTok{)] <-}\StringTok{ }\NormalTok{mat_myC}
\NormalTok{m2_mr.v2 <-}\StringTok{ }\KeywordTok{lm}\NormalTok{(DV }\OperatorTok{~}\StringTok{ }\DecValTok{1} \OperatorTok{+}\StringTok{ }\NormalTok{FA }\OperatorTok{+}\StringTok{ }\NormalTok{FB }\OperatorTok{+}\StringTok{ }\NormalTok{FAxB, }\DataTypeTok{data=}\NormalTok{simdat4)}
\end{Highlighting}
\end{Shaded}

\begin{table}[h]
\begin{center}
\begin{threeparttable}
\caption{\label{tab:table18}Regression analysis with sum contrasts.}
\begin{tabular}{lllll}
\toprule
Predictor & \multicolumn{1}{c}{$Estimate$} & \multicolumn{1}{c}{95\% CI} & \multicolumn{1}{c}{$t(16)$} & \multicolumn{1}{c}{$p$}\\
\midrule
Intercept & 20 & $[15$, $25]$ & 8.94 & < .001\\
A1 & -5 & $[-10$, $0]$ & -2.24 & .040\\
B1 & -10 & $[-15$, $-5]$ & -4.47 & < .001\\
A1 $\times$ B1 & 5 & $[0$, $10]$ & 2.24 & .040\\
\bottomrule
\end{tabular}
\end{threeparttable}
\end{center}
\end{table}

\begin{table}[h]
\begin{center}
\begin{threeparttable}
\caption{\label{tab:table19}Defining sum contrasts using the model.matrix() function.}
\begin{tabular}{lllll}
\toprule
Predictor & \multicolumn{1}{c}{$Estimate$} & \multicolumn{1}{c}{95\% CI} & \multicolumn{1}{c}{$t(16)$} & \multicolumn{1}{c}{$p$}\\
\midrule
Intercept & 20 & $[15$, $25]$ & 8.94 & < .001\\
FA & -5 & $[-10$, $0]$ & -2.24 & .040\\
FB & -10 & $[-15$, $-5]$ & -4.47 & < .001\\
FAxB & 5 & $[0$, $10]$ & 2.24 & .040\\
\bottomrule
\end{tabular}
\end{threeparttable}
\end{center}
\end{table}

When using \textsc{sum contrasts}, the results from the multiple regression models (see Tables \ref{tab:table18} and \ref{tab:table19}) are identical to the results from the ANOVA (Table \ref{tab:table16}). This is visible as the F-value for factor \(A\) is now \(-2.24^2 = 5\), for factor \(B\) \textcolor{black}{it} is \(-4.47^2 = 20\), and for the interaction \textcolor{black}{it} is again \(2.24^2 = 5\). All F-values are now the same as in the ANOVA model.

Next, we reproduce the \(A(2) \times B(2)\) - ANOVA with contrasts specified for the corresponding one-way \(F(4)\) ANOVA\textcolor{black}{, that is by treating the $2 \times 2 = 4$ condition means as four levels of a single factor F}. In other words, we go back to the data frame simulated for the analysis of \textsc{repeated contrasts} \textcolor{black}{(see section} \emph{Further examples of contrasts illustrated with a factor with four levels}). We first define weights for condition means according to our hypotheses, invert this matrix, and use it as the contrast matrix for factor F in a LM. \textcolor{black}{We define weights of $1/4$ and $-1/4$. We do so because (a) we want to compare the mean of two conditions to the mean of two other conditions (e.g., factor A compares $\frac{F1 + F2}{2}$ to $\frac{F3 + F4}{2}$). Moreover, (b) we want to use sum contrasts, where the regression coefficients assess half the difference between means. Together (a+b), this yields weights of $1/2 \cdot 1/2 = 1/4$. The resulting contrast matrix contains contrast coefficients of $+1$ or $-1$, showing that we successfully implemented sum contrasts.} The results, presented in Table \ref{tab:table20}, are identical to the previous models.

\begin{Shaded}
\begin{Highlighting}[]
\KeywordTok{t}\NormalTok{(}\KeywordTok{fractions}\NormalTok{(HcInt <-}\StringTok{ }\KeywordTok{rbind}\NormalTok{(}\DataTypeTok{A  =}\KeywordTok{c}\NormalTok{(}\DataTypeTok{F1=}\DecValTok{1}\OperatorTok{/}\DecValTok{4}\NormalTok{,}\DataTypeTok{F2=} \DecValTok{1}\OperatorTok{/}\DecValTok{4}\NormalTok{,}\DataTypeTok{F3=}\OperatorTok{-}\DecValTok{1}\OperatorTok{/}\DecValTok{4}\NormalTok{,}\DataTypeTok{F4=}\OperatorTok{-}\DecValTok{1}\OperatorTok{/}\DecValTok{4}\NormalTok{),}
                           \DataTypeTok{B  =}\KeywordTok{c}\NormalTok{(}\DataTypeTok{F1=}\DecValTok{1}\OperatorTok{/}\DecValTok{4}\NormalTok{,}\DataTypeTok{F2=}\OperatorTok{-}\DecValTok{1}\OperatorTok{/}\DecValTok{4}\NormalTok{,}\DataTypeTok{F3=} \DecValTok{1}\OperatorTok{/}\DecValTok{4}\NormalTok{,}\DataTypeTok{F4=}\OperatorTok{-}\DecValTok{1}\OperatorTok{/}\DecValTok{4}\NormalTok{),}
                           \DataTypeTok{AxB=}\KeywordTok{c}\NormalTok{(}\DataTypeTok{F1=}\DecValTok{1}\OperatorTok{/}\DecValTok{4}\NormalTok{,}\DataTypeTok{F2=}\OperatorTok{-}\DecValTok{1}\OperatorTok{/}\DecValTok{4}\NormalTok{,}\DataTypeTok{F3=}\OperatorTok{-}\DecValTok{1}\OperatorTok{/}\DecValTok{4}\NormalTok{,}\DataTypeTok{F4=} \DecValTok{1}\OperatorTok{/}\DecValTok{4}\NormalTok{))))}
\end{Highlighting}
\end{Shaded}

\begin{verbatim}
##    A    B    AxB 
## F1  1/4  1/4  1/4
## F2  1/4 -1/4 -1/4
## F3 -1/4  1/4 -1/4
## F4 -1/4 -1/4  1/4
\end{verbatim}

\begin{Shaded}
\begin{Highlighting}[]
\NormalTok{(XcInt <-}\StringTok{ }\KeywordTok{ginv2}\NormalTok{(HcInt))}
\end{Highlighting}
\end{Shaded}

\begin{verbatim}
##    A  B  AxB
## F1  1  1  1 
## F2  1 -1 -1 
## F3 -1  1 -1 
## F4 -1 -1  1
\end{verbatim}

\begin{Shaded}
\begin{Highlighting}[]
\KeywordTok{contrasts}\NormalTok{(simdat3}\OperatorTok{$}\NormalTok{F) <-}\StringTok{ }\NormalTok{XcInt}
\NormalTok{m3_mr <-}\StringTok{ }\KeywordTok{lm}\NormalTok{(DV }\OperatorTok{~}\StringTok{ }\DecValTok{1} \OperatorTok{+}\StringTok{ }\NormalTok{F, }\DataTypeTok{data=}\NormalTok{simdat3)}
\end{Highlighting}
\end{Shaded}

\begin{table}[h]
\begin{center}
\begin{threeparttable}
\caption{\label{tab:table20}Main effects and interaction: Custom-defined sum contrasts (scaled).}
\begin{tabular}{lllll}
\toprule
Predictor & \multicolumn{1}{c}{$Estimate$} & \multicolumn{1}{c}{95\% CI} & \multicolumn{1}{c}{$t(16)$} & \multicolumn{1}{c}{$p$}\\
\midrule
Intercept & 20 & $[15$, $25]$ & 8.94 & < .001\\
FA & -5 & $[-10$, $0]$ & -2.24 & .040\\
FB & -10 & $[-15$, $-5]$ & -4.47 & < .001\\
FAxB & 5 & $[0$, $10]$ & 2.24 & .040\\
\bottomrule
\end{tabular}
\end{threeparttable}
\end{center}
\end{table}

This shows that \textcolor{black}{it is possible to} specify the contrasts not only for each factor (e.g., here in the \(2 \times 2\) design) separately. Instead, \textcolor{black}{one can also} pool all experimental conditions (or design cells) into one large factor (here factor F with \(4\) levels), and specify the contrasts for the main effects and for the interactions in the resulting one large contrast matrix simultaneously.

\hypertarget{nestedEffects}{%
\subsection{Nested effects}\label{nestedEffects}}

\textcolor{black}{One} can specify hypotheses that do not correspond directly to main effects and interaction of the traditional ANOVA. For example, in a \(2 \times 2\) experimental design, where factor \(A\) codes word frequency (low/high) and factor \(B\) is part of speech (noun/verb), \textcolor{black}{one} can test the effect of word frequency within nouns and the effect of word frequency within verbs. Formally, \(A_{B1}\) versus \(A_{B2}\) \textcolor{black}{are} nested within levels of \(B\). Said differently, simple effects of factor \(A\) \textcolor{black}{are tested} for each of the levels of factor \(B\).
In this version, we test whether there is a main effect of part of speech (\(B\); as in traditional ANOVA). However, instead of also testing the second main effect word frequency, \(A\), and the interaction, we test (1) whether the two levels of word frequency, \(A\), differ significantly for the first level of \(B\) (i.e., nouns) and (2) whether the two levels of \(A\) differ significantly for the second level of \(B\) (i.e., verbs). In other words, we test whether there are significant differences for \(A\) in \textcolor{black}{each of} the levels of \(B\). Often researchers have hypotheses about these differences, and not about the interaction.

\begin{Shaded}
\begin{Highlighting}[]
\KeywordTok{t}\NormalTok{(}\KeywordTok{fractions}\NormalTok{(HcNes <-}\StringTok{ }\KeywordTok{rbind}\NormalTok{(}\DataTypeTok{B   =}\KeywordTok{c}\NormalTok{(}\DataTypeTok{F1=} \DecValTok{1}\OperatorTok{/}\DecValTok{2}\NormalTok{,}\DataTypeTok{F2=}\OperatorTok{-}\DecValTok{1}\OperatorTok{/}\DecValTok{2}\NormalTok{,}\DataTypeTok{F3=} \DecValTok{1}\OperatorTok{/}\DecValTok{2}\NormalTok{,}\DataTypeTok{F4=}\OperatorTok{-}\DecValTok{1}\OperatorTok{/}\DecValTok{2}\NormalTok{),}
                           \DataTypeTok{B1xA=}\KeywordTok{c}\NormalTok{(}\DataTypeTok{F1=}\OperatorTok{-}\DecValTok{1}\NormalTok{  ,}\DataTypeTok{F2=} \DecValTok{0}\NormalTok{  ,}\DataTypeTok{F3=} \DecValTok{1}\NormalTok{  ,}\DataTypeTok{F4=} \DecValTok{0}\NormalTok{  ),}
                           \DataTypeTok{B2xA=}\KeywordTok{c}\NormalTok{(}\DataTypeTok{F1=} \DecValTok{0}\NormalTok{  ,}\DataTypeTok{F2=}\OperatorTok{-}\DecValTok{1}\NormalTok{  ,}\DataTypeTok{F3=} \DecValTok{0}\NormalTok{  ,}\DataTypeTok{F4=} \DecValTok{1}\NormalTok{ ))))}
\end{Highlighting}
\end{Shaded}

\begin{verbatim}
##    B    B1xA B2xA
## F1  1/2   -1    0
## F2 -1/2    0   -1
## F3  1/2    1    0
## F4 -1/2    0    1
\end{verbatim}

\begin{Shaded}
\begin{Highlighting}[]
\NormalTok{(XcNes <-}\StringTok{ }\KeywordTok{ginv2}\NormalTok{(HcNes))}
\end{Highlighting}
\end{Shaded}

\begin{verbatim}
##    B    B1xA B2xA
## F1  1/2 -1/2    0
## F2 -1/2    0 -1/2
## F3  1/2  1/2    0
## F4 -1/2    0  1/2
\end{verbatim}

\begin{Shaded}
\begin{Highlighting}[]
\KeywordTok{contrasts}\NormalTok{(simdat3}\OperatorTok{$}\NormalTok{F) <-}\StringTok{ }\NormalTok{XcNes}
\NormalTok{m4_mr <-}\StringTok{ }\KeywordTok{lm}\NormalTok{(DV }\OperatorTok{~}\StringTok{ }\DecValTok{1} \OperatorTok{+}\StringTok{ }\NormalTok{F, }\DataTypeTok{data=}\NormalTok{simdat3)}
\end{Highlighting}
\end{Shaded}

\begin{table}[h]
\begin{center}
\begin{threeparttable}
\caption{\label{tab:table21}Regression model for nested effects.}
\begin{tabular}{lllll}
\toprule
Predictor & \multicolumn{1}{c}{$Estimate$} & \multicolumn{1}{c}{95\% CI} & \multicolumn{1}{c}{$t(16)$} & \multicolumn{1}{c}{$p$}\\
\midrule
Intercept & 20 & $[15$, $25]$ & 8.94 & < .001\\
FB & -20 & $[-29$, $-11]$ & -4.47 & < .001\\
FB1xA & 0 & $[-13$, $13]$ & 0.00 & > .999\\
FB2xA & 20 & $[7$, $33]$ & 3.16 & .006\\
\bottomrule
\end{tabular}
\end{threeparttable}
\end{center}
\end{table}

Regression coefficients (cf.~Table \ref{tab:table21}) estimate the GM, the difference for the main effect of word frequency (\(A\)) and the two differences (for \(B\); i.e., simple main effects) within levels of word frequency (\(A\)).

\textcolor{black}{These custom nested contrasts' columns are scaled versions of the corresponding hypothesis matrix. This is the case because the columns are orthogonal. It illustrates the advantage of orthogonal contrasts for the interpretation of regression coefficients: the underlying hypotheses being tested are already clear from the contrast matrix.}

\textcolor{black}{There is also a} built-in R-formula specification of nested designs. The order of factors in the formula from left to right specifies a top-down order of nesting within levels, i.e., here factor \(A\) (word frequency) is nested within levels of the factor \(B\) (part of speech). This (see Table \ref{tab:table22}) yields the exact same result as our previous result based on custom nested contrasts (cf.~Table \ref{tab:table21}).

\begin{Shaded}
\begin{Highlighting}[]
\KeywordTok{contrasts}\NormalTok{(simdat4}\OperatorTok{$}\NormalTok{A) <-}\StringTok{ }\KeywordTok{c}\NormalTok{(}\OperatorTok{-}\FloatTok{0.5}\NormalTok{,}\OperatorTok{+}\FloatTok{0.5}\NormalTok{)}
\KeywordTok{contrasts}\NormalTok{(simdat4}\OperatorTok{$}\NormalTok{B) <-}\StringTok{ }\KeywordTok{c}\NormalTok{(}\OperatorTok{+}\FloatTok{0.5}\NormalTok{,}\OperatorTok{-}\FloatTok{0.5}\NormalTok{)}
\NormalTok{m4_mr.x <-}\StringTok{ }\KeywordTok{lm}\NormalTok{(DV }\OperatorTok{~}\StringTok{ }\DecValTok{1} \OperatorTok{+}\StringTok{ }\NormalTok{B }\OperatorTok{/}\StringTok{ }\NormalTok{A, }\DataTypeTok{data=}\NormalTok{simdat4)}
\end{Highlighting}
\end{Shaded}

\begin{table}[h]
\begin{center}
\begin{threeparttable}
\caption{\label{tab:table22}Nested effects: R-formula.}
\begin{tabular}{lllll}
\toprule
Predictor & \multicolumn{1}{c}{$Estimate$} & \multicolumn{1}{c}{95\% CI} & \multicolumn{1}{c}{$t(16)$} & \multicolumn{1}{c}{$p$}\\
\midrule
Intercept & 20 & $[15$, $25]$ & 8.94 & < .001\\
B1 & -20 & $[-29$, $-11]$ & -4.47 & < .001\\
BB1 $\times$ A1 & 0 & $[-13$, $13]$ & 0.00 & > .999\\
BB2 $\times$ A1 & 20 & $[7$, $33]$ & 3.16 & .006\\
\bottomrule
\end{tabular}
\end{threeparttable}
\end{center}
\end{table}

In cases such as these, where \(A_{B1}\) vs. \(A_{B2}\) \textcolor{black}{are} nested within levels of \(B\), why \textcolor{black}{is it necessary} to estimate the effect of \(B\) (part of speech) at all, when \textcolor{black}{the interest} might only be in the effect of \(A\) (word frequency) within levels of \(B\) (part of speech)? Setting up a regression model where the main effect of \(B\) (part of speech) is removed yields:

\begin{Shaded}
\begin{Highlighting}[]
\CommentTok{# Extract contrasts as covariates from model matrix}
\NormalTok{mat_myC <-}\StringTok{ }\KeywordTok{model.matrix}\NormalTok{(}\OperatorTok{~}\StringTok{ }\DecValTok{1} \OperatorTok{+}\StringTok{ }\NormalTok{F, simdat3)}
\KeywordTok{fractions}\NormalTok{(}\KeywordTok{as.matrix}\NormalTok{(}\KeywordTok{data.frame}\NormalTok{(mat_myC)))}
\end{Highlighting}
\end{Shaded}

\begin{verbatim}
##    X.Intercept. FB   FB1xA FB2xA
## 1     1          1/2 -1/2     0 
## 2     1          1/2 -1/2     0 
## 3     1          1/2 -1/2     0 
## 4     1          1/2 -1/2     0 
## 5     1          1/2 -1/2     0 
## 6     1         -1/2    0  -1/2 
## 7     1         -1/2    0  -1/2 
## 8     1         -1/2    0  -1/2 
## 9     1         -1/2    0  -1/2 
## 10    1         -1/2    0  -1/2 
## 11    1          1/2  1/2     0 
## 12    1          1/2  1/2     0 
## 13    1          1/2  1/2     0 
## 14    1          1/2  1/2     0 
## 15    1          1/2  1/2     0 
## 16    1         -1/2    0   1/2 
## 17    1         -1/2    0   1/2 
## 18    1         -1/2    0   1/2 
## 19    1         -1/2    0   1/2 
## 20    1         -1/2    0   1/2
\end{verbatim}

\begin{Shaded}
\begin{Highlighting}[]
\CommentTok{# Repeat the multiple regression with covariates }
\NormalTok{simdat3[, }\KeywordTok{c}\NormalTok{(}\StringTok{"GM"}\NormalTok{, }\StringTok{"B"}\NormalTok{, }\StringTok{"B1_A"}\NormalTok{, }\StringTok{"B2_A"}\NormalTok{)] <-}\StringTok{ }\NormalTok{mat_myC}
\NormalTok{m2_mr.myC1 <-}\StringTok{ }\KeywordTok{lm}\NormalTok{(DV }\OperatorTok{~}\StringTok{ }\DecValTok{1} \OperatorTok{+}\StringTok{ }\NormalTok{B }\OperatorTok{+}\StringTok{ }\NormalTok{B1_A }\OperatorTok{+}\StringTok{ }\NormalTok{B2_A, }\DataTypeTok{data=}\NormalTok{simdat3)}
\KeywordTok{summary}\NormalTok{(m2_mr.myC1)}\OperatorTok{$}\NormalTok{sigma }\CommentTok{# standard deviation of residual}
\end{Highlighting}
\end{Shaded}

\begin{verbatim}
## [1] 10
\end{verbatim}

\begin{Shaded}
\begin{Highlighting}[]
\CommentTok{# Run the multiple regression by leaving out C1}
\NormalTok{m2_mr.myC2 <-}\StringTok{ }\KeywordTok{lm}\NormalTok{(DV }\OperatorTok{~}\StringTok{ }\DecValTok{1} \OperatorTok{+}\StringTok{ }\NormalTok{B1_A }\OperatorTok{+}\StringTok{ }\NormalTok{B2_A, }\DataTypeTok{data=}\NormalTok{simdat3)}
\KeywordTok{summary}\NormalTok{(m2_mr.myC2)}\OperatorTok{$}\NormalTok{sigma }\CommentTok{# standard deviation of residual}
\end{Highlighting}
\end{Shaded}

\begin{verbatim}
## [1] 14.6
\end{verbatim}

\begin{table}[h]
\begin{center}
\begin{threeparttable}
\caption{\label{tab:table22a}Nested effects: Full model.}
\begin{tabular}{lllll}
\toprule
Predictor & \multicolumn{1}{c}{$Estimate$} & \multicolumn{1}{c}{95\% CI} & \multicolumn{1}{c}{$t(16)$} & \multicolumn{1}{c}{$p$}\\
\midrule
Intercept & 20 & $[15$, $25]$ & 8.94 & < .001\\
B & -20 & $[-29$, $-11]$ & -4.47 & < .001\\
B1 A & 0 & $[-13$, $13]$ & 0.00 & > .999\\
B2 A & 20 & $[7$, $33]$ & 3.16 & .006\\
\bottomrule
\end{tabular}
\end{threeparttable}
\end{center}
\end{table}

\begin{table}[h]
\begin{center}
\begin{threeparttable}
\caption{\label{tab:table-myC2}Nested effects: Without main effect of B B.}
\begin{tabular}{lllll}
\toprule
Predictor & \multicolumn{1}{c}{$Estimate$} & \multicolumn{1}{c}{95\% CI} & \multicolumn{1}{c}{$t(17)$} & \multicolumn{1}{c}{$p$}\\
\midrule
Intercept & 20 & $[13$, $27]$ & 6.15 & < .001\\
B1 A & 0 & $[-19$, $19]$ & 0.00 & > .999\\
B2 A & 20 & $[1$, $39]$ & 2.17 & .044\\
\bottomrule
\end{tabular}
\end{threeparttable}
\end{center}
\end{table}

The results (Tables \ref{tab:table22a} and \ref{tab:table-myC2}) show that in the fully balanced data-set simulated here, the estimates for \(A_{B1}\) and \(A_{B2}\) are still the same. However, the confidence intervals and associated p-values are now larger. One loses statistical power because the variance explained by factor \(B\) is now not taken into account in the linear regression any more. As a consequence, the unexplained variance increases the standard deviation of the residual from \(10\) to \(14.55\), and increases uncertainty about \(A_{B1}\) and \(A_{B2}\). In unbalanced or nonorthogonal designs, leaving out the effect of \(B\) from the linear regression can lead to dramatic changes in the estimated slopes.

Of course, we can also ask the reverse question: Are the differences significant for part of speech (\(B\)) in the levels of word frequency (\(A\); in addition to testing the main effect of word frequency, \(A\)). That is, do nouns differ from verbs for low-frequency words (\(B_{A1}\)) and do nouns differ from verbs for high-frequency words (\(B_{A2}\))?

\begin{Shaded}
\begin{Highlighting}[]
\KeywordTok{t}\NormalTok{(}\KeywordTok{fractions}\NormalTok{(HcNes2 <-}\StringTok{ }\KeywordTok{rbind}\NormalTok{(}\DataTypeTok{A   =}\KeywordTok{c}\NormalTok{(}\DataTypeTok{F1=}\DecValTok{1}\OperatorTok{/}\DecValTok{2}\NormalTok{,}\DataTypeTok{F2=} \DecValTok{1}\OperatorTok{/}\DecValTok{2}\NormalTok{,}\DataTypeTok{F3=}\OperatorTok{-}\DecValTok{1}\OperatorTok{/}\DecValTok{2}\NormalTok{,}\DataTypeTok{F4=}\OperatorTok{-}\DecValTok{1}\OperatorTok{/}\DecValTok{2}\NormalTok{),}
                            \DataTypeTok{A1_B=}\KeywordTok{c}\NormalTok{(}\DataTypeTok{F1=}\DecValTok{1}\NormalTok{  ,}\DataTypeTok{F2=}\OperatorTok{-}\DecValTok{1}\NormalTok{  ,}\DataTypeTok{F3=} \DecValTok{0}\NormalTok{  ,}\DataTypeTok{F4=} \DecValTok{0}\NormalTok{  ),}
                            \DataTypeTok{A2_B=}\KeywordTok{c}\NormalTok{(}\DataTypeTok{F1=}\DecValTok{0}\NormalTok{  ,}\DataTypeTok{F2=} \DecValTok{0}\NormalTok{  ,}\DataTypeTok{F3=} \DecValTok{1}\NormalTok{  ,}\DataTypeTok{F4=}\OperatorTok{-}\DecValTok{1}\NormalTok{ ))))}
\end{Highlighting}
\end{Shaded}

\begin{verbatim}
##    A    A1_B A2_B
## F1  1/2    1    0
## F2  1/2   -1    0
## F3 -1/2    0    1
## F4 -1/2    0   -1
\end{verbatim}

\begin{Shaded}
\begin{Highlighting}[]
\NormalTok{(XcNes2 <-}\StringTok{ }\KeywordTok{ginv2}\NormalTok{(HcNes2))}
\end{Highlighting}
\end{Shaded}

\begin{verbatim}
##    A    A1_B A2_B
## F1  1/2  1/2    0
## F2  1/2 -1/2    0
## F3 -1/2    0  1/2
## F4 -1/2    0 -1/2
\end{verbatim}

\begin{Shaded}
\begin{Highlighting}[]
\KeywordTok{contrasts}\NormalTok{(simdat3}\OperatorTok{$}\NormalTok{F) <-}\StringTok{ }\NormalTok{XcNes2}
\NormalTok{m4_mr <-}\StringTok{ }\KeywordTok{lm}\NormalTok{(DV }\OperatorTok{~}\StringTok{ }\DecValTok{1} \OperatorTok{+}\StringTok{ }\NormalTok{F, }\DataTypeTok{data=}\NormalTok{simdat3)}
\end{Highlighting}
\end{Shaded}

\begin{table}[h]
\begin{center}
\begin{threeparttable}
\caption{\label{tab:table-m4mr}Factor B nested within Factor A.}
\begin{tabular}{lllll}
\toprule
Predictor & \multicolumn{1}{c}{$Estimate$} & \multicolumn{1}{c}{95\% CI} & \multicolumn{1}{c}{$t(16)$} & \multicolumn{1}{c}{$p$}\\
\midrule
Intercept & 20 & $[15$, $25]$ & 8.94 & < .001\\
FA & -10 & $[-19$, $-1]$ & -2.24 & .040\\
FA1 B & -10 & $[-23$, $3]$ & -1.58 & .133\\
FA2 B & -30 & $[-43$, $-17]$ & -4.74 & < .001\\
\bottomrule
\end{tabular}
\end{threeparttable}
\end{center}
\end{table}

Regression coefficients (cf.~Table \ref{tab:table-m4mr}) estimate the GM, the difference for the main effect of word frequency (\(A\)) and the two part of speech effects (for \(B\)\textcolor{black}{; i.e., simple main effects}) within levels of word frequency (\(A\)).

\hypertarget{interactions-between-contrasts}{%
\subsection{Interactions between contrasts}\label{interactions-between-contrasts}}

\textcolor{black}{We have discussed above that in a $2 \times 2$ experimental design, the results from sum contrasts (see Table}~\ref{tab:table18}\textcolor{black}{) are equivalent to typical ANOVA results (see Table}~\ref{tab:table16}). In addition, we had also run the analysis with treatment contrasts. \textcolor{black}{It was clear} that the results for treatment contrasts (see Table~\ref{tab:table17}) did not correspond to the results from the ANOVA. However, if the results for treatment contrasts do not correspond to the typical ANOVA results, what do they then test? That is, \textcolor{black}{is it still possible to} meaningfully interpret the results from the treatment contrasts in a simple \(2 \times 2\) design?

\textcolor{black}{This leads us to a very important principle in interpreting results from contrasts: When interactions between contrasts are included in a model, then the results of one contrast actually depend on the specification of the other contrast(s) in the analysis! This may be counter-intuitive at first. However, it is very important and essential to keep in mind when interpreting results from contrasts. How does this work in detail?}

\textcolor{black}{The general rule to remember is that the main effect of one contrast measures its effect at the location $0$ of the other contrast(s) in the analysis. What does that mean? Let us consider the example that we use two treatment contrasts in a $2 \times 2$ design (see results in Table}~\ref{tab:table17}\textcolor{black}{). Let's take a look at the main effect of factor A. How can we interpret what this measures or tests? This main effect actually tests the effect of factor A at the "location" where factor B is coded as $0$. Factor B is coded as a treatment contrast, that is, it codes a zero at its baseline condition, which is B1. Thus, the main effect of factor A tests the effect of A nested within the baseline condition of B. We take a look at the data presented in Figure}~\ref{fig:twobytwosimdatFig}\textcolor{black}{, what this nested effect should be. Figure}~\ref{fig:twobytwosimdatFig} shows that the effect of factor A nested in B1 is \(0\). If we now compare this to the results from the linear model, \textcolor{black}{it is indeed clear} that the main effect of factor A (see Table~\ref{tab:table17}\textcolor{black}{) is exactly estimated as $0$. As expected, when factor B is coded as a treatment contrast, the main effect of factor A tests the effect of A nested within the baseline level of factor B.}

Next, consider the main effect of factor B. According to the same logic, this main effect tests the effect of factor B at the \enquote{location} where factor A is \(0\). Factor A is also coded as a treatment contrast, that is, it codes its baseline condition A1 as \(0\). The main effect of factor B tests the effect of B nested within the baseline condition of A. Figure~\ref{fig:twobytwosimdatFig} shows that this effect should be \(10\); this indeed corresponds to the main effect of B as estimated in the regression model for treatment contrasts (see Table \ref{tab:table17}, the \emph{Estimate} for BB2). As we had seen before, the interaction term, however, does not differ between \textcolor{black}{the} treatment contrast and ANOVA (\(t^2 = 2.24^2 = F = 5.00\)).

\textcolor{black}{How do we know what the "location" is, where a contrast applies? For the treatment contrasts discussed here, it is possible to reason this through because all contrasts are coded as $0$ or $1$. However, how is it possible to derive the "location" in general? What we can do is to }
look at the hypotheses tested by the treatment contrasts in the presence of an interaction between them by using the generalized matrix inverse. We go back to the default treatment contrasts. Then we extract the contrast matrix from the design matrix:

\begin{Shaded}
\begin{Highlighting}[]
\KeywordTok{contrasts}\NormalTok{(simdat4}\OperatorTok{$}\NormalTok{A) <-}\StringTok{ }\KeywordTok{contr.treatment}\NormalTok{(}\DecValTok{2}\NormalTok{)}
\KeywordTok{contrasts}\NormalTok{(simdat4}\OperatorTok{$}\NormalTok{B) <-}\StringTok{ }\KeywordTok{contr.treatment}\NormalTok{(}\DecValTok{2}\NormalTok{)}
\NormalTok{XcTr <-}\StringTok{ }\NormalTok{simdat4 }\OperatorTok{%>%}
\StringTok{  }\KeywordTok{group_by}\NormalTok{(A, B) }\OperatorTok{%>%}
\StringTok{  }\KeywordTok{summarise}\NormalTok{() }\OperatorTok{%>%}
\StringTok{  }\KeywordTok{model.matrix}\NormalTok{(}\OperatorTok{~}\StringTok{ }\DecValTok{1} \OperatorTok{+}\StringTok{ }\NormalTok{A}\OperatorTok{*}\NormalTok{B, .) }\OperatorTok{%>%}
\StringTok{  }\KeywordTok{as.data.frame}\NormalTok{() }\OperatorTok{%>%}\StringTok{ }\KeywordTok{as.matrix}\NormalTok{()}
\KeywordTok{rownames}\NormalTok{(XcTr) <-}\StringTok{ }\KeywordTok{c}\NormalTok{(}\StringTok{"A1_B1"}\NormalTok{,}\StringTok{"A1_B2"}\NormalTok{,}\StringTok{"A2_B1"}\NormalTok{,}\StringTok{"A2_B2"}\NormalTok{)}
\NormalTok{XcTr}
\end{Highlighting}
\end{Shaded}

\begin{verbatim}
##       (Intercept) A2 B2 A2:B2
## A1_B1           1  0  0     0
## A1_B2           1  0  1     0
## A2_B1           1  1  0     0
## A2_B2           1  1  1     1
\end{verbatim}

\textcolor{black}{This shows the treatment contrast for factors} \texttt{A} and \texttt{B}\textcolor{black}{, and their interaction. We now apply the generalized inverse. Remember, when we apply the generalized inverse to the contrast matrix, we obtain the corresponding hypothesis matrix (again, we use matrix transpose for better readability):}

\begin{Shaded}
\begin{Highlighting}[]
\KeywordTok{t}\NormalTok{(}\KeywordTok{ginv2}\NormalTok{(XcTr))}
\end{Highlighting}
\end{Shaded}

\begin{verbatim}
##       (Intercept) A2 B2 A2:B2
## A1_B1  1          -1 -1  1   
## A1_B2  0           0  1 -1   
## A2_B1  0           1  0 -1   
## A2_B2  0           0  0  1
\end{verbatim}

\textcolor{black}{As discussed above, the main effect of factor} \texttt{A} \textcolor{black}{tests its effect nested within the baseline level of factor} \texttt{B}\textcolor{black}{. Likewise, the main effect of factor} \texttt{B} \textcolor{black}{tests its effect nested within the baseline level of factor} \texttt{A}.

\textcolor{black}{How does this work for sum contrasts? They do not have a baseline condition that is coded as $0$. In sum contrasts, however, the average of the contrast coefficients is $0$. Therefore, main effects test the average effect across factor levels. This is what is typically also tested in standard ANOVA. Let's look at the example shown in Table}~\ref{tab:table18}\textcolor{black}{: given that factor B has a sum contrast, the main effect of factor A is tested as the average across levels of factor B. Figure}~\ref{fig:twobytwosimdatFig} \textcolor{black}{shows that the effect of factor A in level B1 is $10 - 10 = 0$, and in level B2 it is $20 - 40 = -20$. The average effect across both levels is $(0 - 20)/2 = -10$. Due to the sum contrast coding, we have to divide this by 2, yielding an expected effect of $-10 / 2 = -5$. This is exactly what the main effect of factor A measures (see Table}~\ref{tab:table18}, \emph{Estimate} \textcolor{black}{for A1).}

Similarly, factor B tests its effect at the location \(0\) of factor A. Again, \(0\) is exactly the mean of the contrast coefficients from factor A, which is coded as a sum contrast. Therefore, factor B tests the effect of B averaged across factor levels of A. For factor level A1, factor B has an effect of \(10 - 20 = -10\). For factor level A2, factor B has an effect of \(10 - 40 = -30\). The average effect is \((-10 - 30)/2 = -20\), which \textcolor{black}{again needs to be} divided by \(2\) due to the sum contrast. This yields exactly the estimate of \(-10\) that is also reported in Table~\ref{tab:table18} (\emph{Estimate} \textcolor{black}{for B1).}

\textcolor{black}{Again, we look at the hypothesis matrix for the main effects and the interaction:}

\begin{Shaded}
\begin{Highlighting}[]
\KeywordTok{contrasts}\NormalTok{(simdat4}\OperatorTok{$}\NormalTok{A) <-}\StringTok{ }\KeywordTok{contr.sum}\NormalTok{(}\DecValTok{2}\NormalTok{)}
\KeywordTok{contrasts}\NormalTok{(simdat4}\OperatorTok{$}\NormalTok{B) <-}\StringTok{ }\KeywordTok{contr.sum}\NormalTok{(}\DecValTok{2}\NormalTok{)}
\NormalTok{XcSum <-}\StringTok{ }\NormalTok{simdat4 }\OperatorTok{%>%}
\StringTok{  }\KeywordTok{group_by}\NormalTok{(A, B) }\OperatorTok{%>%}
\StringTok{  }\KeywordTok{summarise}\NormalTok{() }\OperatorTok{%>%}
\StringTok{  }\KeywordTok{model.matrix}\NormalTok{(}\OperatorTok{~}\StringTok{ }\DecValTok{1} \OperatorTok{+}\StringTok{ }\NormalTok{A}\OperatorTok{*}\NormalTok{B, .) }\OperatorTok{%>%}
\StringTok{  }\KeywordTok{as.data.frame}\NormalTok{() }\OperatorTok{%>%}\StringTok{ }\KeywordTok{as.matrix}\NormalTok{()}
\KeywordTok{rownames}\NormalTok{(XcSum) <-}\StringTok{ }\KeywordTok{c}\NormalTok{(}\StringTok{"A1_B1"}\NormalTok{,}\StringTok{"A1_B2"}\NormalTok{,}\StringTok{"A2_B1"}\NormalTok{,}\StringTok{"A2_B2"}\NormalTok{)}
\NormalTok{XcSum}
\end{Highlighting}
\end{Shaded}

\begin{verbatim}
##       (Intercept) A1 B1 A1:B1
## A1_B1           1  1  1     1
## A1_B2           1  1 -1    -1
## A2_B1           1 -1  1    -1
## A2_B2           1 -1 -1     1
\end{verbatim}

\begin{Shaded}
\begin{Highlighting}[]
\KeywordTok{t}\NormalTok{(}\KeywordTok{ginv2}\NormalTok{(XcSum))}
\end{Highlighting}
\end{Shaded}

\begin{verbatim}
##       (Intercept) A1   B1   A1:B1
## A1_B1  1/4         1/4  1/4  1/4 
## A1_B2  1/4         1/4 -1/4 -1/4 
## A2_B1  1/4        -1/4  1/4 -1/4 
## A2_B2  1/4        -1/4 -1/4  1/4
\end{verbatim}

\textcolor{black}{This shows that each of the main effects now does not compute nested comparisons any more, but that they rather test their effect averaged across conditions of the other factor. The averaging involves using weights of $1/2$. Moreover, the regression coefficients in the sum contrast measure half the distance between conditions, leading to weights of $1/2 \cdot 1/2 = 1/4$.}

\textcolor{black}{The general rule to remember from these examples is that when interactions between contrasts are tested, what a main effect of a factor tests depends on the contast coding of the other factors in the design! The main effect of a factor tests the effect nested within the location zero of the other contrast(s) in an analysis. If another contrast is centered, and zero is the average of this other contrasts' coefficients, then the contrast of interest tests the average effect, averaged across the levels of the other factor. Importantly, this property holds only when the interaction between two contrasts is included into a model. If the interaction is omitted and only main effects are tested, then there is no such "action at a distance".}

\textcolor{black}{This may be a very surprising result for interactions of contrasts. However, it is also essential to interpreting contrast coefficients involved in interactions. It is particularly relevant for the analysis of the default \textsc{treatment contrast}, where the main effects test nested effects rather than average effects.}

\hypertarget{a-priori-interaction-contrasts}{%
\subsection{A priori interaction contrasts}\label{a-priori-interaction-contrasts}}

\textcolor{black}{When testing interaction effects, if at least one of the factors involved in an interaction has more than two levels, then omnibus F tests for the interaction are not informative about the source of the interaction. Contrasts are used to test hypotheses about specific differences between differences. Sometimes, researchers may have a hypothesis believed to fully explain the observed pattern of means associated with the interaction between two factors} (Abelson \& Prentice, 1997). In such situations, \textcolor{black}{it is possible to specify} an a priori interaction contrast for testing this hypothesis. Such contrasts are more focused compared to the associated omnibus F test for the interaction term.
Maxwell, Delaney, and Kelley (2018), citing Abelson and Prentice (1997), \textcolor{black}{write that "psychologists have historically failed to capitalize on possible advantages of testing a priori interaction contrasts even when theory dictates the relevance of the approach"} (p.~346).

Abelson and Prentice (1997) \textcolor{black}{demonstrate interesting example cases for such a priori interaction contrasts, which they argue are often useful in psychological studies, including what they term a "matching" pattern. We here illustrate this "matching" pattern, by simulating data from a hypothetical priming study: we assume a $Prime(3) \times Target(3)$ between-subject-Factor design, where factor} \texttt{Prime} \textcolor{black}{indicates three ordinally scaled levels of different prime stimuli} \texttt{c("Prime1",\ "Prime2",\ "Prime3")}\textcolor{black}{, whereas factor} \texttt{Target} \textcolor{black}{indicates three ordinally scaled levels of different target stimuli} \texttt{c("Target1",\ "Target2",\ "Target3")}\textcolor{black}{. We assume a similarity structure in which} \texttt{Prime1} \textcolor{black}{primes} \texttt{Target1}, \texttt{Prime2} \textcolor{black}{primes} \texttt{Target2}\textcolor{black}{, and} \texttt{Prime3} \textcolor{black}{primes} \texttt{Target3}\textcolor{black}{. In addition, we assume some similarity between neighboring categories, such that, e.g.,} \texttt{Prime1} \textcolor{black}{also weakly primes} \texttt{Target2}, \texttt{Prime3} \textcolor{black}{also weakly primes} \texttt{Target2}\textcolor{black}{, and} \texttt{Prime2} \textcolor{black}{also weakly primes} \texttt{Target1} and \texttt{Target2}\textcolor{black}{. We therefore assume the following pattern of mean response times, for which we simulate response times:}

\begin{Shaded}
\begin{Highlighting}[]
\NormalTok{M9 <-}\StringTok{ }\KeywordTok{matrix}\NormalTok{(}\KeywordTok{c}\NormalTok{(}\DecValTok{150}\NormalTok{,}\DecValTok{175}\NormalTok{,}\DecValTok{200}\NormalTok{, }\DecValTok{175}\NormalTok{,}\DecValTok{150}\NormalTok{,}\DecValTok{175}\NormalTok{, }\DecValTok{200}\NormalTok{,}\DecValTok{175}\NormalTok{,}\DecValTok{150}\NormalTok{),}\DataTypeTok{ncol=}\DecValTok{1}\NormalTok{)}
\KeywordTok{matrix}\NormalTok{(M9,}\DataTypeTok{nrow=}\DecValTok{3}\NormalTok{,}\DataTypeTok{dimnames=}\KeywordTok{list}\NormalTok{(}\KeywordTok{paste0}\NormalTok{(}\StringTok{"Prime"}\NormalTok{,}\DecValTok{1}\OperatorTok{:}\DecValTok{3}\NormalTok{),}\KeywordTok{paste0}\NormalTok{(}\StringTok{"Target"}\NormalTok{,}\DecValTok{1}\OperatorTok{:}\DecValTok{3}\NormalTok{)))}
\end{Highlighting}
\end{Shaded}

\begin{verbatim}
##        Target1 Target2 Target3
## Prime1     150     175     200
## Prime2     175     150     175
## Prime3     200     175     150
\end{verbatim}

\begin{Shaded}
\begin{Highlighting}[]
\KeywordTok{set.seed}\NormalTok{(}\DecValTok{1}\NormalTok{)}
\NormalTok{simdat5 <-}\StringTok{ }\KeywordTok{mixedDesign}\NormalTok{(}\DataTypeTok{B=}\KeywordTok{c}\NormalTok{(}\DecValTok{3}\NormalTok{,}\DecValTok{3}\NormalTok{), }\DataTypeTok{W=}\OtherTok{NULL}\NormalTok{, }\DataTypeTok{n=}\DecValTok{5}\NormalTok{, }\DataTypeTok{M=}\NormalTok{M9, }\DataTypeTok{SD=}\DecValTok{50}\NormalTok{, }\DataTypeTok{long =} \OtherTok{TRUE}\NormalTok{) }
\KeywordTok{names}\NormalTok{(simdat5)[}\DecValTok{1}\OperatorTok{:}\DecValTok{2}\NormalTok{] <-}\StringTok{ }\KeywordTok{c}\NormalTok{(}\StringTok{"Prime"}\NormalTok{,}\StringTok{"Target"}\NormalTok{)}
\KeywordTok{levels}\NormalTok{(simdat5}\OperatorTok{$}\NormalTok{Prime) <-}\StringTok{ }\KeywordTok{paste0}\NormalTok{(}\StringTok{"Prime"}\NormalTok{,}\DecValTok{1}\OperatorTok{:}\DecValTok{3}\NormalTok{)}
\KeywordTok{levels}\NormalTok{(simdat5}\OperatorTok{$}\NormalTok{Target) <-}\StringTok{ }\KeywordTok{paste0}\NormalTok{(}\StringTok{"Target"}\NormalTok{,}\DecValTok{1}\OperatorTok{:}\DecValTok{3}\NormalTok{)}
\NormalTok{table5 <-}\StringTok{ }\NormalTok{simdat5 }\OperatorTok{%>%}\StringTok{ }\KeywordTok{group_by}\NormalTok{(Prime, Target) }\OperatorTok{%>%}\StringTok{ }\CommentTok{# plot interaction}
\StringTok{    }\KeywordTok{summarize}\NormalTok{(}\DataTypeTok{N=}\KeywordTok{length}\NormalTok{(DV), }\DataTypeTok{M=}\KeywordTok{mean}\NormalTok{(DV), }\DataTypeTok{SD=}\KeywordTok{sd}\NormalTok{(DV), }\DataTypeTok{SE=}\NormalTok{SD}\OperatorTok{/}\KeywordTok{sqrt}\NormalTok{(N))}
\NormalTok{table5 }\OperatorTok{%>%}\StringTok{ }\KeywordTok{select}\NormalTok{(Prime,Target,M) }\OperatorTok{%>%}\StringTok{ }\KeywordTok{spread}\NormalTok{(}\DataTypeTok{key=}\NormalTok{Target, }\DataTypeTok{value=}\NormalTok{M) }\OperatorTok{%>%}\StringTok{ }
\StringTok{  }\KeywordTok{data.frame}\NormalTok{()}
\end{Highlighting}
\end{Shaded}

\begin{verbatim}
##    Prime Target1 Target2 Target3
## 1 Prime1     150     175     200
## 2 Prime2     175     150     175
## 3 Prime3     200     175     150
\end{verbatim}

\begin{figure}

{\centering \includegraphics{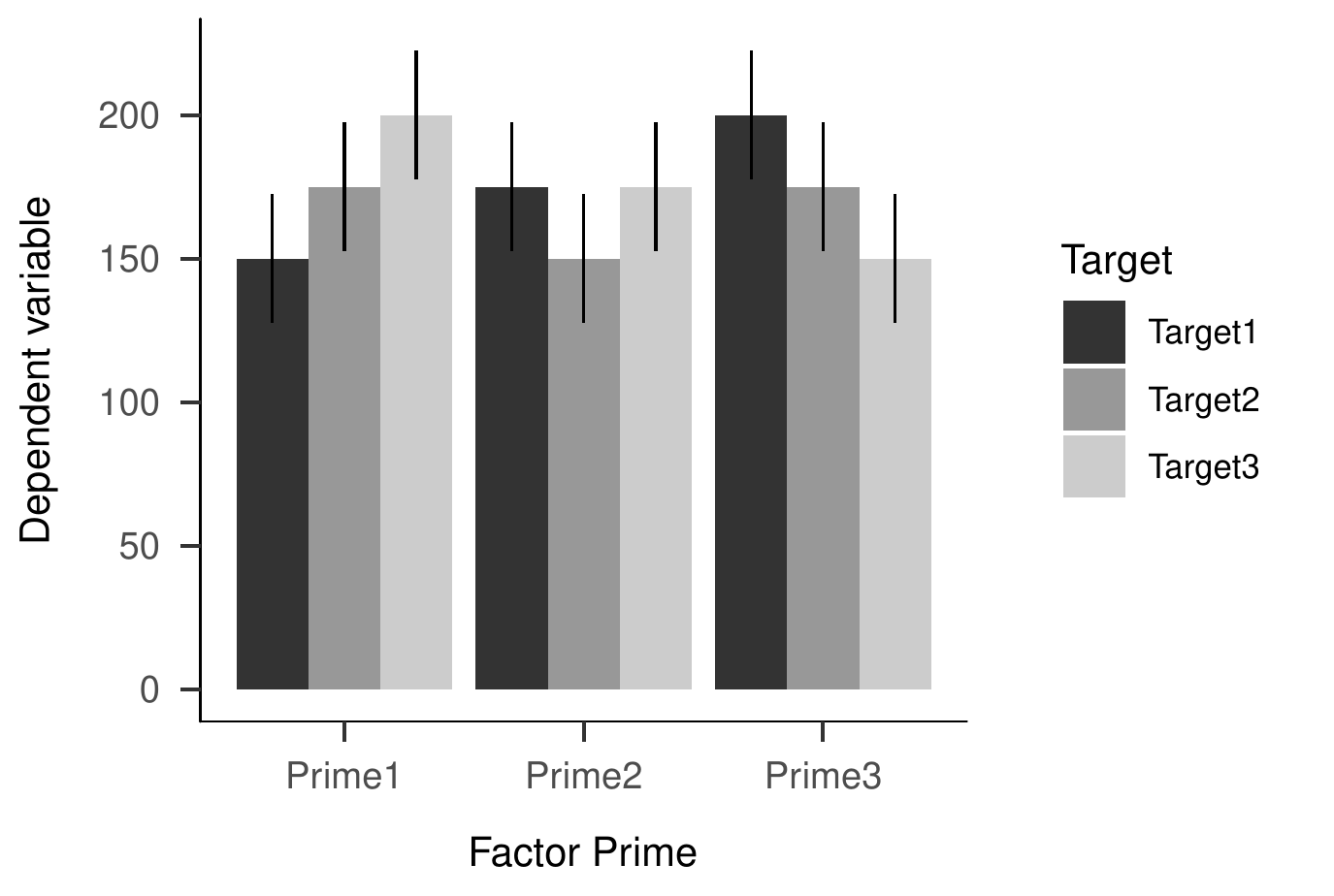} 

}

\caption{Means and error bars (showing standard errors) for a simulated data-set with a 3 x 3 between-subjects factorial design.}\label{fig:3by3simdatFig}
\end{figure}

\textcolor{black}{This reflects a "matching" pattern, where relative match between prime and target is the only factor determining differences in response time between conditions. We specify this hypothesis in an a priori contrast. One way to start for formulating a contrast matrix can be to write down the expected pattern of means. Here we assume that we would a priori hypothesize that fully matching primes and targets speed up response times, and that all other combinations lead to equally slow responses:}

\begin{Shaded}
\begin{Highlighting}[]
\NormalTok{meansExp <-}\StringTok{ }\KeywordTok{rbind}\NormalTok{(}\DataTypeTok{Prime1=}\KeywordTok{c}\NormalTok{(}\DataTypeTok{Target1=}\DecValTok{150}\NormalTok{,}\DataTypeTok{Target2=}\DecValTok{200}\NormalTok{,}\DataTypeTok{Target3=}\DecValTok{200}\NormalTok{), }
                  \DataTypeTok{Prime2=}\KeywordTok{c}\NormalTok{(        }\DecValTok{200}\NormalTok{,        }\DecValTok{150}\NormalTok{,        }\DecValTok{200}\NormalTok{), }
                  \DataTypeTok{Prime3=}\KeywordTok{c}\NormalTok{(        }\DecValTok{200}\NormalTok{,        }\DecValTok{200}\NormalTok{,        }\DecValTok{150}\NormalTok{))}
\end{Highlighting}
\end{Shaded}

\textcolor{black}{We transform this into a contrast matrix by subtracting the mean response time. We also scale the matrix to ease readability.}

\begin{Shaded}
\begin{Highlighting}[]
\NormalTok{(XcS <-}\StringTok{ }\NormalTok{(meansExp}\OperatorTok{-}\KeywordTok{mean}\NormalTok{(meansExp))}\OperatorTok{*}\DecValTok{3}\OperatorTok{/}\DecValTok{50}\NormalTok{)}
\end{Highlighting}
\end{Shaded}

\begin{verbatim}
##        Target1 Target2 Target3
## Prime1      -2       1       1
## Prime2       1      -2       1
## Prime3       1       1      -2
\end{verbatim}

\textcolor{black}{This matrix is orthogonal to the main effects as all the rows and all the columns sum to zero. This is important for contrast matrices for interaction contrasts to ensure that the contrast matrix does not simply capture and test parts of the main effects. If some rows or columns from this matrix do not sum to zero, then it is important to change the interaction contrast matrix to fulfill this requirement. Based on the interaction contrast matrix, we here perform an ANOVA decomposition of the main effects, the omnibus F test for the interaction, and for the effects involved within the interaction:}

\begin{Shaded}
\begin{Highlighting}[]
\NormalTok{simdat5}\OperatorTok{$}\NormalTok{cMatch <-}\StringTok{ }\KeywordTok{ifelse}\NormalTok{(}
    \KeywordTok{as.numeric}\NormalTok{(simdat5}\OperatorTok{$}\NormalTok{Prime)}\OperatorTok{==}\KeywordTok{as.numeric}\NormalTok{(simdat5}\OperatorTok{$}\NormalTok{Target),}\OperatorTok{-}\DecValTok{2}\NormalTok{,}\DecValTok{1}\NormalTok{)}
\NormalTok{mOmn <-}\StringTok{ }\KeywordTok{summary}\NormalTok{(}\KeywordTok{aov}\NormalTok{(DV}\OperatorTok{~}\NormalTok{Prime}\OperatorTok{*}\NormalTok{Target       }\OperatorTok{+}\KeywordTok{Error}\NormalTok{(id), }\DataTypeTok{data=}\NormalTok{simdat5))}
\NormalTok{mCon <-}\StringTok{ }\KeywordTok{summary}\NormalTok{(}\KeywordTok{aov}\NormalTok{(DV}\OperatorTok{~}\NormalTok{Prime}\OperatorTok{*}\NormalTok{Target}\OperatorTok{+}\NormalTok{cMatch}\OperatorTok{+}\KeywordTok{Error}\NormalTok{(id), }\DataTypeTok{data=}\NormalTok{simdat5))}
\end{Highlighting}
\end{Shaded}

The outputs from both models are shown in Table~\ref{tab:tableIntCon}.

\begin{table}[h]
\begin{center}
\begin{threeparttable}
\caption{\label{tab:tableIntCon}Interaction contrast.}
\begin{tabular}{llllrll}
\toprule
Effect & \multicolumn{1}{c}{$F$} & \multicolumn{1}{c}{$\mathit{df}_1$} & \multicolumn{1}{c}{$\mathit{df}_2$} & \multicolumn{1}{c}{$\mathit{Sum}$ $\mathit{Sq}$} & \multicolumn{1}{c}{$p$} & \multicolumn{1}{c}{$\hat{\eta}^2_G$}\\
\midrule
Prime & 0.14 & 2 & 36 & 694 & .871 & .008\\
Target & 0.14 & 2 & 36 & 694 & .871 & .008\\
Prime $\times$ Target & 1.39 & 4 & 36 & 13889 & .257 & .134\\ \midrule
\ \ \ Matching contrast & 4.44 & 1 & 36 & 11111 & .042 & .110\\
\ \ \ Contrast residual & 0.37 & 3 & 36 & 2778 & .775 & .030\\
\bottomrule
\end{tabular}
\end{threeparttable}
\end{center}
\end{table}

\textcolor{black}{The a priori matching contrast is significant, whereas the omnibus F test for the interaction is not. In addition to the matching contrast, there is also a term "contrast residual", which captures all differences in means of the $3 \times 3$ interaction that are not captured by the matching contrast (and not by the main effects). It provides information about how much of the interaction is explained by our a priori interaction matching contrast, and whether there is additional pattern in the data, that is not yet explained by this a priori contrast.}

\textcolor{black}{Formally speaking, the omnibus F test for the interaction term can be additively decomposed into the a priori matching contrast plus some residual variance associated with the interaction. That is, the $4$ degrees of freedom for the omnibus interaction test are decomposed into $1$ degree of freedom for the matching contrast and $3$ degrees of freedom for the contrast residual: $4\;df = 1\;df + 3\;df$. Likewise, the total sum of squares associated with the omnibus interaction test (Sum Sq $=$} 13889\textcolor{black}{) is decomposed into the sum of squares associated with the matching contrast (Sum Sq $=$} 11111\textcolor{black}{) plus the residual contrast sum of squares (Sum Sq $=$} 2778): 13889 \(=\) 11111 \(+\) 2778. \textcolor{black}{Here, the a priori matching contrast explains a large portion of the variance associated with the interaction of $r^2_{alerting} =$} 11111 \(/\) 13889 \(=\) 0.80\textcolor{black}{, whereas the contribution of the contrast residual is small: $r^2_{alerting} =$} 2778 \(/\) 13889 \(=\) 0.20

\textcolor{black}{Interestingly, the results in Table}~\ref{tab:tableIntCon} \textcolor{black}{also show a significance test for the contrast residual, which here is not significant. This suggests that our a priori contrast provides a good account of the interaction, a situation that} Abelson and Prentice (1997) \textcolor{black}{term a} \emph{canonical} \textcolor{black}{outcome. It suggests that the a priori contrast may be a good account of the true pattern of means. Sometimes, the contrast residual term is significant, clearly suggesting that systematic variance is associated with the interaction beyond what is captured in the a priori contrast. In this situation, $r^2_{alerting}$ for the a priori contrast will often be smaller, whereas $r^2_{alerting}$ for the contrast residuals may be larger.} Abelson and Prentice (1997) \textcolor{black}{term this an} \emph{ecumenical} \textcolor{black}{outcome, which suggests that the a priori contrast seems to be missing some of the systematic variance associated with the interaction. For such} \emph{ecumenical} \textcolor{black}{outcomes,} Abelson and Prentice (1997) \textcolor{black}{suggest it is important for researchers to further explore the data and to search for structure in the residuals to inform future studies and future a priori expectations.}

\textcolor{black}{Even if the contrast residual term is not significant, as is the case here, it can still hide systematic structure. For example, in the present simulations, we assumed that priming did not only facilitate processing of targets from the same category, but also facilitated processing of targets from neighboring categories. We know that response times should gradually increase further towards the diagonal for off-diagonal elements of the $3 \times 3$ design, and that additional structure is actually contained in the present data --- even though the contrast residual is not significant. We could try to directly test for this structure here by formulating a contrast that captures this gradual increase in response times off from the diagonal.}

\textcolor{black}{This procedure of testing an a priori interaction contrast in addition to the automatically generated or "residual" interaction term is advantageous (1) because the automatically generated or "residual" interaction term consumes many degrees of freedom and (2) because it is generally difficult to interpret and not particularly scientifically meaningful itself, except for potentially pointing to residual variance in the interaction that needs to be explained.}

\hypertarget{summary}{%
\section{Summary}\label{summary}}

Contrast coding allows us to implement comparisons in a very flexible and general manner. Specific hypotheses can be formulated, coded in \textcolor{black}{hypothesis} matrices, and converted into contrast matrices. As Maxwell et al. (2018) put it: \enquote{{[}S{]}uch tests may preclude the need for an omnibus interaction test and may be more informative than following the more typical path of testing simple effects} (p.~347).

\textcolor{black}{Barring convergence issues or the use of any kind of regularization, all sensible (non-rank-deficient) contrast coding schemes are essentially fitting the same model, since they are all linear transformations of each other. So the utility of contrast coding is that the researcher can pick and choose how she/he wants to} \emph{interpret} \textcolor{black}{the regression coefficients, in a way that's statistically well-founded (e.g., it does not require running many post-hoc tests) without changing the model (in the sense that the fit in data space does not change) or compromising model fit.}

The generalized inverse provides a powerful tool to convert between \textcolor{black}{contrast} matrices \(X_c\) \textcolor{black}{for experimental designs} used in linear models as well as the associated hypothesis matrices \(H_c = X_c^{inv}\), which define the estimation of regression coefficients and definition of hypothesis tests. Understanding these two representations provides a flexible tool for designing custom contrast matrices and for understanding the hypotheses that are being tested by a given contrast matrix.

\hypertarget{from-omnibus-f-tests-in-anova-to-contrasts-in-regression-models}{%
\subsubsection{From omnibus F tests in ANOVA to contrasts in regression models}\label{from-omnibus-f-tests-in-anova-to-contrasts-in-regression-models}}

ANOVAs can be reformulated as multiple regressions. This has the advantage that hypotheses about comparisons between specific condition means can be tested. In R, contrasts are flexibly defined for factors via contrast matrices. Functions are available to specify a set of pre-defined contrasts including \textsc{treatment contrasts} (\texttt{contr.treatment()}), \textsc{sum contrasts} (\texttt{contr.sum()}), \textsc{repeated contrasts} (\texttt{MASS::contr.sdif()}), and \textsc{polynomial contrasts} (\texttt{contr.poly()}). \textcolor{black}{An additional contrast that we did not cover in detail is the \textsc{Helmert contrast} (\texttt{contr.helmert()}).} These functions generate contrast matrices for a desired number of factor levels. The generalized inverse (function \texttt{ginv()}) is used to obtain hypothesis matrices that inform about how each of the regression coefficients are estimated from condition means and about which hypothesis each of the regression coefficients tests. Going beyond pre-defined contrast functions, the hypothesis matrix is also used to flexibly define custom contrasts tailored to the specific hypotheses of interest. A priori interaction contrasts moreover allow us to formulate and test specific hypotheses about interactions in a design, and to assess their relative importance for understanding the full interaction pattern.\}

\hypertarget{further-readings}{%
\subsubsection{Further readings}\label{further-readings}}

There are many introductions to contrast coding. For further reading we recommend Abelson and Prentice (1997), Baguley (2012) (chapter 15), and Rosenthal et al. (2000). It is also informative to revisit the exchange between Rosnow and Rosenthal (1989), Rosnow and Rosenthal (1996), Abelson (1996), and Petty, Fabrigar, Wegener, and Priester (1996) and the earlier one between Rosnow and Rosenthal (1995) and Meyer (1991). R-specific background about contrasts can be found in section 2.3 of Chambers and Hastie (1992), section 8.3 of Fieller (2016), and section 6.2 of Venables and Ripley (2002). Aside from the default functions in base R, there are also the \textsc{contrast} (Kuhn, Weston, Wing, \& Forester, 2013) and \textsc{multcomp} (Hothorn, Bretz, \& Westfall, 2008) packages in R.

\hypertarget{acknowledgements}{%
\section{Acknowledgements}\label{acknowledgements}}

We thank Thom Baguley, Adrian Staub, and Dave Kleinschmidt for helpful comments. Thanks to Titus von der Malsburg, Amy Goodwin Davies, and Jan Vanhove for suggestions on improvement, and to Atri Vasishth for help in designing Figure \ref{fig:leastsquares}. This work was partly funded by the Deutsche Forschungsgemeinschaft (DFG), Sonderforschungsbereich 1287, project number 317633480 (Limits of Variability in Language), and the Forschergruppe 1617 (grant SCHA 1971/2) at the University of Potsdam, and by the Volkwagen Foundation (grant 89 953).

\newpage

\newpage

\hypertarget{references}{%
\section{References}\label{references}}

\begingroup
\setlength{\parindent}{-0.5in}
\setlength{\leftskip}{0.5in}

\hypertarget{refs}{}
\leavevmode\hypertarget{ref-Abelson1996}{}%
Abelson, R. P. (1996). Vulnerability of contrast tests to simpler interpretations: An addendum to Rosnow and Rosenthal. \emph{Psychological Science}, \emph{7}(4), 242--246.

\leavevmode\hypertarget{ref-AbelsonPrentice1997}{}%
Abelson, R. P., \& Prentice, D. A. (1997). Contrast tests of interaction hypothesis. \emph{Psychological Methods}, \emph{2}(4), 315.

\leavevmode\hypertarget{ref-Baayen2008lme}{}%
Baayen, R. H., Davidson, D. J., \& Bates, D. M. (2008). Mixed-effects modeling with crossed random effects for subjects and items. \emph{Journal of Memory and Language}, \emph{59}(4), 390--412.

\leavevmode\hypertarget{ref-Baguley2012}{}%
Baguley, T. (2012). \emph{Serious stats: A guide to advanced statistics for the behavioral sciences}. Palgrave Macmillan.

\leavevmode\hypertarget{ref-bates2015parsimonious}{}%
Bates, D. M., Kliegl, R., Vasishth, S., \& Baayen, R. H. (2015). Parsimonious mixed models. \emph{arXiv Preprint arXiv:1506.04967}.

\leavevmode\hypertarget{ref-batesetal2014a}{}%
Bates, D. M., Maechler, M., Bolker, B., \& Walker, S. (2014). \emph{lme4: Linear mixed-effects models using Eigen and S4}. Retrieved from \url{http://CRAN.R-project.org/package=lme4}

\leavevmode\hypertarget{ref-Bolker2018}{}%
Bolker, B. (2018). Https://github.com/bbolker/mixedmodels-misc/blob/master/notes/contrasts.rmd.

\leavevmode\hypertarget{ref-ChambersHastie1992}{}%
Chambers, J. M., \& Hastie, T. J. (1992). \emph{Statistical models in S}. New York: Wadsworth \& Brooks/Cole. Retrieved from \url{http://www.stats.ox.ac.uk/pub/MASS4}

\leavevmode\hypertarget{ref-dobson2011introduction}{}%
Dobson, A. J., \& Barnett, A. (2011). \emph{An introduction to generalized linear models}. CRC press.

\leavevmode\hypertarget{ref-fieller2016}{}%
Fieller, N. (2016). \emph{Basics of matrix algebra for statistics with R}. Boca Raton: CRC Press.

\leavevmode\hypertarget{ref-heplots}{}%
Friendly, M. (2010). HE plots for repeated measures designs. \emph{Journal of Statistical Software}, \emph{37}(4), 1--40.

\leavevmode\hypertarget{ref-friendly_matlib}{}%
Friendly, M., Fox, J., \& Chalmers, P. (2018). \emph{Matlib: Matrix functions for teaching and learning linear algebra and multivariate statistics}. Retrieved from \url{https://CRAN.R-project.org/package=matlib}

\leavevmode\hypertarget{ref-hays1973statistics}{}%
Hays, W. L. (1973). \emph{Statistics for the social sciences} (Vol. 410). Holt, Rinehart; Winston New York.

\leavevmode\hypertarget{ref-heister2012analysing}{}%
Heister, J., Würzner, K.-M., \& Kliegl, R. (2012). Analysing large datasets of eye movements during reading. \emph{Visual Word Recognition}, \emph{2}, 102--130.

\leavevmode\hypertarget{ref-HothornBretzWestfall2008}{}%
Hothorn, T., Bretz, F., \& Westfall, P. (2008). Simultaneous inference in general parametric models. \emph{Biometrical Journal}, \emph{50}(3), 346--363.

\leavevmode\hypertarget{ref-kliegl2010linear}{}%
Kliegl, R., Masson, M., \& Richter, E. (2010). A linear mixed model analysis of masked repetition priming. \emph{Visual Cognition}, \emph{18}(5), 655--681.

\leavevmode\hypertarget{ref-Kuhn2013}{}%
Kuhn, M., Weston, S., Wing, J., \& Forester, J. (2013). The contrast package.

\leavevmode\hypertarget{ref-MaxwellDelaney2018}{}%
Maxwell, S. E., Delaney, H. D., \& Kelley, K. (2018). \emph{Designing experiments and analyzing data: A model comparison perspective} (3rd ed.). New York: Routledge.

\leavevmode\hypertarget{ref-Meyer1991}{}%
Meyer, D. L. (1991). Misinterpretation of interaction effects: A reply to Rosnow and Rosenthal. \emph{Psychological Bulletin}, \emph{110}(3), 571--573.

\leavevmode\hypertarget{ref-Petty1996}{}%
Petty, R. E., Fabrigar, L. R., Wegener, D. T., \& Priester, J. R. (1996). Understanding data when interactions are present or hypothesized. \emph{Psychological Science}, \emph{7}(4), 247--252.

\leavevmode\hypertarget{ref-pinheiro2000linear}{}%
Pinheiro, J. C., \& Bates, D. M. (2000). Linear mixed-effects models: Basic concepts and examples. \emph{Mixed-Effects Models in S and S-Plus}, 3--56.

\leavevmode\hypertarget{ref-Rproject}{}%
R Core Team. (2018). \emph{R: A language and environment for statistical computing}. Vienna, Austria: R Foundation for Statistical Computing. Retrieved from \url{https://www.R-project.org/}

\leavevmode\hypertarget{ref-RosenthalRosnowRubin2000}{}%
Rosenthal, R., Rosnow, R. L., \& Rubin, D. B. (2000). \emph{Contrasts and correlations in behavioral research}. New York: Cambridge University Press.

\leavevmode\hypertarget{ref-RosnowRosenthal1995}{}%
Rosnow, R. L., \& Rosenthal, R. (1989). Definition and interpretations of interaction effects. \emph{Psychological Science}, \emph{105}(1), 143--146.

\leavevmode\hypertarget{ref-RosnowRosenthal1989}{}%
Rosnow, R. L., \& Rosenthal, R. (1995). ``Some things you learn aren't so'': Cohen's paradox, Asch's paradigm, and the interpretation of interaction. \emph{Psychological Bulletin}, \emph{6}(1), 3--9.

\leavevmode\hypertarget{ref-RosnowRosenthal1996}{}%
Rosnow, R. L., \& Rosenthal, R. (1996). Contrasts and interactions redux: Five easy pieces. \emph{Psychological Science}, \emph{7}(4), 253--257.

\leavevmode\hypertarget{ref-snedecor1967statistical}{}%
Snedecor, G. W., \& Cochran, W. G. (1967). \emph{Statistical methods}. Ames, Iowa: Iowa State University Press.

\leavevmode\hypertarget{ref-R-MASS}{}%
Venables, W. N., \& Ripley, B. D. (2002). \emph{Modern applied statistics with S PLUS} (Fourth.). New York: Springer. Retrieved from \url{http://www.stats.ox.ac.uk/pub/MASS4}

\endgroup

\clearpage
\makeatletter
\efloat@restorefloats
\makeatother

\begin{appendix}
\hypertarget{app:Glossary}{%
\section{Glossary}\label{app:Glossary}}

\begin{table}
\begin{center}
\caption{Glossary.}
\begin{tabular}{ lp{13cm} }
\hline            
Centered contrasts & \textcolor{black}{The coefficients of a centered contrast sum to zero. If all contrasts (and covariates) in an analysis are centered, then the intercept assesses the grand mean.} \\
Condition mean & \textcolor{black}{The mean of the dependent variable for one factor level or one design cell.} \\
Contrast coefficients & \textcolor{black}{Each contrast consists of a list (vector) of contrast coefficients - there is one contrast coefficient ($c_i$) for each factor level $i$, which encodes how this factor level contributes in computing the comparison between conditions or bundles of conditions tested by the contrast.} \\
Contrast matrix & Contains contrast coefficients either for one factor, or for all factors in the experimental design; Each condition/group or design cell is represented once. The contrast coefficients indicate the numeric predictor values that are to be used as covariates in the linear model to encode differences between factor levels. \\
Design / model matrix & A matrix where each data point yields one row and each column codes one predictor variable in a linear model. Specifically, the first column usually contains a row of 1s, which codes the intercept. Each of the other columns contains one contrast or one covariate. \\
Grand mean & Average of the means of all experimental conditions/groups \\
Hypothesis matrix & Each column codes one condition/group and each row codes one hypothesis. \textcolor{black}{(We mostly show rows of the hypothesis matrix as columns in this paper.)} Each hypothesis consists of weights of how different condition/group means are combined/compared. \\
Orthogonal contrasts & The coefficients of two contrasts are orthogonal to each other if they have a correlation of zero across conditions. They then represent mutually independent hypotheses about the data. \\
\hline  
\end{tabular}
\end{center}
\end{table}

\begin{table}
\begin{center}
\caption{Glossary (continued).}
\begin{tabular}{ lp{13cm} }
\hline    
Regression coefficients & \textcolor{black}{Estimates of how strongly a predictor variable or a certain contrast influences the dependent variable. They estimate the effects associated with a predictor variable or a contrast.} \\
Weights & \textcolor{black}{(in a hypothesis matrix); Each row in the hypothesis matrix encodes the hypothesis tested by one contrast. Each row (i.e., each hypothesis) is defined as a list (vector) of weights, where each factor level / condition mean has one associated weight. The weight weighs the contribution of this condition mean to defining the null hypothesis of the associated contrast. For this, the weight is multiplied with its associated condition mean.} \\
\hline  
\end{tabular}
\end{center}
\end{table}

\newpage

\hypertarget{app:ContrastNames}{%
\section{Contrast names}\label{app:ContrastNames}}

\begin{table}
\begin{center}
\caption{Contrast names.}
\begin{tabular}{ lp{4.5cm}p{8cm} }
\hline            
R-function & contrast names & description \\
\hline            
contr.treatment() & \textbf{treatment contrasts}; dummy contrasts & Compares each of p-1 conditions to a baseline condition. \\
contr.sum() & \textbf{sum contrasts}; deviation contrasts & Compares each of p-1 conditions to the average across conditions. In the case of only 2 factor levels, it tests half the difference between the two conditions. \\
contr.sum()/2 & \textbf{scaled sum contrasts}; effect coding & In the case of two groups, the scaled sum contrast estimates the difference between groups. \\
contr.sdif() & \textbf{repeated contrasts}; successive difference contrasts; sliding difference contrasts; simple difference contrasts & Compares neighbouring factor levels, i.e., 2-1, 3-2, 4-3. \\
contr.poly() & \textbf{polynomial contrasts}; orthogonal polynomial contrasts & Tests e.g., linear or quadratic trends. \\
contr.helmert() & \textbf{Helmert contrasts}; difference contrasts & The first contrast compares the first two conditions. The second contrast compares the average of the first two conditions to the third condition. \\
& \textbf{custom contrasts} & Customly defined contrasts, not from the standard set. \\
& \textbf{effect coding} & All centered contrasts, which give ANOVA-like test results. \\
\hline  
\end{tabular}
\end{center}
\end{table}

\newpage

\hypertarget{app:mixedDesign}{%
\section{The R-function mixedDesign()}\label{app:mixedDesign}}

The function \texttt{mixedDesign()} can be downloaded from the OSF
repository \url{https://osf.io/7ukf6/}. It allows the researcher to
flexibly simulate data from a wide variety of between- and
within-subject designs, which are commonly analyzed with (generalized)
linear models (ANOVA, ANCOVA, multiple regression analysis) or
(generalized) linear mixed-effects models. It can be used to simulate
data for different response distributions such as Gaussian, Binomial,
Poisson, and others. It involves defining the number of between- and
within-subject factors, the number of factor levels, means and standard
deviations for each design cell, the number of subjects per cell, and
correlations of the dependent variable between within-subject factor
levels and how they vary across between-subject factor levels. The
following lines of code show a very simple example for simulating a data
set for an experiment with one between-subject factor with two levels.

\begin{Shaded}
\begin{Highlighting}[]
\NormalTok{(M <-}\StringTok{ }\KeywordTok{matrix}\NormalTok{(}\KeywordTok{c}\NormalTok{(}\DecValTok{300}\NormalTok{, }\DecValTok{250}\NormalTok{), }\DataTypeTok{nrow=}\DecValTok{2}\NormalTok{, }\DataTypeTok{ncol=}\DecValTok{1}\NormalTok{, }\DataTypeTok{byrow=}\OtherTok{FALSE}\NormalTok{))}
\end{Highlighting}
\end{Shaded}

\begin{verbatim}
##      [,1]
## [1,]  300
## [2,]  250
\end{verbatim}

\begin{Shaded}
\begin{Highlighting}[]
\KeywordTok{set.seed}\NormalTok{(}\DecValTok{1}\NormalTok{)}
\NormalTok{simexp <-}\StringTok{ }\KeywordTok{mixedDesign}\NormalTok{(}\DataTypeTok{B=}\DecValTok{2}\NormalTok{, }\DataTypeTok{W=}\OtherTok{NULL}\NormalTok{, }\DataTypeTok{n=}\DecValTok{5}\NormalTok{, }\DataTypeTok{M=}\NormalTok{M,  }\DataTypeTok{SD=}\DecValTok{20}\NormalTok{) }
\KeywordTok{str}\NormalTok{(simexp)}
\end{Highlighting}
\end{Shaded}

\begin{verbatim}
## 'data.frame':    10 obs. of  3 variables:
##  $ B_A: Factor w/ 2 levels "A1","A2": 1 1 1 1 1 2 2 2 2 2
##  $ id : Factor w/ 10 levels "1","2","3","4",..: 1 2 3 4 5 6 7 8 9 10
##  $ DV : num  320 305 291 270 315 ...
\end{verbatim}

\begin{Shaded}
\begin{Highlighting}[]
\NormalTok{simexp}
\end{Highlighting}
\end{Shaded}

\begin{verbatim}
##    B_A id  DV
## 1   A1  1 320
## 2   A1  2 305
## 3   A1  3 291
## 4   A1  4 270
## 5   A1  5 315
## 6   A2  6 228
## 7   A2  7 230
## 8   A2  8 271
## 9   A2  9 266
## 10  A2 10 256
\end{verbatim}

The first and second argument of the \texttt{mixedDesign()} function
specify the numbers and levels of the between- (\texttt{B\ =}) and
within- (\texttt{W\ =}) subject factors. The arguments take a vector of
numbers (integers). The length of the vector defines the number of
factors, and each individual entry in the vector indicates the number of
levels of the respective factor. For example, a \(2 \times 3 \times 4\)
between-subject design with three between-subject factors, where the
first factor has \(2\) levels, the second has \(3\) levels, and the
third has \(4\) levels, would be coded as
\texttt{B\ =\ c(2,3,4),\ W\ =\ NULL}. A \(3 \times 3\) within-subject
design with two within-subjects factors with each \(3\) levels would be
coded as \texttt{B\ =\ NULL,\ W\ =\ c(3,3)}. A 2 (between) \(\times\) 2
(within) \(\times\) 3 (within) design would be coded
\texttt{B\ =\ 2,\ W\ =\ c(2,3)}.

The third argument (\texttt{n\ =}) takes a single integer value
indicating the number of simulated subjects for each cell of the
between-subject design.

The next necessary argument (\texttt{M\ =}) takes as input a matrix
containing the table of means for the design. The number of rows of this
matrix of means is the product of the number of levels of all
between-subject factors. The number of columns of the matrix is the
product of the number of levels of the within-subject factors. In the
present example, the matrix \texttt{M} has just two rows each containing
the mean of the dependent variable for one between-subject factor level,
that is \(300\) and \(250\). Because there is no within-subject factor,
the matrix has just a single column. The second data set simulated in
the paper contains a between-subject factor with three levels.
Accordingly, we specify three means. The \texttt{mixedDesign} function
generates a data frame with one factor F with three levels, one factor
id with 3 (between-subject factor levels) \(\times\) 4 (n) = 12 levels,
and a dependent variable with 12 entries.

\begin{Shaded}
\begin{Highlighting}[]
\NormalTok{(M <-}\StringTok{ }\KeywordTok{matrix}\NormalTok{(}\KeywordTok{c}\NormalTok{(}\DecValTok{500}\NormalTok{, }\DecValTok{450}\NormalTok{, }\DecValTok{400}\NormalTok{), }\DataTypeTok{nrow=}\DecValTok{3}\NormalTok{, }\DataTypeTok{ncol=}\DecValTok{1}\NormalTok{, }\DataTypeTok{byrow=}\OtherTok{FALSE}\NormalTok{))}
\end{Highlighting}
\end{Shaded}

\begin{verbatim}
##      [,1]
## [1,]  500
## [2,]  450
## [3,]  400
\end{verbatim}

\begin{Shaded}
\begin{Highlighting}[]
\KeywordTok{set.seed}\NormalTok{(}\DecValTok{1}\NormalTok{)}
\NormalTok{simdat2 <-}\StringTok{ }\KeywordTok{mixedDesign}\NormalTok{(}\DataTypeTok{B=}\DecValTok{3}\NormalTok{, }\DataTypeTok{W=}\OtherTok{NULL}\NormalTok{, }\DataTypeTok{n=}\DecValTok{4}\NormalTok{, }\DataTypeTok{M=}\NormalTok{M, }\DataTypeTok{SD=}\DecValTok{20}\NormalTok{) }
\KeywordTok{names}\NormalTok{(simdat2)[}\DecValTok{1}\NormalTok{] <-}\StringTok{ "F"}  \CommentTok{# Rename B_A to F(actor)/F(requency)}
\KeywordTok{levels}\NormalTok{(simdat2}\OperatorTok{$}\NormalTok{F) <-}\StringTok{ }\KeywordTok{c}\NormalTok{(}\StringTok{"low"}\NormalTok{, }\StringTok{"medium"}\NormalTok{, }\StringTok{"high"}\NormalTok{)}
\KeywordTok{str}\NormalTok{(simdat2)}
\end{Highlighting}
\end{Shaded}

\begin{verbatim}
## 'data.frame':    12 obs. of  3 variables:
##  $ F : Factor w/ 3 levels "low","medium",..: 1 1 1 1 2 2 2 2 3 3 ...
##  $ id: Factor w/ 12 levels "1","2","3","4",..: 1 2 3 4 5 6 7 8 9 10 ...
##  $ DV: num  497 474 523 506 422 ...
\end{verbatim}

\begin{Shaded}
\begin{Highlighting}[]
\NormalTok{simdat2}
\end{Highlighting}
\end{Shaded}

\begin{verbatim}
##         F id  DV
## 1     low  1 497
## 2     low  2 474
## 3     low  3 523
## 4     low  4 506
## 5  medium  5 422
## 6  medium  6 467
## 7  medium  7 461
## 8  medium  8 450
## 9    high  9 414
## 10   high 10 412
## 11   high 11 402
## 12   high 12 371
\end{verbatim}

For a 2 (between) \(\times\) 2 (within) subject design, we need a 2
\(\times\) 2 matrix. For multiple between-subject factors, \texttt{M}
has one row for each cell of the between-subject design, describing the
mean for this design cell. For example, in case of a
\(2 \times 3 \times 4\) design, the matrix has
\(2 \cdot 3 \cdot 4 = 24\) rows. The same is true for multiple
within-subject factors, where \texttt{M} has one column for each cell of
the between-subject design, e.g., in a 4 (within) \(\times\) 2 (within)
design it has \(4 \cdot 2 = 8\) columns.

The levels of multiple between-subject factors are sorted such that the
levels of the first factor vary most slowly, whereas the levels of the
later factors vary more quickly. E.g.,

\begin{equation}
\begin{bmatrix}
B\_A1 & B\_B1 & B\_C1 \\
B\_A1 & B\_B1 & B\_C2 \\
B\_A1 & B\_B2 & B\_C1 \\
B\_A1 & B\_B2 & B\_C2 \\
B\_A2 & B\_B1 & B\_C1 \\
B\_A2 & B\_B1 & B\_C2 \\
\dots & \dots & \dots \\
\end{bmatrix}
\end{equation}

\noindent The same logic also applies to the levels of the
within-subject factors, which are sorted across columns.

The argument \texttt{SD\ =} takes either a single number specifying the
standard deviation of the dependent variable for all design cells, or it
takes a table of standard deviations represented as a matrix of the same
dimension as the matrix of means, where each element defines the
standard deviation of the dependent variable for the given design cell.

For designs with at least one within-subject factor it is necessary to
define how the dependent variable correlates between pairs of
within-subject factor levels across subjects. For example, when for each
subject in a lexical decision task response times are measured once for
content words and once for function words, yielding two levels of a
within-subject factor, it is necessary to define how the two
measurements correlate: that is, for subjects with a very slow response
time for content words, will the response times for function words also
be very slow, reflecting a high correlation? Or might it be fast as
well, such that the individual's response time for content words is
uninformative for it's response times for function words, reflecting a
correlation close to zero? When \(n_k\) within-subject design cells are
present in a design, the design defines \(n_r = n_k \cdot (n_k -1) / 2\)
correlations. These can be identical across levels of the
between-subject factors, but they can also differ. The optional argument
\texttt{R\ =} defines these correlations. When assuming that all
correlations in a design are identical, then it is possible to define
\texttt{R} as a single number (between -1 and +1) specifying this
correlation. Under the assumption that correlations differ, \texttt{R}
takes as input a list of correlation matrices. This list contains one
correlation matrix for each combination of between-subject factor
levels. Each of these matrices is of dimension \(n_k \cdot n_k\). The
diagonal elements in correlation matrices need to be \(1\), and the
upper and lower triangles must contain the exact same entries and are
thus symmetrical.

\newpage

\hypertarget{app:LinearAlgebra}{%
\section{Matrix notation of the linear model and the generalized matrix
inverse}\label{app:LinearAlgebra}}

This section requires knowledge of some linear algebra; for a review of
the important facts, see Fieller (2016) and Van de Geer (1971). Here, we
briefly show the derivation for estimating regression coefficients in
linear models using matrix notation. The question we address here is:
given the data \(y\) and some predictor(s) \(x\), how to estimate the
parameters \(\beta\)? The next section is adapted from lectures note by
Vasishth (2018), available from https://osf.io/ces89/.

Consider a deterministic function \(\phi(\mathbf{x},\beta)\) which takes
as input some variable values \(x\) and some fixed values \(\beta\). A
simple example would be

\begin{equation}
y = \beta x
\end{equation}

Another example with two fixed values \(\beta_0\) and \(\beta_1\) is:

\begin{equation}
y = \beta_0 + \beta_1 x
\end{equation}

We can rewrite the above equation as follows.

\begin{equation}
\begin{split}
y=& \beta_0 + \beta_1 x\\
=& \beta_0\times 1 + \beta_1 x\\
=& \begin{pmatrix}
1 & x\\
\end{pmatrix}
\begin{pmatrix}
\beta_0 \\
\beta_1 \\
\end{pmatrix}\\
y =& \phi(x, \beta)\\
\end{split}
\end{equation}

In a statistical model, we don't expect an equation like
\(y=\phi(x,\beta)\) to fit all the points exactly. For example, we could
come up with an equation that, given a person's weight, we can compute
their height:

\begin{equation}
\hbox{height} = \beta_0 + \beta_1 \hbox{weight}
\end{equation}

Given any single value of the weight of a person, we will probably not
get a perfectly correct prediction of the height of that person. This
leads us to a non-deterministic version of the above function:

\begin{equation}
y=\phi(x,\beta,\varepsilon)=\beta_0+\beta_1x+\varepsilon
\end{equation}

Here, \(\varepsilon\) is an error random variable which is assumed to
have some probability density function (specifically, the normal
distribution) associated with it. It is assumed to have expectation
(mean) 0, and some standard deviation (to be estimated from the data)
\(\sigma\). We can write this statement in compact form as
\(\varepsilon \sim N(0,\sigma)\).

The \textbf{general linear model} is a non-deterministic function like
the one above (\(^T\) is the transformation operation on a matrix, which
converts the rows of a matrix into the columns, and vice versa):

\begin{equation}
Y=f(x)^T\beta +\varepsilon 
\end{equation}

The matrix formulation will be written as below. \(n\) refers to the
number of data points (that is, \(Y=y_1,\dots,y_n\)), and the index
\(i\) ranges from \(1\) to \(n\).

\begin{equation}
Y = X\beta + \varepsilon \Leftrightarrow y_i = f(x_i)^T \beta + \varepsilon_i, i=1,\dots,n
\end{equation}

To make this concrete, suppose we have three data points, i.e., \(n=3\).
Then, the matrix formulation is

\begin{equation}
\begin{split}
\begin{pmatrix}
y_1 \\
y_2\\
y_3 \\
\end{pmatrix}
=
\begin{pmatrix}
1 & x_1 \\
1 & x_2 \\
1 & x_3 \\
\end{pmatrix}
\begin{pmatrix}
\beta_0 \\
\beta_1 \\
\end{pmatrix}+ \varepsilon\\
Y =& X \beta + \varepsilon \\
\end{split}
\end{equation}

Here, \(f(x_1)^T = (1~x_1)\), and is the first row of the matrix \(X\),
\(f(x_2)^T = (1~x_2)\) is the second row, and \(f(x_3)^T = (1~x_3)\) is
the third row.

The expectation of \(Y\), \(E[Y]\), is \(X\beta\). \(\beta\) is a
\(p\times 1\) matrix, and \(X\), the \textbf{design matrix}, is an
\(n\times p\) matrix.

We now provide a geometric argument for least squares estimation of the
\(\beta\) parameters. When we have a deterministic model
\(y=\phi(f(x)^T,\beta)=\beta_0+\beta_1x\), this implies a perfect fit to
all data points. This is like solving the equation \(Ax=b\) in linear
algebra: we solve for \(\beta\) in \(X\beta=y\) using, e.g., Gaussian
elimination (Lay, 2005).

But when we have a non-deterministic model
\(y=\phi(f(x)^T,\beta,\varepsilon)\), there is no solution. Now, the
best we can do is to get \(Ax\) to be as close an approximation as
possible to b in \(Ax=b\). In other words, we try to minimize the
absolute distance between b and Ax: \(\mid b-Ax\mid\).

The goal is to estimate \(\beta\); we want to find a value of \(\beta\)
such that the observed Y is as close to its expected value \(X\beta\) as
possible. In order to be able to identify \(\beta\) from \(X\beta\), the
linear transformation \(\beta \rightarrow X\beta\) should be one-to-one,
so that every possible value of \(\beta\) gives a different \(X\beta\).
This in turn requires that \(X\) be of full rank \(p\). (Rank refers to
the number of linearly independent columns or rows. The row rank and
column rank of an \(m\times n\) matrix will be the same, so we can just
talk of rank of a matrix. An \(m\times n\) matrix \(X\) with
rank(X)=min(m,n) is called full rank.)

So, if X is an \(n\times p\) matrix, then it is necessary that
\(n\geq p\). There must be at least as many observations as parameters.
If this is not true, then the model is said to be
\textbf{over-parameterized}.

\begin{figure}

{\centering \includegraphics{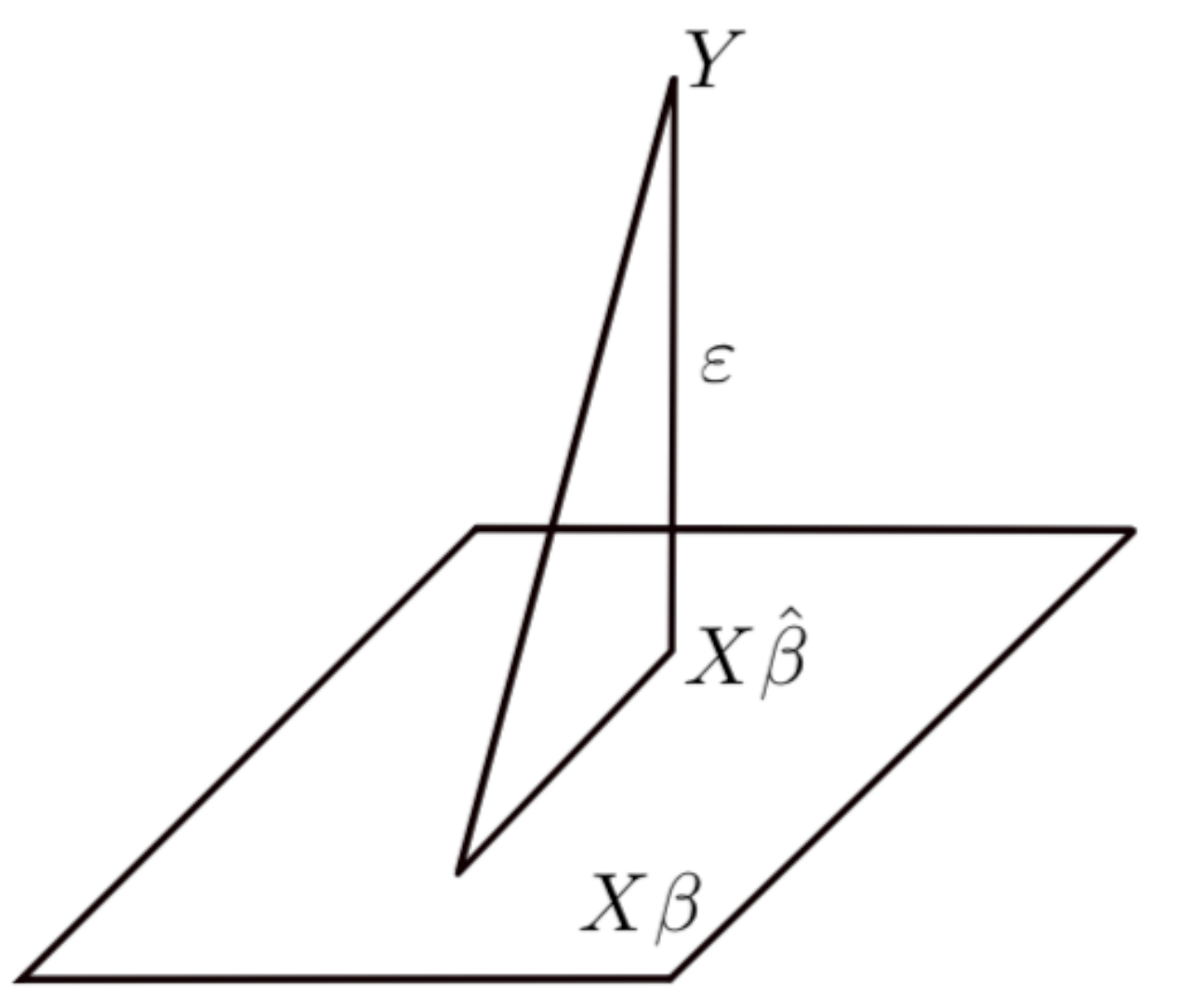} 

}

\caption{Geometric interpretation of least squares.}\label{fig:leastsquares}
\end{figure}

Assuming that \(X\) is of full rank, and that \(n>p\), \(Y\) can be
considered a point in \(n\)-dimensional space and the set of candidate
\(X\beta\) is a \(p\)-dimensional subspace of this space; see Figure
\ref{fig:leastsquares}. There will be one point in this subspace which
is closest to \(Y\) in terms of Euclidean distance. The unique \(\beta\)
that corresponds to this point is the \textbf{least squares estimator}
of \(\beta\); we will call this estimator \(\hat \beta\).

\FloatBarrier

Notice that \(\varepsilon=(Y - X\hat\beta)\) and \(X\beta\) are
perpendicular to each other. Because the dot product of two
perpendicular (orthogonal) vectors is \(0\), we get the result:

\begin{equation}
(Y- X\hat\beta)^T X \beta = 0 \Leftrightarrow (Y- X\hat\beta)^T X = 0 
\end{equation}

Multiplying out the terms, we proceed as follows. One result that we use
here is that \((AB)^T = B^T A^T\) (see Fieller, 2016).

\begin{equation}
\begin{split}
~& (Y- X\hat\beta)^T X = 0  \\
~& (Y^T- \hat\beta^T X^T)X = 0\\
\Leftrightarrow& Y^T X - \hat\beta^TX^T X = 0 \quad  \\
\Leftrightarrow& Y^T X = \hat\beta^TX^T X \\
\Leftrightarrow& (Y^T X)^T = (\hat\beta^TX^T X)^T \\
\Leftrightarrow& X^T Y = X^TX\hat\beta\\
\end{split}
\end{equation}

This gives us the important result:

\begin{equation}
\hat\beta = (X^TX)^{-1}X^T Y
\label{eq:beta}
\end{equation}

This is a key and famous result for linear models. Here, the
transformation of the design matrix \(X\) into \((X^T X)^{-1} X^T\)
plays a central role for estimating regression coefficients. Indeed,
this term \((X^T X)^{-1} X^T\) is exactly the generalized matrix inverse
of the design matrix \(X\), which we can compute in R using the command
\texttt{MASS::ginv()} or \texttt{matlib::Ginv()}.

Conceptually, it converts contrast matrices between two representations,
where one representation defines the design matrix \(X\) used to define
independent variables in the linear regression model \(y = X \beta\),
whereas the other representation, namely the generalized matrix inverse
of the design matrix \(X^{+} = (X^T X)^{-1} X^T = X^{inv}\) defines
weights for how observed data are combined to obtain estimated
regression coefficients via \(\hat{\beta} = X^{+} y\) and for what
formal hypotheses are tested via the contrasts. Given the important
property of the generalized matrix inverse that applying it twice yields
back the original matrix, the generalized matrix inversion operation can
be used to flip back and forth between these two representations of the
design matrix. Fieller (2016) provides a detailed and accessible
introduction to this topic.

As one important aspect, the generalized inverse of the design matrix,
\(X^{+}\), defines weights for each of the \(k\) factor levels of the
independent variables. As estimating coefficients for all \(k\) factor
levels in addition to the intercept is redundant, the design matrix
\(X\) defines coefficients only for \(k - 1\) comparisons. The
generalized matrix inversion can be seen as transforming between these
two spaces (Venables \& Ripley, 2002). A vignette explaining the
generalized inverse is available for the \texttt{Ginv()} function in the
\href{https://cran.r-project.org/web/packages/matlib/vignettes/ginv.html}{\texttt{matlib}
package} (Friendly et al., 2018a).

\newpage

\hypertarget{app:InverseOperation}{%
\section{Inverting orthogonal contrasts}\label{app:InverseOperation}}

\textcolor{black}{When we have a given contrast matrix and we apply the generalized inverse---how exactly does the generalized inverse compute the weights in the hypothesis matrix? That is, what is the formula for computing a single hypothesis weight? We look at this question here for the restricted case that all contrasts in a given design are orthogonal. For this case, we write down the equations for computing each single weight when the generalized inverse is applied. We start out with the definition of the contrast matrix. We use a design with a single factor with three levels with two orthogonal and centered contrasts $x_1$ and $x_2$.
\begin{equation}
X_c =
\left(\begin{array}{cc}
x_{1,1} & x_{2,1} \\
x_{1,2} & x_{2,2} \\
x_{1,3} & x_{2,3}
\end{array}
\right)
\end{equation}
Under the assumption that the contrasts are orthogonal and centered, we obtain the following formulas for the weights:
\begin{equation}
H_c = X_c^{inv} =
\left( \begin{array}{rrr}
\frac{x_{1,1}}{\sum x_1^2} & \frac{x_{1,2}}{\sum x_1^2} & \frac{x_{1,3}}{\sum x_1^2} \\
\frac{x_{2,1}}{\sum x_2^2} & \frac{x_{2,2}}{\sum x_2^2} & \frac{x_{2,3}}{\sum x_2^2}
\end{array} \right)
\end{equation}
That is, here each single weight in the hypothesis matrix is computed as $h_{i,j} = \frac{x_{j,i}}{\sum x_j^2}$.}

\newpage

\hypertarget{information-about-program-version}{%
\section{Information about program
version}\label{information-about-program-version}}

We used R (Version 3.6.0; R Core Team, 2018) and the R-packages
\emph{bindrcpp} (Müller, 2017), \emph{car} (Version 3.0.2; Fox \&
Weisberg, 2011), \emph{dplyr} (Version 0.8.1; Wickham et al., 2017a),
\emph{forcats} (Version 0.4.0; Wickham, 2017a), \emph{ggplot2} (Version
3.1.1; Wickham, 2009), \emph{knitr} (Version 1.22; Xie, 2015),
\emph{MASS} (Version 7.3.51.4; Venables \& Ripley, 2002), \emph{matlib}
(Friendly et al., 2018b), \emph{papaja} (Version 0.1.0.9842; Aust \&
Barth, 2018), \emph{png} (Version 0.1.7; Urbanek, 2013), \emph{purrr}
(Version 0.3.2; Henry \& Wickham, 2017), \emph{readr} (Version 1.3.1;
Wickham et al., 2017b), \emph{sjPlot} (Version 2.6.3; Lüdecke, 2018),
\emph{stringr} (Version 1.4.0; Wickham, 2018), \emph{tibble} (Version
2.1.2; Müller \& Wickham, 2018), \emph{tidyr} (Version 0.8.3; Wickham \&
Henry, 2018), \emph{tidyverse} (Version 1.2.1; Wickham, 2017b),
\emph{XML} (Lang \& CRAN Team, 2017), and \emph{xtable} (Version 1.8.4;
Dahl, 2016) for all our analyses.

\hypertarget{refs}{}
\leavevmode\hypertarget{ref-R-papaja}{}%
Aust, F., \& Barth, M. (2018). \emph{papaja: Create APA manuscripts with
R Markdown}. Retrieved from \url{https://github.com/crsh/papaja}

\leavevmode\hypertarget{ref-R-xtable}{}%
Dahl, D. B. (2016). \emph{Xtable: Export tables to latex or html}.
Retrieved from \url{https://CRAN.R-project.org/package=xtable}

\leavevmode\hypertarget{ref-fieller2016}{}%
Fieller, N. (2016). \emph{Basics of matrix algebra for statistics with
R}. Boca Raton: CRC Press.

\leavevmode\hypertarget{ref-R-car}{}%
Fox, J., \& Weisberg, S. (2011). \emph{An R companion to applied
regression} (Second.). Thousand Oaks CA: Sage. Retrieved from
\url{http://socserv.socsci.mcmaster.ca/jfox/Books/Companion}

\leavevmode\hypertarget{ref-friendly_matlib}{}%
Friendly, M., Fox, J., \& Chalmers, P. (2018a). \emph{Matlib: Matrix
functions for teaching and learning linear algebra and multivariate
statistics}. Retrieved from
\url{https://CRAN.R-project.org/package=matlib}

\leavevmode\hypertarget{ref-R-matlib}{}%
Friendly, M., Fox, J., \& Chalmers, P. (2018b). \emph{Matlib: Matrix
functions for teaching and learning linear algebra and multivariate
statistics}. Retrieved from
\url{https://CRAN.R-project.org/package=matlib}

\leavevmode\hypertarget{ref-R-purrr}{}%
Henry, L., \& Wickham, H. (2017). \emph{Purrr: Functional programming
tools}. Retrieved from \url{https://CRAN.R-project.org/package=purrr}

\leavevmode\hypertarget{ref-R-XML}{}%
Lang, D. T., \& CRAN Team. (2017). \emph{XML: Tools for parsing and
generating xml within r and s-plus}. Retrieved from
\url{https://CRAN.R-project.org/package=XML}

\leavevmode\hypertarget{ref-lay2005linear}{}%
Lay, D. C. (2005). \emph{Linear algebra and its applications} (Third.).
Addison Wesley.

\leavevmode\hypertarget{ref-R-sjPlot}{}%
Lüdecke, D. (2018). \emph{SjPlot: Data visualization for statistics in
social science}. Retrieved from
\url{https://CRAN.R-project.org/package=sjPlot}

\leavevmode\hypertarget{ref-R-bindrcpp}{}%
Müller, K. (2017). \emph{Bindrcpp: An 'rcpp' interface to active
bindings}. Retrieved from
\url{https://CRAN.R-project.org/package=bindrcpp}

\leavevmode\hypertarget{ref-R-tibble}{}%
Müller, K., \& Wickham, H. (2018). \emph{Tibble: Simple data frames}.
Retrieved from \url{https://CRAN.R-project.org/package=tibble}

\leavevmode\hypertarget{ref-R-base}{}%
R Core Team. (2018). \emph{R: A language and environment for statistical
computing}. Vienna, Austria: R Foundation for Statistical Computing.
Retrieved from \url{https://www.R-project.org/}

\leavevmode\hypertarget{ref-R-png}{}%
Urbanek, S. (2013). \emph{Png: Read and write png images}. Retrieved
from \url{https://CRAN.R-project.org/package=png}

\leavevmode\hypertarget{ref-van1971introduction}{}%
Van de Geer, J. P. (1971). \emph{Introduction to multivariate analysis
for the social sciences}. Freeman.

\leavevmode\hypertarget{ref-lmlecturenotesSV}{}%
Vasishth, S. (2018). \emph{Linear modeling: Lecture notes}. Retrieved
from \url{https://osf.io/ces89/}

\leavevmode\hypertarget{ref-R-MASS}{}%
Venables, W. N., \& Ripley, B. D. (2002). \emph{Modern applied
statistics with S PLUS} (Fourth.). New York: Springer. Retrieved from
\url{http://www.stats.ox.ac.uk/pub/MASS4}

\leavevmode\hypertarget{ref-R-ggplot2}{}%
Wickham, H. (2009). \emph{Ggplot2: Elegant graphics for data analysis}.
Springer-Verlag New York. Retrieved from \url{http://ggplot2.org}

\leavevmode\hypertarget{ref-R-forcats}{}%
Wickham, H. (2017a). \emph{Forcats: Tools for working with categorical
variables (factors)}. Retrieved from
\url{https://CRAN.R-project.org/package=forcats}

\leavevmode\hypertarget{ref-R-tidyverse}{}%
Wickham, H. (2017b). \emph{Tidyverse: Easily install and load the
'tidyverse'}. Retrieved from
\url{https://CRAN.R-project.org/package=tidyverse}

\leavevmode\hypertarget{ref-R-stringr}{}%
Wickham, H. (2018). \emph{Stringr: Simple, consistent wrappers for
common string operations}. Retrieved from
\url{https://CRAN.R-project.org/package=stringr}

\leavevmode\hypertarget{ref-R-dplyr}{}%
Wickham, H., Francois, R., Henry, L., \& Müller, K. (2017a).
\emph{Dplyr: A grammar of data manipulation}. Retrieved from
\url{https://CRAN.R-project.org/package=dplyr}

\leavevmode\hypertarget{ref-R-tidyr}{}%
Wickham, H., \& Henry, L. (2018). \emph{Tidyr: Easily tidy data with
'spread()' and 'gather()' functions}. Retrieved from
\url{https://CRAN.R-project.org/package=tidyr}

\leavevmode\hypertarget{ref-R-readr}{}%
Wickham, H., Hester, J., \& Francois, R. (2017b). \emph{Readr: Read
rectangular text data}. Retrieved from
\url{https://CRAN.R-project.org/package=readr}

\leavevmode\hypertarget{ref-R-knitr}{}%
Xie, Y. (2015). \emph{Dynamic documents with R and knitr} (2nd ed.).
Boca Raton, Florida: Chapman; Hall/CRC. Retrieved from
\url{https://yihui.name/knitr/}
\end{appendix}

\end{document}